\documentclass{aa}  
\usepackage{graphicx}
\usepackage{longtable,lscape}
\usepackage{url,fancyvrb}
\usepackage{enumitem}
\usepackage{txfonts}

\usepackage[usenames]{xcolor}
\usepackage[pdftex,bookmarks,colorlinks,breaklinks]{hyperref}
\hypersetup{linkcolor=blue,citecolor=blue,filecolor=black,urlcolor=blue}

\usepackage{natbib}
\usepackage{amssymb,aastexsym,eztean-commands-phd}

\begin{document}

   \titlerunning{Stellar clusters in the inner Galaxy}
   \authorrunning{E. F. E. Morales et al.}

   \title{Stellar clusters in the inner Galaxy and their correlation with cold dust emission\thanks{The full catalog of 695 stellar clusters within the ATLASGAL Galactic range is only available in electronic form at the CDS via anonymous ftp to \texttt{cdsarc.u-strasbg.fr (130.79.128.5)} or via \protect\url{http://cdsweb.u-strasbg.fr/cgi-bin/qcat?J/A+A/}}}

   \author{Esteban F. E. Morales
          \inst{1,2}
          \and
          Friedrich Wyrowski
          \inst{2}
          \and
          Frederic Schuller
          \inst{2,3}
          \and
          Karl M. Menten
          \inst{2}
          }

   \institute{Max-Planck-Institut f{\"u}r Astronomie,
              K\"{o}nigstuhl 17, 69117 Heidelberg, Germany\\
              \email{morales@mpia.de}
          \and
              Max-Planck-Institut f{\"u}r Radioastronomie,
              Auf dem H{\"u}gel 69, 53121 Bonn, Germany
          \and
              European Southern Observatory,
              Alonso de C\'ordova 3107, Casilla 19001, Santiago, Chile
             }

  \date{Received 3 April 2013 / Accepted 16 September 2013}

 
  \abstract
  {Stars are born within dense clumps of giant molecular clouds, constituting young stellar agglomerates known as embedded clusters, which only evolve into bound open clusters under special conditions.}  
   {We statistically study all embedded clusters (ECs) and open clusters (OCs) known so far in the inner Galaxy, investigating particularly their interaction with the surrounding molecular environment and the differences in their evolution.}
   {We first compiled a merged list of 3904 clusters from optical and infrared clusters catalogs in the literature, including 75 new (mostly embedded) clusters discovered by us in the GLIMPSE survey. From this list, 695 clusters are within the Galactic range $|\ell| \le 60\degr$ and $|b| \le 1.5\degr$ covered by the ATLASGAL survey, which was used to search for correlations with submm dust continuum emission tracing dense molecular gas. We defined an evolutionary sequence of five morphological types: deeply embedded cluster (EC1), partially embedded cluster (EC2), emerging open cluster (OC0), OC still associated with a submm clump in the vicinity (OC1), and OC without correlation with ATLASGAL emission (OC2). Together with this process, we performed a thorough literature survey of these 695 clusters, compiling a considerable number of physical and observational properties in a catalog that is publicly available.}
   {We found that an OC defined observationally as OC0, OC1, or OC2 and confirmed as a real cluster is equivalent to the physical concept of OC (a bound exposed cluster) for ages in excess of $\sim 16$~Myr. Some observed OCs younger than this limit can actually be unbound associations.  We found that our OC and EC samples are roughly complete up to $\sim 1$~kpc and $\sim 1.8$~kpc from the Sun, respectively, beyond which the completeness decays exponentially. Using available age estimates for a few ECs, we derived an upper limit of 3~Myr for the duration of the embedded phase. Combined with the OC age distribution within 3~kpc of the Sun, which shows an excess of young exposed clusters compared to a theoretical fit that considers classical disruption mechanisms, we computed an embedded and young cluster dissolution fraction of $88 \pm 8\%$. This high fraction is thought to be produced by several factors and not only by the classical paradigm of fast gas expulsion.}
   {}


   \maketitle

\section{Introduction}
\label{sec:introduction}

Stars form by gravitational collapse of high-density fluctuations in the interstellar molecular gas, which are generated by supersonic turbulent motions \citep[e.g.,][]{Klessen2011-lectures}. Following the nomenclature of \citet{Williams2000}, star formation takes place in dense ($n \gtrsim 10^4$~\dens) \emph{clumps}, which are in turn fragmented into denser ($n \gtrsim 10^5$~\dens) \emph{cores}, in which individual stars or small multiple systems are born. Given this nature of the star formation process, stars are born correlated in space and time, with typical scales of 1~pc and 1~Myr, respectively \citep[see][]{Kroupa2011}, constituting young stellar agglomerates known as \emph{embedded clusters} (ECs). \citet{Bressert2010} studied the spatial distribution of star formation within 500~pc from the Sun and found that, in fact, most of the young stellar objects (YSOs) in their sample are found in regions with number densities greater than $\sim 2\,{\rm pc}^{-3}$, which is more than an order of magnitude higher than the density of field stars in the Galactic disk, $0.13\,{\rm pc}^{-3}$ \citep{Chabrier2001}.

Many of the ECs defined in this way, however, are not gravitationally bound and will not become classical open clusters (OCs), i.e., bound stellar agglomerates that are free of gas and have lifetimes on the order of 100~Myr. It is very important to make the distinction from the start because there is often some confusion about this in the literature. In the definition used throughout this work (see Section~\ref{sec:cluster-definition}), ECs are \emph{not} necessarily the direct progenitors of bound OCs, but just the natural outcome of the star formation process, which is ``clustered'' with respect to the field stars.

The dynamical evolution of an EC is quite complex and can progress in several possible ways, depending on both the characteristics of the recently born stellar population and the physical properties of the parent molecular cloud. A gravitationally unbound molecular cloud or an unbound region of a molecular complex might still be able to form stars in subregions that are locally bound \citep[e.g.,][]{Bonnell2011}, but the resulting EC born there is globally unbound and quickly disperses into the field. On the other hand, within a molecular complex, especially in bound regions, many ECs might merge and form a few large entities \citep{Maschberger2010}. If a certain EC (once born or after merging) manages to remain gravitationally bound in the gas potential, at some point the effect of stellar feedback starts to influence the parent molecular material in the vicinity. These feedback mechanisms include protostellar outflows, evaporation driven by non-ionizing ultraviolet radiation, photoionization and subsequent \ion{H}{ii} region expansion, stellar winds, radiation pressure and, eventually, supernovae. Again, the relative importance of a certain dissipation process is determined by the physical conditions of the system and the environment \citep{Fall2010}.

The energy and momentum introduced by stellar feedback eventually disrupts the clump and sweeps up the residual gas out of the cluster volume. The stars of this emerging cluster are now tied to each other uniquely by the stellar gravitational potential, which might not be sufficient to keep the stars together, so that the cluster dissolves. This is the classical ``infant mortality'' paradigm established by \citet{LadaLada2003}. However, \citet{Kruijssen2011} argue that this effect is only important in low-density regions, and by analyzing the dynamical state of the ECs arising from star formation hydrodynamic simulations, they find that in dense regions the formed clusters are actually bound and even close to virial equilibrium. They propose that those clusters are instead destroyed via tidal shocks from the surrounding dense gas. An alternative disruption mechanism for small-$N$ systems or larger clusters with a hierarchical substructure has recently been studied by \citet{Moeckel2012}, who find through $N$-body simulations that those clusters undergo a quick expansion owing fast internal relaxation. Bound exposed clusters are therefore the few survivors of all these processes and represent the remnants of originally more massive ECs.

The observational study of ECs is fundamental to account for most of the newly formed stellar population in the Galaxy and to investigate the interaction with its parent molecular material through stellar feedback. In the past decade, thanks to the development of all-sky infrared imaging surveys, such as 2MASS and GLIMPSE (see Section~\ref{sec:galactic-surveys}), many new ECs have been discovered in the Galaxy \citep[e.g.,][]{Dutra2003-2mass,Bica2003-2mass,Mercer2005,Borissova2011}, significantly increasing the number of known systems. However, so far there have only been a few systematic studies of the whole current sample of ECs and OCs in a significant fraction of the Galactic plane \citep[e.g.,][]{BonattoBica2011,Kharchenko2012}, and none of these studies has distinguished clearly the embedded population from the OC sample (see below). The main goal of this paper is to fill this gap.

Here, we statistically study all OCs and ECs known so far in the inner Galaxy from different cluster catalogs in the literature, after compiling a considerable number of physical and observational properties of these objects, particularly their degree of correlation with the surrounding molecular environment, if present. We take advantage of the recently completed ATLASGAL submm continuum survey (see Section~\ref{sec:galactic-surveys}), which provides a spatially unbiased view of the distribution of the dense molecular material in the Milky Way. While the distinction of ECs from OCs in these catalogs has primarily been made via correlations with known \ion{H}{ii} regions or nebulae seen in the infrared, the ATLASGAL survey allows us to objectively tell\footnote{In combination with distance information for cases of ambiguous physical relation.} whether or not these objects are associated with dense molecular gas, as well as to possibly detect the presence of stellar feedback via simple morphological criteria.

This paper is organized as follows. In the remainder of this introduction, we shortly present the main observational data and the nomenclature used throughout this work (Sections~\ref{sec:galactic-surveys} and \ref{sec:cluster-definition}, respectively). In Section~\ref{sec:catalogs-summary}, we describe the literature compilation of a merged list of Galactic OCs and ECs, including a new search for ECs we conducted on the GLIMPSE survey; more details about the literature cluster lists used here are given in Appendix~\ref{sec:catalogs-long}. Section~\ref{sec:huge-table} summarizes the construction of an extensive catalog for the cluster sample within the Galactic range covered by ATLASGAL, with many pieces of information, including: characteristics of the submm and mid-infrared emission, correlation with known objects, distances (kinematic and/or stellar), ages, and membership in big molecular complexes. A more detailed description of all the assumptions and procedures made when organizing this information in the catalog is given in Appendix~\ref{sec:huge-table-details}. In Section~\ref{sec:analysis}, we report the results of a statistical analysis performed on this catalog, in which we delineate a morphological evolutionary sequence with decreasing correlation with ATLASGAL emission, classify the sample in ECs and OCs, and separately study their distance distribution, completeness, and age distribution. Finally, Section~\ref{sec:conclusions} summarizes the main conclusions of this paper.

\subsection{Observations: Galactic surveys}
\label{sec:galactic-surveys}

The APEX Telescope Large Area Survey of the Galaxy \citep[ATLASGAL,][]{Schuller2009} is the first unbiased submm continuum survey of the whole inner Galactic disk, covering a total of 360 square degrees of the sky with Galactic coordinates in the range $|\ell| \le 60\degr$ and $|b| \le 1.5\degr$. The observations were carried out at 870~\micron\ using the Large APEX Bolometer Camera \citep[LABOCA;][]{Siringo2009} of the APEX telescope \citep{Guensten2006}, located on Llano de Chajnantor, Chile, at 5100~m of altitude. With an antenna diameter of 12~m, the observations reach an angular resolution\footnote{Throughout this paper, we will refer as angular resolution to the full width at half-maximum of the point-spread function (or telescope beam).} of $19.2''$ at this wavelength. The submm continuum emission mainly represents thermal radiation from cool dust, which is generally optically thin and, therefore, an excellent tracer of the amount of interstellar material on the line of sight. The ATLASGAL survey reaches an average rms noise level of $\sim 50$~mJy/beam, which translates in a $3\sigma$ detection limit of $\sim 4~M_{\sun}$ of total molecular mass (for a nominal distance of 2~kpc and a dust temperature of $T_\rd = 20$~K).

In the infrared, we primarily use two large scale surveys that cover the inner Galactic plane: The Two Micron All Sky Survey \citep[2MASS,][]{Skrutskie2006} which provides near-infrared (NIR) images of the whole sky, in the $J$ (1.25~\micron), $H$ (1.65~\micron), and $K_s$ (2.16~\micron) filters, with an angular resolution of $\sim 2.5''$; and the Galactic Legacy Infrared Mid-Plane Survey Extraordinaire \citep[GLIMPSE,][]{Benjamin2003,Churchwell2009}, which is a set of various mid-infrared (MIR) surveys of the Galactic plane carried out with the InfraRed Array Camera \citep[IRAC,][]{Fazio2004}, on board of the \emph{Spitzer Space Telescope} \citep{Werner2004}. Here we use the GLIMPSE~I and II surveys which cover the $(\ell,b)$ ranges: $5\degr < |\ell| \le 65\degr$ and $|b| \le 1\degr$; $2\degr < |\ell| \le 5\degr$ and $|b| \le 1.5\degr$; $|\ell| \le 2\degr$ and $|b| \le 2\degr$, comprising a total of 274 square degrees. The IRAC camera provides images at four filters centered at wavelengths 3.6, 4.5, 5.6, and 8.0~\micron, with an angular resolution of $\sim 2''$.

The GLIMPSE surveys have revealed very peculiar structures in star-forming regions \citep[a summary is provided in Section~2 of][]{Churchwell2009}. The 8.0~\micron\ filter is particularly useful to detect the presence of bright fluorescent emission from polycyclic aromatic hydrocarbons (PAHs), which are excited by the stellar far ultraviolet (UV) field, but are destroyed by the harder UV radiation present within ionized gas regions. Thus, PAH emission is often observed from \emph{IR bubbles}, which appear projected as ring-like structures and in many cases are tracing molecular material swept up by the expansion of \ion{H}{ii} regions created by the ionizing radiation from massive stars \citep{Deharveng2010}. On the other hand, \emph{infrared dark clouds} (IRDCs), already found in previous MIR surveys, are seen as extinction features against the bright and diffuse mid-infrared Galactic background. They represent the densest and coldest condensations within giant molecular clouds and are the most likely sites of future star formation.

For a few regions within the ATLASGAL Galactic range not covered by the GLIMPSE survey, we use data from the Wide-field Infrared Survey Explorer \citep[WISE,][]{Wright2010}, which mapped the entire sky in four infrared bands centered
at 3.4, 4.6, 12, and 22~\micron, with an angular resolution of $\sim 6''$ in the first three bands. Despite the lower sensitivity and coarser resolution as compared with GLIMPSE, bright PAH emission and prominent IRDCs can still be identified in the WISE images, specially at 12~\micron\ (see Section~\ref{sec:MIR-morphology}).

\subsection{``Stellar cluster'' definitions}
\label{sec:cluster-definition}

In this paper, we define:
\begin{itemize}[label=\textbullet]
 \item an \emph{embedded cluster} (EC) as any stellar group recently born and still containing an important fraction of residual gas within and surrounding its volume, keeping in mind that it may never become a bound open cluster on its own. Since star formation takes place in molecular clouds, this definition is equivalent to the concept of a \emph{correlated star formation event} introduced by \citet{Kroupa2011}; we keep the term ``cluster'' in order to match older designations in the literature.
 \item an \emph{open cluster} (OC) as any agglomerate of spatially correlated stars, and relatively free of the remaining gas. We use this observational definition of OC (see also Section~\ref{sec:classification-oc-ec}) in order to account for those objects that observationally appear like classical OCs, but whose dynamical state is unknown, in some cases they can actually be gravitationally unbound.
 \item a \emph{physical OC} as a gravitationally bound OC (i.e., a classical OC).
 \item an \emph{association} as an unbound OC.
\end{itemize}
In this work, we sometimes use the term ``star clusters'' generically for all the classes defined above, especially when concerning observations. Bound, exposed star clusters, however, will be always be referred to explicitly as \emph{physical OCs}.

\section{Compilation of cluster lists}
\label{sec:catalogs-summary}

Although the number of known OCs and ECs in the Galaxy has considerably increased over the last years, the current cluster sample is still far from being complete. As we discuss in Section~\ref{sec:completeness}, the detection of a stellar cluster in the inner Galactic plane is particularly difficult, due to the high extinction and the crowded stellar background, making the cluster sample severely incomplete for distances larger than a few kpc from the Sun. If we are able to quantify this incompleteness, however, all the statistical results can properly be corrected, as we do in this work. Of course, the more complete the cluster sample, the smaller the corresponding uncertainties.

We thus performed an extensive compilation of all Galactic star cluster catalogs from the literature. For completeness, this compilation was initially not restricted to the ATLASGAL Galactic range; we only did it afterwards for the comparison with ATLASGAL emission and all the subsequent analysis. The catalogs are listed in the first three columns of Table~\ref{tab:catalogs}, where we give, respectively, an ID used throughout this work, the corresponding reference, and its category according to the wavelength at which the clusters are detected: \emph{optical}, \emph{NIR} or \emph{MIR}. Optical clusters are taken mostly from the current version (3.1, from November, 2010) of the catalog by \citet{Dias2002}. NIR cluster catalogs are compilations, or lists from visual and automated searches mainly performed on the 2MASS survey. MIR clusters represent the objects detected by \citet{Mercer2005} in the GLIMPSE data, and the new clusters discovered by us using a different search method on the same survey, which were missing in the \citet{Mercer2005} list (see Section~\ref{sec:newglimpse}). In our total sample, we also included individual star clusters from the literature not listed in the previous catalogs (referred to as ``Not cataloged clusters'' in Table~\ref{tab:catalogs}). A more detailed description of the diverse catalogs and references used to construct our cluster sample is given in Appendix~\ref{sec:catalogs-long}. This literature compilation has been updated till August, 2011.

Since we are dealing with different cluster catalogs which were constructed independently, a specific object can be present in more than one list. We therefore implemented a simple merging procedure to finally have an unique sample of stellar clusters. The first condition to identify one repetition, i.e., the same object in two different catalogs, was that the angular distance between the two given center positions were less than both listed angular diameters. We checked all merged objects under this criterion looking for the corresponding cluster names, when available, and confirmed a repetition when the names coincided. Otherwise (names not available or different), two clusters were considered the same object when the angular distance was less than both angular radii, which were also required to agree within a factor of 5. The last condition was imposed to account for the case when a compact infrared cluster shares the same field of view of a (different) optical cluster with a large angular size. This cross-identification process was not intended to be perfect, but good enough to not affect the statistical results of the whole cluster sample. Within the ATLASGAL Galactic range, a much more thorough revision was done (see Section~\ref{sec:huge-table}), further refining the cross-identifications, and even recognizing a few duplications and spurious clusters which were excluded from the final sample (see Section~\ref{sec:spurious}).

\begin{table*}[!t]
\renewcommand{\arraystretch}{1.1}
\caption{Number of clusters for every catalog used in this work.}
\label{tab:catalogs}
\centering
\begin{tabular}{lllrr|rr|rr}
\hline\hline
\multicolumn{2}{l}{Catalog}  & Type & \multicolumn{2}{c}{Total} & 
\multicolumn{2}{c}{ATLASGAL} & \multicolumn{2}{c}{ATLASGAL} \\
\multicolumn{2}{c}{     } &  & \multicolumn{2}{c}{     } & 
\multicolumn{2}{c}{range\tablefootmark{a}}
   & \multicolumn{2}{c}{emission\tablefootmark{b}} \\
ID & Reference & & $N_{\rm cl}$ & $N_{\rm cl}^*$
                 & $N_{\rm cl}$ & $N_{\rm cl}^*$
                 & $N_{\rm cl}$ & $N_{\rm cl}^*$\\
\hline
01 & \citet[][ver. 3.1]{Dias2002}\tablefootmark{c} & \emph{Optical} & 2117 & 2117 & 216 & 216 &  29 &  29 \\
02 & \citet{Kronberger2006}\tablefootmark{d}       & \emph{Optical} &  239 &  130 &  29 &  11 &   5 &   4 \\
03 & \citet{DutraBica2000}                         & \emph{NIR}     &   22 &    8 &  18 &   8 &   8 &   2 \\
04 & \citet{Bica2003-lit}\tablefootmark{e}         & \emph{NIR}     &  275 &  264 &  30 &  28 &  28 &  26 \\
05 & \citet{Dutra2003-2mass}                       & \emph{NIR}     &  174 &  167 &  81 &  80 &  78 &  77 \\
06 & \citet{Bica2003-2mass}                        & \emph{NIR}     &  163 &  155 &  69 &  68 &  63 &  62 \\
07 & \citet{LadaLada2003}                          & \emph{NIR}     &   76 &   12 &   4 &   0 &   4 &   0 \\
08 & \citet{Porras2003}                            & \emph{NIR}     &   73 &   21 &   0 &   0 &   0 &   0 \\
09 & \citet{Mercer2005}                            & \emph{MIR}     &   90 &   86 &  83 &  81 &  55 &  54 \\
10 & \citet{Kumar2006}                             & \emph{NIR}     &   54 &   20 &   0 &   0 &   0 &   0 \\
11 & \citet{Froebrich2007}\tablefootmark{d}        & \emph{NIR}     &  998 &  676 &  44 &  21 &   2 &   0 \\
12 & \citet{Faustini2009}                          & \emph{NIR}     &   23 &   16 &   9 &   9 &   9 &   9 \\
13 & \citet{Glushkova2010}                         & \emph{NIR}     &  194 &   32 &  12 &   4 &   1 &   0 \\
14 & \citet{Borissova2011}                         & \emph{NIR}     &   96 &   96 &  85 &  85 &  65 &  65 \\
15 & Not cataloged (NIR)\tablefootmark{f}          & \emph{NIR}     &   26 &   26 &  12 &  12 &  10 &  10 \\
16 & Not cataloged (MIR)\tablefootmark{f}          & \emph{MIR}     &    3 &    3 &   3 &   3 &   0 &   0 \\
17 & New GLIMPSE (this work)                       & \emph{MIR}     &  111 &   75 & 103 &  69 &  94 &  67 \\
\hline
   & Total \emph{Optical}                          &                & 2247 & 2247 & 227 & 227 &  33 &  33 \\
   & Total \emph{NIR}                              &                & 1950 & 1493 & 356 & 315 & 265 & 251 \\
   & Total \emph{MIR}                              &                &  197 &  164 & 182 & 153 & 144 & 121 \\
\hline
\end{tabular}
\tablefoot{
$N_{\rm cl}$ is the absolute number of entries in every catalog, whereas $N_{\rm cl}^*$, for a given reference, is the number of objects not present in any of the catalogs listed before it (see Section~\ref{sec:catalogs-summary} for details). Absolute numbers for whole categories ($N_{\rm cl}$ for last three lines) take into account repetitions inside the category, naturally. All numbers in this table are after removing a few spurious objects (listed in Table~\ref{tab:spurious}).\\
\tablefoottext{a}{Clusters with galactic coordinates within the ATLASGAL range: $|\ell| \le 60\degr$ and $|b| \le 1.5\degr$.}
\tablefoottext{b}{Clusters associated with ATLASGAL emission (see Section~\ref{sec:evolutionary-sequence}).}
\tablefoottext{c}{Version 3.1 is from November, 2010.}
\tablefoottext{d}{Only provides cluster \emph{candidates}.}
\tablefoottext{e}{Includes clusters from \citet{DutraBica2001}.}
\tablefoottext{f}{Individual clusters studied in the infrared (NIR or MIR), which are not listed in the previous lists; references for objects within the ATLASGAL range are given in our on-line catalog.}
}
\end{table*}

In Table~\ref{tab:catalogs}, for a given reference, we represent as $N_{\rm cl}$ the absolute (original) number of clusters in the catalog, whereas $N_{\rm cl}^*$ is the number of different entries with respect to all catalogs listed before it (i.e., after merging). The optical catalogs were put first, so that any cluster visible in the optical is considered an \emph{optical} cluster. The infrared lists (including the \emph{NIR} and \emph{MIR} clusters) were positioned afterwards in chronological order, and therefore following roughly the discovery time. Absolute and after-merging numbers are presented for the total sky range of every list, the ATLASGAL Galactic range ($|\ell| \le 60\degr$ and $|b| \le 1.5\degr$), and finally for only those associated with ATLASGAL emission according to the criterion explained in Section~\ref{sec:evolutionary-sequence}. We warn that the number of clusters given there are after removing a few spurious objects and globular clusters (listed in Table~\ref{tab:spurious}).

After cross-identifications, we ended up with a final sample of 3904 stellar clusters, of which 2247 are \emph{optical}, 1493 \emph{NIR}, and 164 \emph{MIR} clusters. Taking into account the repetitions within each category, but not between them, the numbers of objects are 2247 for \emph{optical}, 1950 for \emph{NIR}, and 197 for \emph{MIR}. Note that the low number of \emph{MIR} clusters is due to the confined Galactic range of the GLIMPSE survey; actually, when only considering the ATLASGAL range, which is similar to the GLIMPSE range, the numbers of objects are of the same order for the different categories: 227 \emph{optical}, 315 \emph{NIR}, and 153 \emph{MIR} clusters, after merging.

As argued in Section~\ref{sec:spurious}, for ECs (as defined in this work) we expect a minimal contamination by spurious detections, whereas for OCs that have not been confirmed by follow-up studies, we estimate a spurious contamination rate of $\sim 50\%$, following \citet{Froebrich2007}.

\subsection{New search for ECs in GLIMPSE}
\label{sec:newglimpse}

The GLIMPSE on-line viewer\footnote{\url{http://www.alienearths.org/glimpse/glimpse.php}} from the Space Science Institute represents a very useful tool to quickly examine color images constructed from data collected in the four 3.6, 4.5, 5.8 and 8.0~\micron\ IRAC filters, of the whole survey. By inspecting some specific regions with this viewer, we noticed that some heavily ECs are still missing in the \citet{Mercer2005} list. An EC consists mostly of YSOs, which are intrinsically redder than field stars due to thermal emission from circumstellar dust, so that they are distinguished from background/foreground stars mainly by their red colors. Such a cluster would therefore produce a clearer spatial overdensity of stars in a point source catalog previously filtered by a red-color criterion, and would be more likely missed in a search of overdensities considering the totality of point sources, due the high number of field stars. We believe that this is the principal reason which would explain the incompleteness of the \citet{Mercer2005} catalog. 

We then implemented a very simple automated algorithm using the GLIMP\-SE point source catalog to find the locations of EC candidates. First, we selected all point sources satisfying a red-color criterion: $[4.5] - [8.0] \ge 1$, following \citet{Robitaille2008}, who applied this condition to create their catalog of GLIMPSE intrinsically red sources. As already explained in that work, the use of these specific IRAC bands is supported by the fact that the interstellar extinction law is approximately flat between 4.5 and 8.0~\micron, and therefore the contamination by extinguished field stars in this selection is reduced compared to other red-color criteria. By applying this condition to the entire GLIMPSE catalog, 268\,513 sources were selected. We did not impose the additional brightness and quality restrictions used by \citet{Robitaille2008} because we favor the number of sources (and therefore higher sensitivity to possible YSO overdensities) rather than strict completeness and photometric reliability, which are not needed to only detect the locations of potential ECs. With the 268\,513 selected sources, a stellar surface density map was constructed by counting the number of sources within boxes of 0.01$\degr$ ($=36\arcsec$), in steps of 0.002$\degr$ ($=7.2\arcsec$). This significant oversampling was adopted in order to detect density enhancements that would have fallen into two or more boxes if we had used not overlapping bins. The bin size correspond to the typical angular dimension of some ECs serendipitously found using the on-line GLIMPSE viewer. To account for larger overdensities, a second stellar density map was produced with a bin size of 0.018$\degr$ ($=64.8\arcsec$), using the same step size of 0.002$\degr$.

The red-source density maps were checked in a test field, and we found that thresholds of 5 sources for the small bin, and 7 sources for the large bin, are needed to detect the positions of all clusters which can be identified by-eye using the GLIMPSE on-line viewer within that area, although at the same time these low thresholds yield the detection of many spurious red-source overdensities that do not contain clusters. We decided to keep these thresholds in order not to miss any real cluster that might have a low number of members listed in the point source catalog, and perform a visual inspection of the images after the automated search to filter all spurious detections. It was also noticed that using the GLIMPSE point source archive instead of the catalog is roughly equivalent to utilizing the catalog with a lower threshold, so as long as we choose a correct threshold, the use of the more reliable GLIMPSE catalog (with respect to the archive) is justified. Within the whole GLIMPSE area, we detected 702 independent positions of overdensities (bins containing not-intersecting subsets of red sources), corresponding to 172 bins of 36$\arcsec$ with densities $\geq 5$~sources/bin, 195 bins of 64.8$\arcsec$ with densities $\geq 7$~sources/bin, and 335 locations satisfying the thresholds for both bin sizes. It should be noted that since the red-color criterion produced density maps with low crowding and therefore the local background density is always close to zero, a more sophisticated algorithm is not needed. In fact, the red-source density maps have a mean and a standard deviation of 0.039 and 0.21 sources/bin for the small bin, and 0.13 and 0.43 sources/bin for the large bin, which means that the used thresholds are above the $15\sigma$ level. Again, we emphasize that the automated search was only used to find possible locations of ECs; we did not intend to catch the complete YSO population for a given cluster in this process.

\begin{longtab}
\renewcommand{\arraystretch}{1.1}
\begin{longtable}{lrrccrrcl}
\caption{New GLIMPSE stellar clusters identified in this work.}
\label{tab:newglimpse}\\
\hline\hline
G3CC & $\ell$      & $b$      & $\alpha$ & $\delta$ & Diam.       & $N_{\rm circ}$ & Det. & Flags \\
     & ($\degr$)& ($\degr$)& (J2000)  & (J2000)  & ($\arcsec$) &                &      &       \\
(1)  &   (2)    &    (3)   &   (4)    &   (5)    &   (6)       & (7)            &  (8) &  (9)  \\
\hline
\endfirsthead
\caption[]{continued.}\\
\hline\hline
G3CC & $l$      & $b$      & $\alpha$ & $\delta$ & Diam.       & $N_{\rm circ}$ & Det. & Flags \\
     & ($\degr$)& ($\degr$)& (J2000)  & (J2000)  & ($\arcsec$) &                &      &       \\
(1)  &   (2)    &    (3)   &   (4)    &   (5)    &   (6)       & (7)            &  (8) &  (9)  \\
\hline
\endhead
\hline
\endfoot
 1 & 295.151 & $-$0.587 & 11:43:24.9 & $-$62:25:36 &  98 &       16 & A &           C8,E8,S \\
 2 & 299.014 &    0.128 & 12:17:24.9 & $-$62:29:04 &  60 &        4 & V &              B,E8 \\
 3 & 299.051 &    0.181 & 12:17:47.9 & $-$62:26:12 &  81 &       14 & A &                C8 \\
 4\tablefootmark{a} & 299.337 & $-$0.319 & 12:19:43.1 & $-$62:58:08 &  51 &        9 & A &             BR,E8 \\
 5 & 300.913 &    0.887 & 12:34:16.2 & $-$61:55:04 &  76 &       10 & A &             C8,E8 \\
 6 & 301.643 & $-$0.240 & 12:40:02.6 & $-$63:05:01 &  67 &        9 & A &           DC,E8,S \\
 7 & 301.947 &    0.313 & 12:42:53.7 & $-$62:32:32 &  65 &       12 & A &                E8 \\
 8 & 303.927 & $-$0.687 & 13:00:22.2 & $-$63:32:30 & 107 &       14 & A &             C8,E8 \\
 9 & 304.002 &    0.464 & 13:00:40.3 & $-$62:23:17 &  82 & $\cdots$ & A &           BR,E8,S \\
10 & 304.887 &    0.635 & 13:08:12.3 & $-$62:10:23 &  41 &        7 & A &             DC,E4 \\
11 & 307.083 &    0.528 & 13:26:58.8 & $-$62:03:25 &  71 &        8 & A &        C8,DC,E8,S \\
12 & 309.421 & $-$0.621 & 13:48:38.1 & $-$62:46:11 &  48 &       10 & A &                DC \\
13 & 309.537 & $-$0.742 & 13:49:51.6 & $-$62:51:42 &  38 &        7 & A &          C8,DC,E8 \\
14 & 309.968 &    0.302 & 13:51:25.6 & $-$61:44:51 &  40 &        6 & A &             DC,E8 \\
15 & 309.996 &    0.507 & 13:51:15.8 & $-$61:32:30 &  88 &        8 & A &             E8,DC \\
16 & 313.762 & $-$0.860 & 14:24:58.6 & $-$61:44:56 &  80 &       15 & A & BR,C8,DC,E4,E8,U8 \\
17 & 314.203 &    0.213 & 14:25:15.4 & $-$60:35:22 &  86 &       12 & A &          C8,E8,U8 \\
18 & 314.269 &    0.092 & 14:26:06.6 & $-$60:40:43 &  87 &        8 & A &     C8,DC,E8,S,V2 \\
19 & 317.466 & $-$0.401 & 14:51:19.3 & $-$59:50:46 &  45 &        7 & A &          DC,E4,E8 \\
20 & 317.884 & $-$0.253 & 14:53:45.6 & $-$59:31:34 &  74 &       15 & A &          DC,E4,E8 \\
21 & 318.049 &    0.088 & 14:53:42.2 & $-$59:08:49 &  88 &       20 & A &          C8,DC,U8 \\
22 & 318.777 & $-$0.144 & 14:59:33.5 & $-$59:00:59 & 105 &        8 & A &           B,E8,V2 \\
23 & 319.336 &    0.912 & 14:59:31.0 & $-$57:49:18 &  65 &       12 & A &                   \\
24\tablefootmark{a} & 321.937 & $-$0.006 & 15:19:43.2 & $-$57:18:04 &  33 &        9 & A &          C8,DC,E8 \\
25 & 321.952 &    0.014 & 15:19:44.6 & $-$57:16:35 &  37 &       10 & A &                E8 \\
26 & 326.476 &    0.699 & 15:43:18.0 & $-$54:07:23 &  81 &       12 & A &       C8,DC,E4,U8 \\
27 & 326.796 &    0.385 & 15:46:20.3 & $-$54:10:35 &  54 &       10 & A &             DC,E4 \\
28 & 328.165 &    0.587 & 15:52:42.6 & $-$53:09:48 &  31 &        6 & A &             E4,U8 \\
29\tablefootmark{a} & 328.252 & $-$0.531 & 15:57:58.9 & $-$53:58:02 &  58 &        9 & A &       C8,DC,E4,E8 \\
30 & 328.809 &    0.635 & 15:55:48.4 & $-$52:43:00 &  82 &        9 & V &          C8,DC,E4 \\
31 & 329.184 & $-$0.313 & 16:01:47.0 & $-$53:11:40 &  73 &        8 & A &          DC,E4,U8 \\
32\tablefootmark{a} & 330.031 &    1.043 & 16:00:09.4 & $-$51:36:52 &  56 &        6 & A &           DC,E8,S \\
33 & 335.061 & $-$0.428 & 16:29:23.5 & $-$49:12:25 &  63 &        6 & A &          C8,DC,E4 \\
34 & 337.153 & $-$0.393 & 16:37:48.5 & $-$47:38:53 &  49 &        4 & A &          DC,U8,V2 \\
35 & 338.396 & $-$0.406 & 16:42:43.2 & $-$46:43:36 &  65 &        8 & A &          C8,DC,E4 \\
36 & 338.922 &    0.390 & 16:41:15.7 & $-$45:48:23 &  97 &       11 & A &           C8,E8,S \\
37 & 338.930 & $-$0.495 & 16:45:08.6 & $-$46:22:50 &  80 &       11 & A &       C8,DC,E8,U8 \\
38 & 339.584 & $-$0.127 & 16:45:59.1 & $-$45:38:44 &  53 &        9 & A &          DC,E4,E8 \\
39\tablefootmark{a} & 344.221 & $-$0.569 & 17:04:06.6 & $-$42:18:57 &  51 &       11 & A &          BR,E4,E8 \\
40 & 344.996 & $-$0.224 & 17:05:09.7 & $-$41:29:26 &  75 &       15 & A &          DC,E4,U8 \\
41 & 347.883 & $-$0.291 & 17:14:27.3 & $-$39:12:35 &  62 &        6 & V &             C8,E8 \\
42 & 348.180 &    0.483 & 17:12:08.1 & $-$38:30:54 &  38 &        7 & A &                E8 \\
43 & 348.584 & $-$0.920 & 17:19:11.6 & $-$39:00:08 &  52 &       10 & A &             C8,E4 \\
44 & 350.105 &    0.085 & 17:19:26.7 & $-$37:10:48 & 167 &       25 & A &          C8,E8,V2 \\
45 & 350.930 &    0.753 & 17:19:04.7 & $-$36:07:16 &  90 &       14 & A &        C8,DC,E8,S \\
46 & 351.776 & $-$0.538 & 17:26:43.1 & $-$36:09:18 &  93 &       14 & A &       C8,DC,E4,E8 \\
47 & 352.489 &    0.797 & 17:23:15.6 & $-$34:48:53 &  84 &        7 & A &             C8,E8 \\
48 & 358.386 & $-$0.482 & 17:43:37.5 & $-$30:33:51 &  57 &        5 & A &    C8,DC,E4,E8,V2 \\
49 &   0.675 & $-$0.046 & 17:47:23.7 & $-$28:22:59 & 140 &       23 & A &           C8,E8,S \\
50 &   4.001 &    0.335 & 17:53:34.5 & $-$25:19:57 &  56 &       12 & A &          BR,C8,E8 \\
51 &   5.636 &    0.239 & 17:57:33.9 & $-$23:58:05 &  65 &        7 & A &          C8,DC,E8 \\
52 &   6.797 & $-$0.256 & 18:01:57.6 & $-$23:12:26 &  50 &       11 & A &       C8,DC,E4,U8 \\
53 &   8.492 & $-$0.633 & 18:06:59.3 & $-$21:54:55 & 126 &       28 & A &              DC,S \\
54 &   9.221 &    0.166 & 18:05:31.3 & $-$20:53:21 &  42 &        7 & A &             DC,E8 \\
55 &  14.113 & $-$0.571 & 18:18:12.4 & $-$16:57:18 &  57 &        9 & A &             DC,E8 \\
56 &  14.341 & $-$0.642 & 18:18:55.2 & $-$16:47:15 & 124 &       15 & A &       C8,DC,E4,E8 \\
57 &  17.168 &    0.815 & 18:19:08.4 & $-$13:36:29 &  61 &       12 & A &                DC \\
58 &  25.297 &    0.309 & 18:36:20.5 & $-$06:38:57 &  39 &        8 & A &                E8 \\
59 &  26.507 &    0.284 & 18:38:40.0 & $-$05:35:06 &  49 &        7 & A &             C8,DC \\
60 &  31.158 &    0.047 & 18:48:02.1 & $-$01:33:26 &  50 &        8 & A &                E8 \\
61 &  34.403 &    0.229 & 18:53:18.4 &    01:24:47 &  91 &        8 & A &             DC,E4 \\
62 &  39.497 & $-$0.993 & 19:07:00.0 &    05:23:05 &  53 &        7 & A &             C8,V2 \\
63 &  43.040 & $-$0.451 & 19:11:38.7 &    08:46:40 &  52 &        6 & A &          C8,E4,E8 \\
64 &  43.893 & $-$0.785 & 19:14:26.8 &    09:22:44 &  63 &        7 & A &             C8,E8 \\
65 &  47.874 &    0.309 & 19:18:04.1 &    13:24:41 &  68 &       11 & A &             C8,E8 \\
66 &  49.912 &    0.369 & 19:21:47.7 &    15:14:20 &  55 &       11 & V &          BR,C8,E8 \\
67 &  50.053 &    0.064 & 19:23:11.3 &    15:13:10 & 107 &       14 & A &              DC,S \\
68 &  52.570 & $-$0.955 & 19:31:54.7 &    16:56:44 &  44 &        9 & A &             E4,E8 \\
69 &  53.147 &    0.071 & 19:29:18.0 &    17:56:41 & 119 &       13 & A &           C8,DC,S \\
70 &  53.237 &    0.056 & 19:29:32.3 &    18:00:57 &  76 &       19 & A &              DC,S \\
71 &  56.961 & $-$0.234 & 19:38:16.7 &    21:08:02 &  58 &        8 & A &             C8,E8 \\
72 &  58.471 &    0.432 & 19:38:58.4 &    22:46:32 &  73 &       10 & A &             C8,E8 \\
73 &  59.783 &    0.071 & 19:43:09.9 &    23:44:14 & 120 &       11 & A &       C8,E4,E8,V2 \\
74 &  62.379 &    0.298 & 19:48:02.4 &    26:05:51 &  47 &        7 & A &                   \\
75 &  64.272 & $-$0.425 & 19:55:09.4 &    27:21:18 &  55 &       10 & A &                BR \\
\end{longtable}
\tablefoot{
Units of right ascension are hours, minutes, and seconds, and units of declination are degrees, arcminutes, and arcseconds. Column 6 gives the estimated angular diameter. Column 7 gives the estimated number of stellar members within the assumed radius, considered as a lower limit due to possible non-detection of low mass stars. Column 8 indicates the detection method: automated search (A), or on-line viewer (V). Column 9 lists different flags determined after visual inspection of the GLIMPSE three-color images, indicating: association with extended 8.0~\micron\ emission (E8) or localized diffuse 4.5~\micron\ emission (E4); cluster embedded in an infrared dark cloud (DC); cluster composed of red sources and additional bright normal stars (BR); cluster composed of bright normal stars alone (B); presence of additional probable YSOs, identified as sources uniquely detected at 8.0~\micron\ (U8), or compact 8.0~\micron\ objects not listed in the point source catalog or archive (C8); sparse, not centrally condensed morphology (S); cluster identified by-eye in a nearby location of an automatically detected overdensity, but not exactly at the same position (V2).
\tablefoottext{a}{Contained within a cluster from the \citet{Majaess2013} catalog: clusters with IDs 147, 161, 167, 169, and 177 contain clusters G3CC 4, 24, 29, 32, and 39, respectively.}

}
\end{longtab}

\begin{figure*}[!t]
\centering
\includegraphics[width=0.9\textwidth]{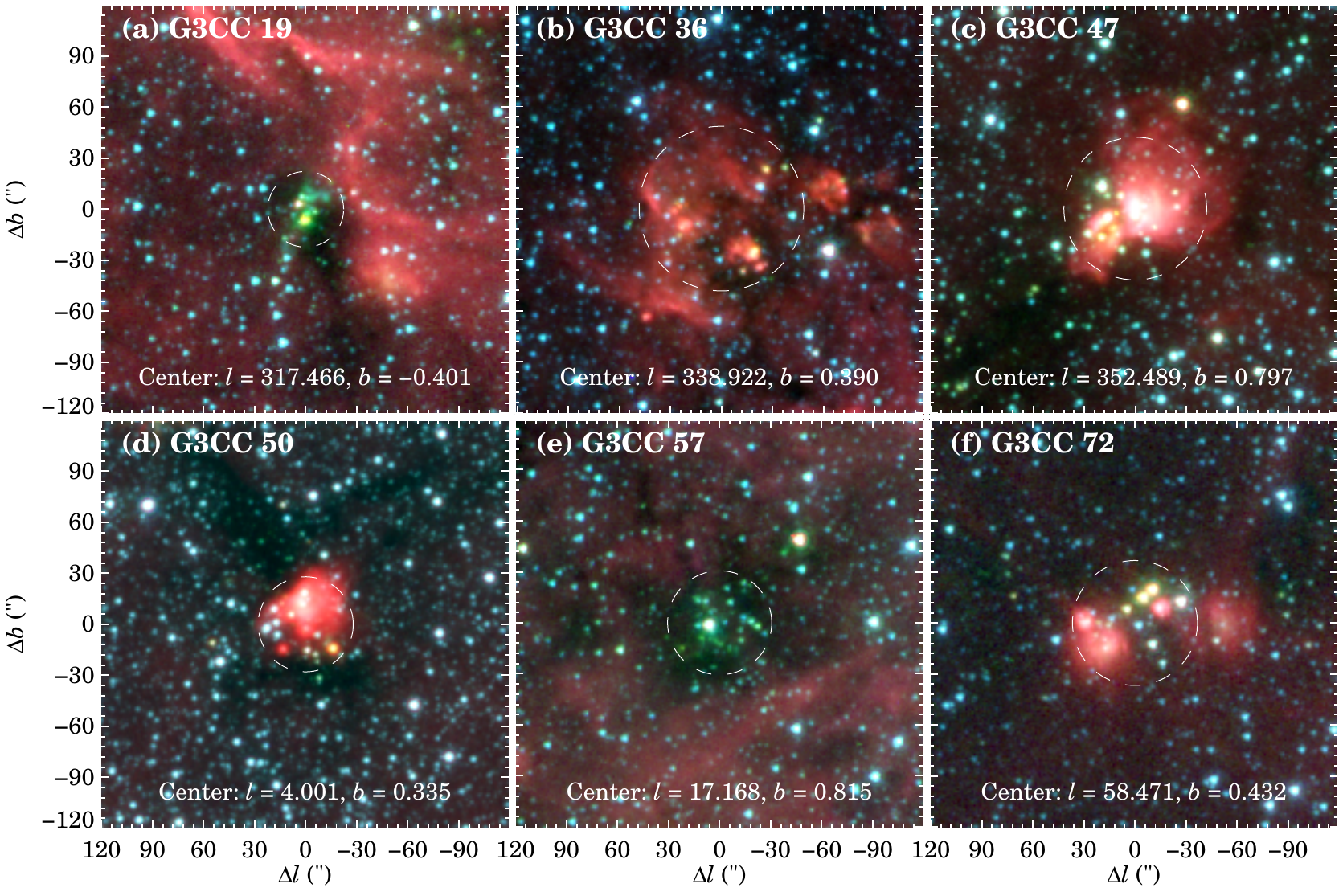}
\caption[Examples of new the GLIMPSE ECs discovered in this work]{\emph{Spitzer}-IRAC three-color images made with the 3.6 (blue), 4.5 (green) and 8.0 \micron\ (red) bands, of six (out of 75) new ECs discovered in this work, using the GLIMPSE survey. The dashed circles represent the estimated angular sizes. The images are in Galactic coordinates and the given offsets are with respect to the cluster center, indicated at the bottom of each panel.
}
\label{fig:newglimpse-examples}
\end{figure*}

As pointed out above, a subsequent visual selection was performed by examining the GLIMPSE images, based on a series of criteria which are explained in the following. Because the GLIMPSE on-line viewer has limited pixel resolution and is not efficient to inspect a high number of specific locations, we downloaded original GLIMPSE cutouts around these 702 positions and constructed three-color images using the 3.6 (blue), 4.5 (green) and 8.0~\micron\ (red) IRAC bands. This by-eye inspection led us to finally select 88 overdensities as locations of clusters, 17 of which are identified as known clusters from our literature compilation presented before. The remaining 71 new objects are listed in Table~\ref{tab:newglimpse}. The adopted identification is a record number (column 1) preceded by the acronym ``G3CC'' (GLIMPSE 3-Color Cluster\footnote{Referring to the fact that the clusters were finally selected on the GLIMPSE three-color images}). The final coordinates and the angular diameter (column 6) were estimated by eye on the GLIMPSE three-color images fitting circles interactively with the display software \emph{SAO Image DS9}\footnote{\url{http://hea-www.harvard.edu/RD/ds9/}}. The visual criteria applied to select the 88 overdensities are identified for each new object as flags in the last column of Table~\ref{tab:newglimpse}. Figure~\ref{fig:newglimpse-examples} shows GLIMPSE three-color images of 6 clusters, illustrating these different criteria. An almost ubiquitous characteristic of the selected clusters (present in 82 cases) is their association with typical mid-infrared star formation signposts (see Section~\ref{sec:galactic-surveys}), namely: extended 8.0~\micron\ emission in the immediate surroundings (flag E8, see Fig.~\ref{fig:newglimpse-examples}(a,b,c,d,f)), likely representing radiation from UV-excited PAHs or warm dust; more localized extended 4.5~\micron\ emission within the cluster area (flag E4, Fig.~\ref{fig:newglimpse-examples}(a)), which might trace shocked gas by outflowing activity from protostars \citep[see][and references therein]{Cyganowski2008}; and presence of an infrared dark cloud in which the cluster is embedded (flag DC, Fig.~\ref{fig:newglimpse-examples}(a,e)). We also indicate whether a cluster appears to have more stellar members than those identified by the red-color criterion, including the following situations: cluster composed of red sources and additional bright normal (not reddened) stars (flag BR, Fig.~\ref{fig:newglimpse-examples}(d)), suggesting that the cluster is in a more evolved phase, probably emerging from the molecular cloud; cluster exclusively composed of bright normal stars (flag B, but only two cases, in conjunction with flag V2, see below); and presence of additional probable YSOs within the cluster, identified as sources uniquely detected at 8.0~\micron\ (flag U8, representing extreme cases of red color), or compact 8.0~\micron\ objects not listed in the point source catalog or archive (flag C8, Fig.~\ref{fig:newglimpse-examples}(b,c,d,f)), due to the bright and variable extended emission at this wavelength, saturation for bright sources, or localized diffuse emission around a particular source which makes its apparent size larger than a point-source. The other flags indicate when the cluster shows up as a sparse, not centrally condensed set of sources (flag S, Fig.~\ref{fig:newglimpse-examples}(b)), or if the cluster was noticed by-eye on the GLIMPSE images in a nearby location of an automatically detected overdensity, but not exactly at the same position (flag V2).

The remaining positions were rejected as clusters, and typically correspond to background stars extinguished by dark clouds or seen behind foreground 8.0~\micron\ diffuse emission, producing a red-source density enhancement by chance, sometimes together in the same line of sight with a couple of intrinsically red sources (YSOs) which however do not represent a cluster by their own. Quantitatively, we found that, in general, most of the rejected positions are overdensities with fewer elements than the ones selected as clusters. In fact, if we choose stricter thresholds of 8 sources for the small bin, and 10 sources for the large bin, instead of the originally used 5 and 7, respectively, the total set of overdensities decreases from 702 to just 87 independent positions, 37 of which represent our clusters.
This would mean an improved ``success'' rate of $37/87 = 43\%$ for the automated method rather than the original $88/702 = 13\%$. Furthermore, if we consider the \emph{effective} number of elements in the 88 bins originally selected as being locations of clusters, i.e., summing possible additional stellar members (flags BR,C8,U8) within the bins, we find that 61 of our clusters satisfy the new threshold. We emphasize, however, that the additional stellar members of each cluster were recognized after detailed inspection of the GLIMPSE images, so that the use of low density thresholds in the automated method was necessary to identify the initial cluster locations, despite of the consequent detection of many spurious red-source overdensities. If we had used from the beginning the stricter thresholds, we would have missed $88-37=51$ clusters. Column 7 of Table~\ref{tab:newglimpse} lists for every cluster the estimated number of stellar members within the assumed radius, $N_{\rm circ}$, counting the YSOs selected by the red-color criterion and the additional members identified in the images (flags BR,C8,U8). Note that this number represents a lower limit, especially in distant clusters, since lower mass members could still be undetected due to the limited angular resolution and sensitivity of the GLIMPSE data.

We note that, because our simple automated method to find YSO overdensities is based on the GLIMPSE point source catalog, it is unavoidably biased towards young ECs that are not yet associated with very bright extended emission, which would hide many of the cluster members from the point source detection algorithm. Fortunately, it is quite likely that those bright nebulae were already looked for the presence of clusters by previous by-eye searches (see Section~\ref{sec:completeness}), so probably a few of them are really missing in our total compiled sample. We tried anyway to complete our list of new clusters by performing a systematic visual inspection with the on-line viewer over the entire area surveyed by GLIMPSE, including also fully exposed clusters that appear bright at 3.6~\micron\ (equivalent to flag `B'). We found from this process 23 additional clusters, of which, however, only 4 are new discoveries with respect to our literature compilation. They are marked in column 8 of Table~\ref{tab:newglimpse} with a `V', while the ones detected by the automated method are indicated with an `A'. We remark that, of the 17 known clusters we rediscovered from the red-source overdensities, only 3 are from the \citet{Mercer2005} list. This practically null overlap between the two detection methods demonstrates that our search is fully complementary and particularly useful to detect ECs, confirming the ideas we presented at the beginning of this Section.

Although our literature compilation of clusters is up to date until August, 2011, it is interesting to cross-check our list of new GLIMPSE clusters with the ECs recently discovered by \citet{Majaess2013}, who applied a combination of color and spectral index criteria to find YSO candidates using the WISE and 2MASS catalogs, and then looked for clusters by visually inspecting the YSOs spatial distribution. We found that only 5 new GLIMPSE clusters (they are indicated in Table~\ref{tab:newglimpse}) are associated with objects from the published list by \citet{Majaess2013}, in particular these 5 clusters are \emph{contained} within the corresponding objects identified by \citet{Majaess2013}, which cover a much larger area. Due to the coarser angular resolution of WISE data with respect to GLIMPSE data, the typical stellar densities in our ECs are probably too high to make all the individual members detectable at the WISE resolution, and consequently they are hidden in the \citet{Majaess2013} YSO selection.

\section{Properties of the cluster sample}
\label{sec:huge-table}

The next step of this work was to characterize the ATLASGAL emission, if present, at the positions of the star clusters compiled in Section~\ref{sec:catalogs-summary}, and to compare this emission with NIR and MIR images. Hereafter, our study is naturally restricted to the ATLASGAL Galactic range ($|\ell| \le 60\degr$ and $|b| \le 1.5\degr$), and we refer to the list of the 695 stellar clusters within that range as the ``whole cluster sample'' (or simply as the ``cluster sample''), unless noted. Together with this process, we performed a critical literature revision in order to add and update distances and ages for an important fraction of the sample, as well as to look for connections with known \ion{H}{ii} regions, IRDCs, and IR bubbles. We organize all this information in an unique catalog, whose construction is summarized in the following, and described in more detail in Appendix~\ref{sec:huge-table-details}. The catalog is only available in electronic form at the CDS, together with a companion list of all the references with the corresponding identification numbers used throughout the table. For illustration, an excerpt of the catalog is given in Appendix~\ref{sec:catalog-excerpt}.

\subsection{ATLASGAL and MIR emission}
\label{sec:atlasgal-and-mir}

\begin{figure}[!t]
\centering
\includegraphics[width=0.49\textwidth]{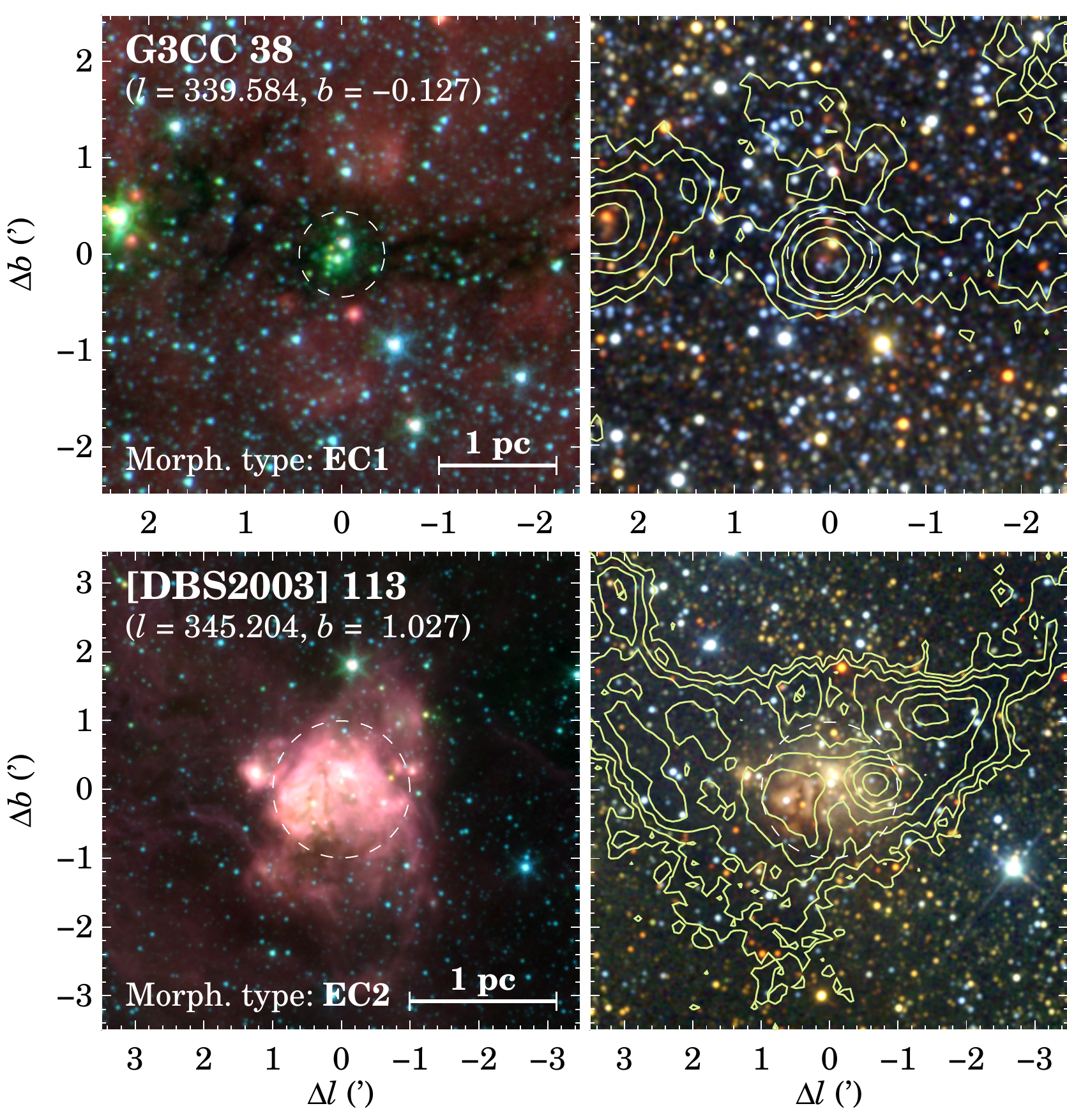}
\caption[Examples of the two morphological types of ECs]{Examples of the two morphological types defined for ECs (see Section~\ref{sec:evolutionary-sequence}): The cluster G3CC~38 of type EC1 (top panels), and the cluster $[$DBS2003$]$~113 of type EC2 (bottom panels). The left panels show \emph{Spitzer}-IRAC three-color images made with the 3.6 (blue), 4.5 (green) and 8.0 \micron\ (red) bands. The right panels present 2MASS three-color images of the same field of view, constructed with the $J$ (blue), $H$ (green), and $K_s$ (red) bands. The overlaid contours on the 2MASS images represent ATLASGAL emission (870~\micron); the contour levels are $\{5,8.8,15,25,46,88,170\}\times \sigma$, where $\sigma$ is the local rms noise level ($\sigma = 45$~mJy/beam for G3CC~38, and $\sigma = 42$~mJy/beam for $[$DBS2003$]$~113). The images are in Galactic coordinates and the given offsets are with respect to the cluster center, indicated in the left panels below the cluster name. The dashed circles represent the estimated angular sizes from the original cluster catalogs (see Section~\ref{sec:basic-information}). The 1~pc scale-bar was estimated using the corresponding distance adopted in our catalog.}
\label{fig:EC-examples}
\end{figure}

\begin{figure}[!t]
\centering
\includegraphics[width=0.49\textwidth]{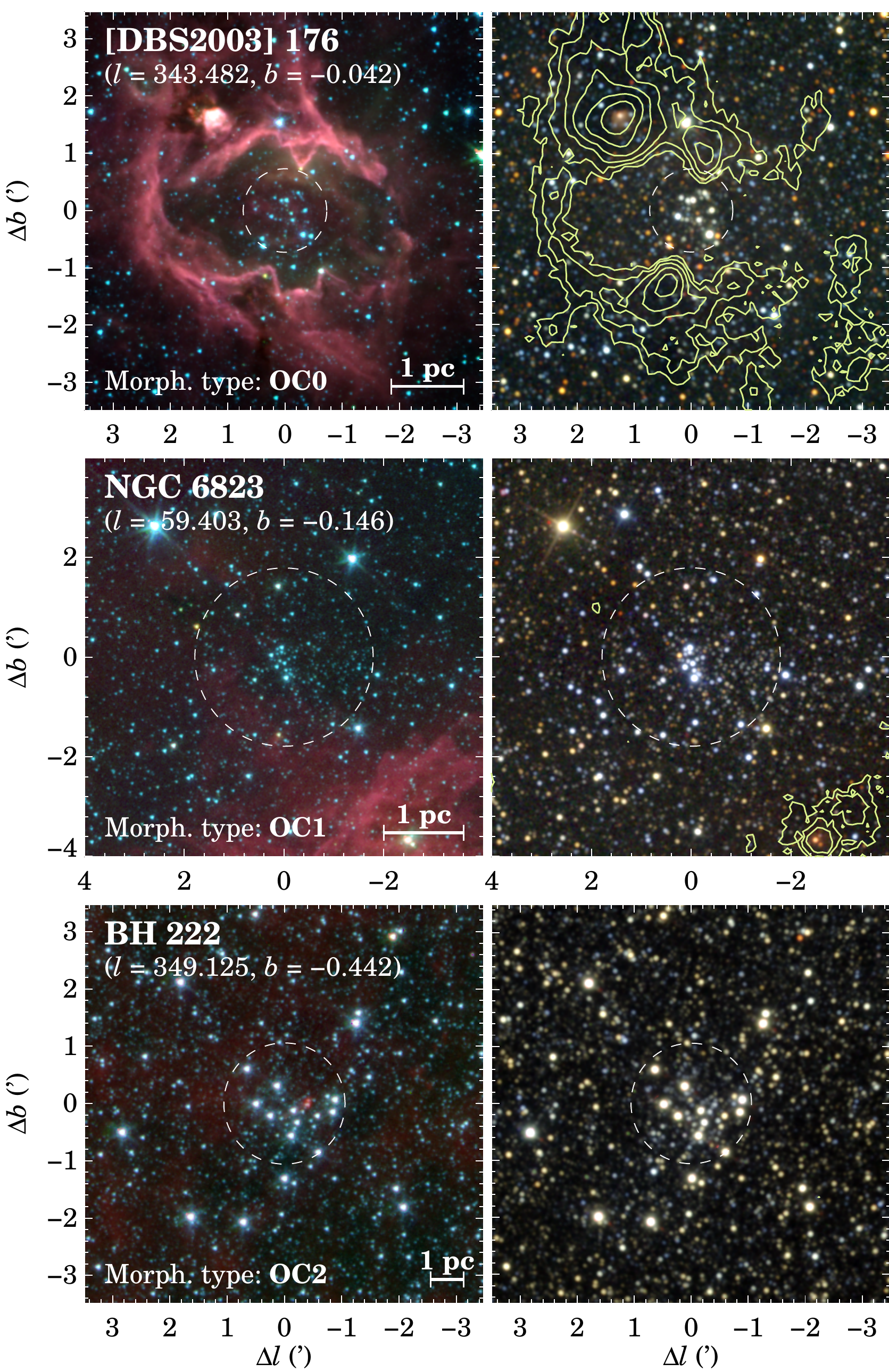}
\caption[Examples of the three morphological types of OCs]{Examples of the three morphological types defined for OCs (see Section~\ref{sec:evolutionary-sequence}): The cluster $[$DBS2003$]$~176 of type OC0 (top panels), the cluster NGC~6823 of type OC1 (middle panels), and the cluster BH~222 of type OC2 (bottom panels). The local rms noise level of the ATLASGAL emission is, respectively, 36, 46, and 29~mJy/beam. See caption of Figure~\ref{fig:EC-examples} for more details of the images.}
\label{fig:OC-examples}
\end{figure}

In order to search for submm dust continuum emission tracing molecular gas likely associated with the clusters, we examined the ATLASGAL emission around the cluster positions. The column \verb|Morph| is a text flag that gives information about the morphology of the detected ATLASGAL emission versus the IR emission. It is composed of two parts separated by a period. The first part tells about how the ATLASGAL emission is distributed throughout the immediate star cluster area, including the following cases: 
\begin{itemize}[label=\textbullet]
 \item \verb|emb|: cluster fully embedded, with its center matching the submm clump peak (Fig.~\ref{fig:EC-examples}, \emph{top}).
 \item \verb|p-emb|: cluster partially embedded, whose area is not completely covered, or the submm clump peak is significantly shifted from the (proto-) stars locations (Fig.~\ref{fig:EC-examples}, \emph{bottom}).
 \item \verb|surr|: possibly associated submm emission surrounding the cluster or close to its boundaries (Fig.~\ref{fig:OC-examples}, \emph{top}).
 \item \verb|few|: one or a few ATLASGAL clumps within the cluster area (mostly for optical clusters having a large angular size), not necessarily physically related with the cluster.
 \item \verb|few*|: the same morphology as before, but now the clump(s) is (are) likely associated with the star cluster according to previous studies in the literature, or because the kinematic distance derived from molecular lines agrees with the stellar distance. See Section~\ref{sec:distance-and-ages} for a brief description of the distance determinations.
 \item \verb|exp|: exposed cluster, without ATLASGAL emission in immediate surroundings (Fig.~\ref{fig:OC-examples}, \emph{middle} and \emph{bottom}).
 \item \verb|exp*|: cluster that is physically exposed, but presents submm emission within the cluster area which appears in the same line of sight, but with a kinematic distance discrepant from the stellar distance (the cluster would be categorized as \verb|few| or \verb|surr| if no distance information were available).
\end{itemize}

We indicate in the second part of the column \verb|Morph| (after the period) details about the mid-infrared morphology of each cluster, after visually inspecting GLIMPSE three-color images made with the 3.6 (blue), 4.5 (green) and 8.0 \micron\ (red) bands. For a few clusters with no coverage in the GLIMPSE survey (7\% of the cluster sample), we instead examined WISE three-color images using the 3.4, 4.6 and 12~\micron\ filters. This flag includes the following cases:
\begin{itemize}[label=\textbullet]
 \item \verb|bub-cen|: presence of an IR bubble which seems to be produced by the cluster through stellar feedback, and appears in the images centered near the cluster position (Fig.~\ref{fig:OC-examples}, \emph{top}).
 \item \verb|bub-cen-trig|: the same situation than before, together with the presence of possible YSOs at the periphery of the bubble identified by their reddened appearance in the images, suggesting triggered star formation generated by the cluster (see also Fig.~\ref{fig:OC-examples}, \emph{top}).
 \item \verb|bub-edge|: in this case, the cluster itself appears at the edge of an IR bubble, suggesting that it was probably formed by triggering from an independent cluster or massive star.
 \item \verb|pah|: presence of bright and irregular emission at 8.0~\micron\ (12~\micron\ for WISE) which seems to be produced by the cluster through stellar radiative feedback (Fig.~\ref{fig:EC-examples}, \emph{bottom}); it is attributed to radiation from UV excited PAHs or warm dust, but is not clearly identified as an IR bubble (though it sometimes shows bubble-like borders)\footnote{This situation is conceptually different from the one indicated by the flag E8 for G3CC objects (see Section~\ref{sec:newglimpse}), where any extended 8.0~\micron\ emission in the vicinity of the cluster is flagged. Here, the emission has to be located throughout most of the cluster area and appear as produced by the whole cluster.}.
\end{itemize}

\subsection{Correlation with known objects}
\label{sec:known-objects}

Associated IR bubbles that are listed in the catalogs by \citet{Churchwell2006,Churchwell2007} are identified in the table column \verb|Bub|. On the GLIMPSE three-color images and on the 8.0~\micron\ images (WISE three-color and 12~\micron\ images when GLIMPSE data were not available), we also identified the presence of an infrared dark cloud in which the cluster appears to be embedded (column \verb|IRDC|; see Fig.~\ref{fig:EC-examples}, \emph{top}), and we give the designation from the catalogs by \citet{Simon2006} or \citet{PerettoFuller2009} when the object is listed there. Finally, we searched in the literature for associated \ion{H}{ii} regions (column \verb|HII_reg|), and we flagged the sources that have been classified in the literature as ultra compact (UC) \ion{H}{ii} regions.

\subsection{Distance and age}
\label{sec:distance-and-ages}

An important part of this work was to assign distances to as many clusters as possible. In this regard, we took advantage of the fact that many of the ATLASGAL clumps at the locations or in the vicinity of the stellar clusters have measurements of molecular line LSR velocities \citep[e.g.,][]{Wienen2012, Bronfman1996, Urquhart2008}. Using these velocities and a combined rotation curve based on the models by \citet{BrandBlitz1993} and \citet{Levine2008}, we computed kinematic distances for the clumps (column \verb|KDist|) and, therefore, for the corresponding clusters when they were assumed to be physically associated. The kinematic distance ambiguity (KDA) was disentangled mainly by searching for previous resolutions in the literature \citep[e.g.][]{CaswellHaynes1987, Faundez2004, AndersonBania2009, Roman-Duval2009}, for the clumps themselves or nearby \ion{H}{ii} regions in the phase space. A total of 424 clusters have kinematic distance estimates for the ATLASGAL clumps, 92\% of which have available KDA solutions. The uncertainties (column \verb|e_KDist|) have been determined by shifting the LSR velocities by $\pm 7$~\kms\ to account for random motions, following \citet{Reid2009}, who suggest this value as the typical virial velocity dispersion of a massive star-forming region.

We also compiled values for the stellar distance (column \verb|SDist|) and age (column \verb|Age|), estimated from studies of the stellar population of the clusters. These data were obtained from the original cluster catalogs or from new references found in SIMBAD. To prevent underestimation of the uncertainties (provided in columns \verb|e_SDist| and \verb|e_Age|), we imposed minimum errors depending on the computation method for the stellar distance, and on the range for the age \citep[the latter following][]{BonattoBica2011}. Stellar distances are available for 222 clusters (32\% of the sample), and ages for 209 clusters (30\% of the sample). The most common method for stellar distance and age determination is isochrone fitting \citep[e.g.,][]{Loktin2001}, which implies that these parameters are available mainly for exposed clusters (see Section~\ref{sec:age-distribution}).

The final adopted distance for each cluster (column \verb|Dist|)  was chosen to be the available distance estimate with the lowest uncertainty. In some cases, we adopted independent distance estimates from the literature if they were more accurate than \verb|SDist| and \verb|KDist| \citep[e.g., from maser parallax measurements; see][and references therein]{Reid2009}. Clusters within a particular complex (identified in the column \verb|Complex|) were assumed to be all located at the same distance, determined from the literature, or kinematically from an average position and velocity.

\begin{figure}[!t]
\centering
\includegraphics[width=0.45\textwidth]{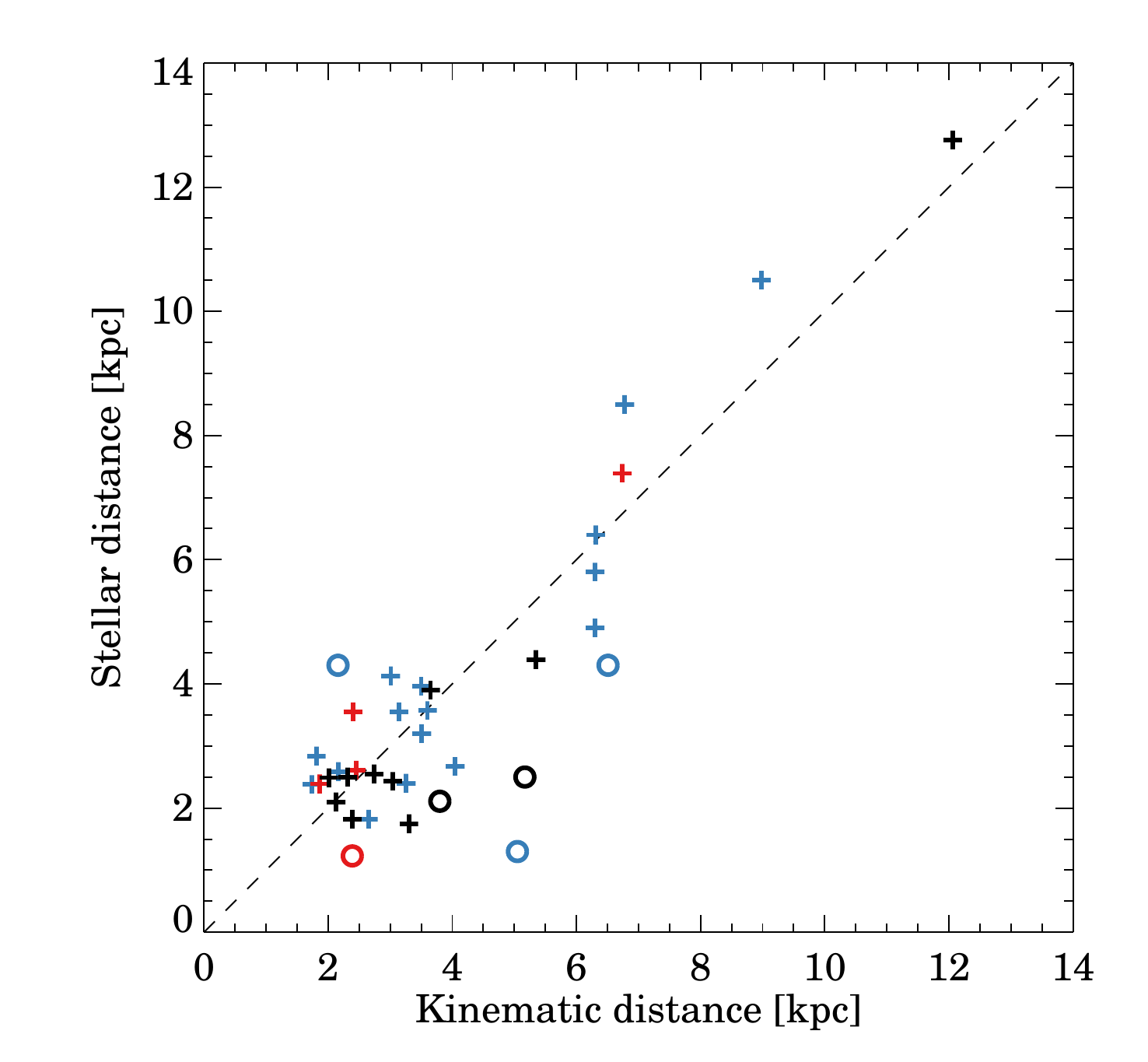}
\caption[Comparison of kinematic and stellar distances]{Comparison of kinematic and stellar distances for the 38 clusters of our sample with both estimations available. Plus signs (+) indicate agreement within the errors, and circles mark the discrepant cases. Colors indicate which distance estimate was finally adopted in our catalog: stellar (\emph{red}), kinematic (\emph{blue}), and other (\emph{black}). The dashed line is the identity.}
\label{fig:dist-comparison}
\end{figure}

In total, there are distance determinations (\verb|Dist|) for 538 clusters, i.e., for 77\% of our sample. Naturally, there is a dichotomy in the distance estimation method depending on whether or not the cluster is associated with an ATLASGAL source with available velocity, so that most exposed clusters uniquely have stellar distances, whereas the distances for ECs are mainly kinematic or from associations with complexes. However, it is still possible to compare stellar and kinematic determinations for a subsample of 38 clusters (mostly embedded) which have distances available from both methods. This comparison is shown in Figure~\ref{fig:dist-comparison}, where plus symbols mean agreement between stellar and kinematic distances within the corresponding uncertainties, and circles are the cases in which there is a discrepancy between both techniques; the color indicates which distance estimate was finally adopted in our catalog: stellar (\emph{red}), kinematic (\emph{blue}), and other (\emph{black}). The plot reveals that in our cluster sample, both methods are quite consistent with each other, with a 84\% of agreement (32 out of 38 objects). We note that among the discrepant cases, there are two ECs (points $(2.16, 4.30)$~kpc and $(5.05, 1.30)$~kpc in the plot) whose method for age and (stellar) distance estimation was found to be particularly inaccurate (see Section~\ref{sec:age-embedded}).

The rms between the stellar and kinematic distances compared in Figure~\ref{fig:dist-comparison} is 1.28~kpc, which represents the combined error, for this particular subsample, of both stellar and kinematic distances added in quadrature. If we compute this error from the estimated uncertainties \verb|e_KDist| and \verb|e_SDist| averaged over the subsample, we obtain a value of 1.59~kpc, which means that we slightly overestimated some of the uncertainties, probably because we were quite conservative in determining the minimum errors for the stellar distances (see Section~\ref{sec:physical-parameters}). The average uncertainties are $\langle\verb|e_KDist|\rangle = 0.67$~kpc and $\langle\verb|e_SDist|\rangle = 1.45$~kpc for the subsample of the 38 clusters used for comparison, and $\langle\verb|e_KDist|\rangle = 0.68$~kpc and $\langle\verb|e_SDist|\rangle = 0.58$~kpc for the whole sample. The high average error for the stellar distance in the subsample with respect to the whole sample is due to the fact that most of these clusters have stellar distances estimated from the spectrophotometric method, which is more inaccurate than, e.g., main sequence or isochrone fitting (see Section~\ref{sec:physical-parameters}). The average estimated uncertainty in the adopted distance is $\langle\verb|e_Dist|\rangle = 0.51$~kpc for the whole sample (and 0.52~kpc for the subsample).

\section{Analysis}
\label{sec:analysis}

\subsection{Morphological evolutionary sequence}
\label{sec:evolutionary-sequence}

Here, we use the characterization of the ATLASGAL emission found throughout each cluster's area and/or environment (described in Section~\ref{sec:atlasgal-and-mir}) to define main morphological types and delineate an evolutionary sequence. First, in order to test our visual ATLASGAL morphological flags specified above (corresponding to the first part of the column \verb|Morph|, and represented hereafter by \verb|m|$_0$), we compared them against the more quantitative parameter $s \equiv \verb|Clump_sep|$ of our catalog, which is the projected distance of the nearest ATLASGAL emission pixel, normalized to the cluster angular radius. We found a reasonable correlation: $s = 0$ for all deeply ECs (\verb|m|$_0$ = \verb|emb|), $s < 0.42$ for partially ECs (\verb|m|$_0$ = \verb|p-emb|),  $0.40 < s < 1.97$ for clusters surrounded by submm emission (\verb|m|$_0$ = \verb|surr|), and $s > 0.94$ for exposed clusters (\verb|m|$_0$ = \verb|exp|). Exposed clusters with $s < 1$ only comprise a few cases with a large angular size and very faint emission close to their borders. The remaining morphological flags are very specific and we do not expect any correlation with the quantity \verb|Clump_sep|.

Denoting by \verb|Cf|$_0$ the first digit of the flag \verb|Clump_flag| from our catalog (a value $>0$ means that the nearest ATLASGAL clump is likely associated with the cluster), and using the logical operators $\land$, $\lor$ and $\lnot$ (`and', `or', and `not', respectively), we define five morphological types as follows:
\begin{itemize}[label=\textbullet]
 \item EC1: $\verb|m|_0 = \verb|emb|$
 \item EC2: $\verb|m|_0 = \verb|p-emb|$
 \item OC0: $\verb|m|_0 = \verb|surr| ~\lor~ \verb|m|_0 = \verb|few*| ~\lor~ 
             (\verb|m|_0 = \verb|few| ~\land~  \verb|Cf|_0 > 0)$
 \item OC1: $\verb|m|_0 = \verb|exp| ~\land~ (\verb|Cf|_0 > 0 ~\lor~ 
              \verb|KDist| \simeq \verb|SDist|)$
 \item OC2: ($\verb|m|_0 = \verb|exp| ~\lor~ \verb|m|_0 = \verb|exp*| ~\lor~ 
             \verb|m|_0 = \verb|few|) ~\land~ \lnot(\rm{OC1} ~\lor~ \rm{OC2})$
\end{itemize}

The morphological type for each cluster is given in the column \verb|Morph_type| of our catalog. Figures \ref{fig:EC-examples} and \ref{fig:OC-examples} present one example cluster for each morphological type, shown in GLIMPSE three-color images, and 2MASS three-color images overlaid with ATLASGAL contours. In simpler words, given that star clusters are expected to be less and less associated with molecular gas as time evolves, due to gas dispersal driven by stellar feedback, we have defined above a morphological evolutionary sequence, with decreasing correlation with ATLASGAL emission. EC1 are deeply ECs (Fig.~\ref{fig:EC-examples}, \emph{top}), EC2 are partially ECs (Fig.~\ref{fig:EC-examples}, \emph{bottom}), OC0 are emerging exposed clusters (Fig.~\ref{fig:OC-examples}, \emph{top}), and finally there are two kinds of totally exposed clusters: OC1 are still physically associated with molecular gas in their surrounding neighborhood (an ATLASGAL clump at a projected distance of \verb|Clump_sep| times the cluster radius, see Fig.~\ref{fig:OC-examples}, \emph{middle}), whereas OC2 are all the remaining exposed clusters, which present no correlation with ATLASGAL emission (Fig.~\ref{fig:OC-examples}, \emph{bottom}).

\SaveVerb{refconf}|ref_Conf|
\begin{table*}[!t]
\renewcommand{\arraystretch}{1.1}
\caption{Number of clusters in each morphological type.}
\label{tab:morph-types}
\centering
\begin{tabular}{cccccc}
\hline\hline
Type & $N_{\rm cl}$ & $N_{\rm cl}$($D$ avail.) & 
$N_{\rm cl}(\le D_{\rm rep})$ & $N_{\rm cl}^{\rm conf}(\le D_{\rm rep})$ &
$N_{\rm cl}^{\rm tot}(\le D_{\rm rep})$\\
(1) & (2) & (3) & (4) & (5) & (6) \\
\hline
EC1 & 132 & 125 &  44 &  16 &  56 \\
EC2 & 195 & 177 &  54 &  25 &  68 \\
OC0 &  56 &  49 &  17 &  10 &  36 \\
OC1 &  22 &  22 &   6 &   3 &  11 \\
OC2 & 290 & 167 & 136 & 133 & 475 \\
\hline
\end{tabular}
\tablefoot{
The given numbers are for the whole sample (Column 2), clusters with available distances (Column 3), clusters with distances $\le D_{\rm rep}$ (Column 4), confirmed (\UseVerb{refconf} not empty) clusters with $D \le D_{\rm rep}$ (Column 5), and finally we give the estimated number of clusters with $D \le D_{\rm rep}$ in an ideally complete sample (Column 6). The distance $D_{\rm rep} = 3.0$~kpc defines what we call the \emph{representative} sample (see Section~\ref{sec:representative-sample} for details).}
\end{table*}

Note that, however, this classification is not perfect. For example, although the gas velocity and stellar distance data are quite extensive, they are not complete to identify all the $\verb|m|_0 = \verb|few*|$, $\verb|m|_0 = \verb|exp*|$ and $\verb|KDist| \simeq \verb|SDist|$ cases, so that some misclassification might occur in the type OC2. Similarly, the physical link between the submm emission and the ECs was based on the morphology seen in the images, and some chance alignments might still be present in a few cases (estimated to be about 5\%, see Section~\ref{sec:chance-alignments}). Therefore, the defined morphological types should primarily be considered in a statistical way, and for individual objects they must be treated with caution. Column 2 of Table~\ref{tab:morph-types} lists how many objects fall in each morphological type for the whole cluster sample. Note that the low number of OC1 clusters could be partially due to the observational difficulty in identifying an exposed cluster physically associated with molecular gas in their surroundings, as remarked before. Column 3 gives the number of clusters with available distances, and the remaining columns will be described in Section~\ref{sec:representative-sample}.

With this morphological classification, it is easy to determine (again, statistically) which clusters are associated with ATLASGAL emission: simply as those with types EC1,EC2,OC0 or OC1. These clusters are counted for every catalog in the last two columns of Table~\ref{tab:catalogs}, as absolute and after-merging numbers of objects ($N_{\rm cl}$ and $N_{\rm cl}^*$, respectively). As expected, optical clusters are rarely associated with ATLASGAL emission (only $\sim 15\%$ of them, most of which are of type OC0 or OC1), since otherwise they would be barely visible at optical wavelengths due to dust extinction. On the other hand, the majority of the NIR and MIR clusters are physically related with submm dust radiation ($\sim 79\%$ and 74\% of them, respectively). Although this is also expected because infrared emission is much less affected by dust extinction than visible light, these high percentages might partially be a consequence of the detection method of the infrared cluster catalogs, which in most cases tried to intentionally highlight the EC population. For example, the 2MASS by-eye searches by \citet{Dutra2003-2mass} and \citet{Bica2003-2mass} were done towards known radio/optical nebulae, and our new GLIMPSE cluster candidates were detected after applying a red-color criterion (see Section~\ref{sec:newglimpse}). In these particular catalogs, almost the totality of objects are associated with ATLASGAL emission.

\subsection{Chance alignments}
\label{sec:chance-alignments}

We computed the probability of chance alignments of our stellar clusters with ATLASGAL clumps, and the different known objects looked for spatial correlation in our catalog (see Section~\ref{sec:known-objects}), in order to test the validity of the assumption of physical relation, when this is only based on the position of the objects on the sky. For a given sample of objects, this probability was estimated semi-analytically by assuming that the objects within $|b| \le 1\degr$ (where most sources are located for all samples used) and the longitude range originally covered, are uniformly distributed over that area, and that their angular sizes are distributed according to the observed sizes. We first calculated the probability of overlap of each cluster with one or more objects from this hypothetical sample, and then we averaged these probabilities over two different sets of clusters: morphological types EC1 and EC2 together (hereafter EC-); and types OC0, OC1 and OC2 together (hereafter OC-).

For ATLASGAL clumps, we adopted a total number of 6451 objects within $330\degr \le \ell \le 21\degr$ and $|b| \le 1\degr$, from the compact source catalog by \citet{Contreras2013}, which, together with their estimated effective radii, gives an average chance alignment probability of 8.8\% for clusters with types EC-, and 32\% for clusters with types OC-. Considering that the submm and infrared morphologies of deeply ECs (type EC1) usually support the real physical relation with molecular gas (e.g., matching peaks of submm emission and stellar density), and that partially ECs (type EC2) are generally associated with more than one ATLASGAL clump, in practice the fraction of chance alignments of EC- clusters with ATLASGAL compact sources is likely below 5\%, which is low enough to not affect the statistics of this work. Due to their larger angular sizes, clusters of types OC- are more prone to be aligned with ATLASGAL clumps by chance, and therefore our additional requirements to assume that an exposed cluster is associated with ATLASGAL emission are justified (morphological criteria or matching distances for types OC0 and OC1).

For the known objects considered in our catalog, we assume that there are 4936 IR bubbles in the range $|\ell| \le 60\degr$ and $|b| \le 1\degr$ \citep{Simpson2012}\footnote{This is a recent catalog of IR bubbles which is much more complete than the \citet{Churchwell2006,Churchwell2007} catalogs, but was not used in this work because it was published after our cluster catalog was constructed. In any case, we searched for IR bubbles by eye at every cluster position to describe the MIR morphology (see Section~\ref{sec:atlasgal-and-mir}).}, $17,364$ IRDCs within $10\degr \le |\ell| \le 60\degr$ and $|b| \le 1\degr$ \citep[from the catalogs by][]{Simon2006,PerettoFuller2009}, and 944~\ion{H}{ii} regions in the range $343\degr \le \ell \le 60\degr$ and $|b| \le 1\degr$ \citep[from the recently discovered and previously known \ion{H}{ii} regions listed in][]{Anderson2011}. In this case, to compute the chance alignment probability of each cluster with the objects of a given sample, we also required that the objects were larger than half the size of the cluster and that the distance between the object's position and the cluster center were less than the sum of both radii divided by two, so that the alignment really mimics a physical relation misidentified by eye. The averaged probabilities are quite similar for clusters with types EC- and OC-, and they are all low: $\sim 2\%$ for IR bubbles, $\sim 3.5\%$ for IRDCs, and $\sim 0.3\%$ for \ion{H}{ii} regions.

\subsection{Observational classification of OCs and ECs}
\label{sec:classification-oc-ec}

We can also use the morphological evolutionary sequence established in Section~\ref{sec:evolutionary-sequence} to observationally define in our sample the concepts of EC and OC. Since any stellar agglomerate that appears deeply or partially embedded in ATLASGAL emission would satisfy our physical definition of EC presented in Section~\ref{sec:cluster-definition}, we simply use as observational definition the embedded morphological types: EC =  EC1 $\lor$ EC2. We consider the remaining morphological types as OCs, but excluding those objects that have not been confirmed by follow-up studies, since we expect for them a high contamination rate by spurious candidates (see Section~\ref{sec:spurious}): OC = (OC0 $\lor$ OC1 $\lor$ OC2) $\land$ (\verb|ref_Conf| not empty), where \verb|ref_Conf| is the column in the catalog indicating the reference for cluster confirmation (see Section~\ref{sec:physical-parameters}). 

However, this observational definition of OC does not necessarily mean that the cluster is bound by its own gravity, and therefore, is not fully equivalent to the concept of \emph{physical OC} defined in Section~\ref{sec:cluster-definition}. To investigate under which conditions both definitions agree, we can apply the empirical criterion proposed by \citet{GielesPortegies2011} which distinguishes between physical OCs and associations by comparing the age of the object with its crossing time, $t_{\rm cross}$, computed as if it were in virial equilibrium. In useful physical units, Equation (1) of \citet{GielesPortegies2011} becomes\footnote{Before converting to physical units, we corrected a mistake in the original equation by \citet{GielesPortegies2011}: the transformation from virial radius to projected half-light radius is just $16/(3\pi)$ for a Plummer model, so that the constant in their equation is $[32/(3\pi)]^{3/2} = 6.26$ instead of 10.}
\begin{equation}
\label{eq:tcross-units-GP11}
t_{\rm cross} = 9.33 \left(\frac{100\, M_{\sun}}{M}\right)^{1/2}
\left(\frac{R_{\rm eff}}{\rm pc}\right)^{3/2} \, {\rm Myr},
\end{equation}
where $M$ and $R_{\rm eff}$ are, respectively, the mass and the observed 2D projected half-light radius of the cluster. Unfortunately, mass estimates and accurate structural parameters are usually not directly available in the OC catalogs; in particular, there are no mass data in the \citet{Dias2002} catalog, and the given sizes come from individual studies compiled there and are mostly derived from visual inspection. We therefore used the masses and radii determined by \citet{Piskunov2007}, who fitted a three-parameter King's profile \citep{King1962} to the observed stellar surface density distribution of 236 objects taken from an homogeneous sample of 650 optical clusters in the solar neighborhood \citep{Kharchenko2005-known,Kharchenko2005-new}, which is a subset of the current version of the \citet{Dias2002} catalog. \citet{Piskunov2007} estimated the masses from the tidal radii, and the effective radius $R_{\rm eff}$ entering in Equation~(\ref{eq:tcross-units-GP11}) can be derived from both the core and tidal radius \citep[we used Equation (B1) of][]{Wolf2010}. Because only 14 of the clusters analyzed by \citet{Piskunov2007} are within the ATLASGAL sky coverage, in order to improve the statistics we applied the \citet{GielesPortegies2011} criterion to the 236 studied objects, under the assumption that they are all OCs as observationally defined by us. This supposition is quite acceptable since they are optically-detected clusters and indeed within the ATLASGAL range almost all of them (13 out of 14) are classified as OCs.

\begin{figure}[!t]
\centering
\includegraphics[width=0.48\textwidth]{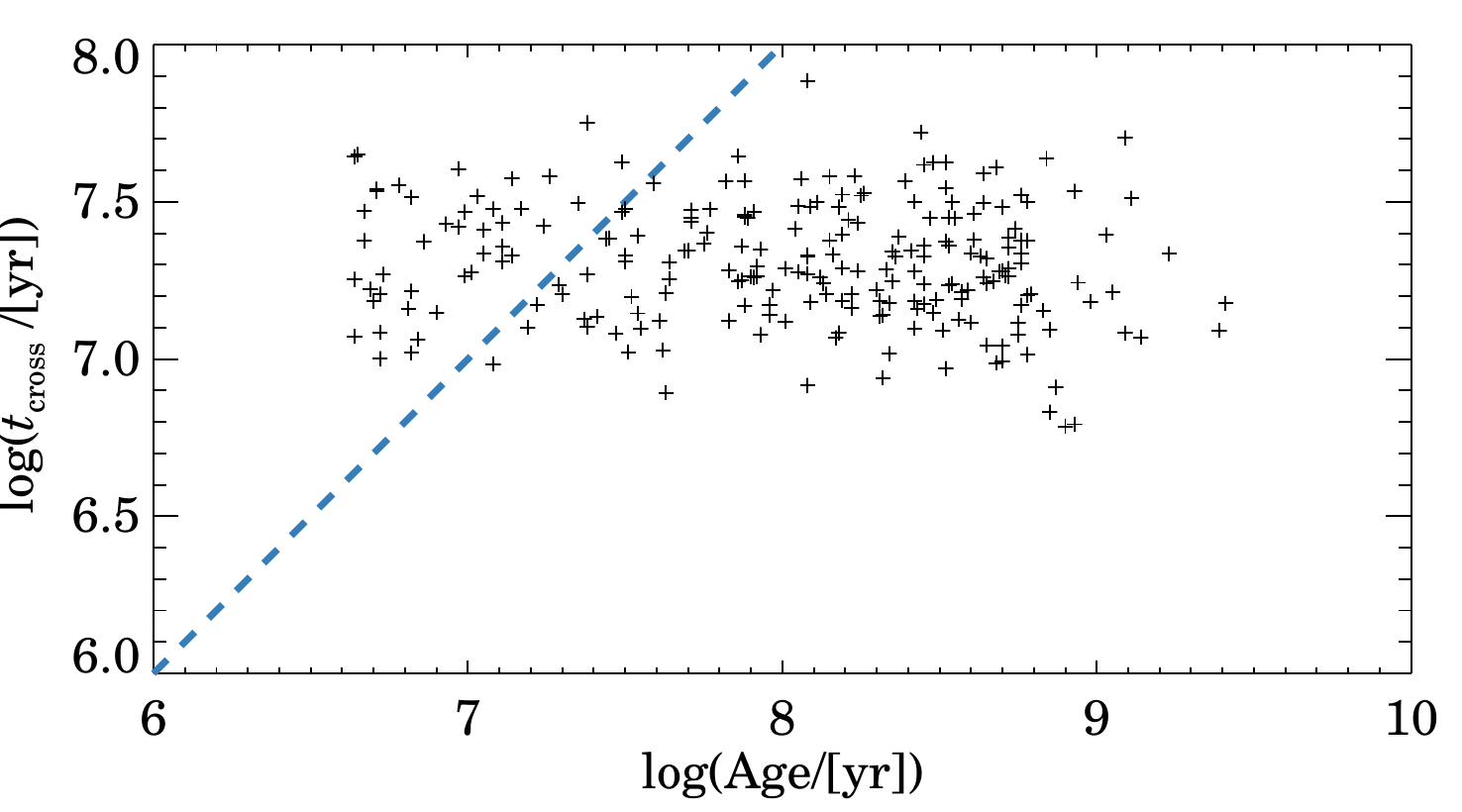}
\caption[Crossing time vs. age in an solar neighborhood OC sample]{Crossing time vs. age for an all-sky sample of 236 clusters \citep{Piskunov2006} taken from an homogeneous catalog of 650 optical clusters in the solar neighborhood \citep{Kharchenko2005-known,Kharchenko2005-new}. The dashed line is the identity $t_{\rm cross} =$~Age, which divides the physical OCs ($t_{\rm cross} \leq$~Age) from associations ($t_{\rm cross} >$~Age) according to the criterion proposed by \citet{GielesPortegies2011}.}
\label{fig:clusters-vs-associations}
\end{figure}

We computed the crossing times using Equation~(\ref{eq:tcross-units-GP11}), and in Figure~\ref{fig:clusters-vs-associations} they are plotted versus the corresponding ages available from the \citet{Kharchenko2005-known,Kharchenko2005-new} catalogs. The dashed line is the identity $t_{\rm cross} =$~Age, which divides the physical OCs ($t_{\rm cross} \leq$~Age) from associations ($t_{\rm cross} >$~Age). It can be seen in the plot that, because the resulting crossing times are relatively short ($\log(t_{\rm cross}/{\rm yr}) \lesssim 7.6$), the majority of the objects studied by \citet{Piskunov2007} are physical OCs for ages in excess of 10~Myr. In fact, for $\log({\rm Age}/{\rm yr}) > 7.2$, which is the threshold above which the age distribution can uniquely be explained through classical cluster disruption mechanisms (see Section~\ref{sec:age-distribution-fit}), only 2.6\% of the objects are formally associations. We thus conclude that our observational definition of OC agrees with the physical one provided by \citet[][what we call a \emph{physical OC}]{GielesPortegies2011} for ages greater than $\sim 16$~Myr, which corresponds to the 74\% of our OC sample within the ATLASGAL range. Younger OCs can be either associations, as a result of early dissolution, or already physical OCs.

\subsection{Spatial distribution}
\label{sec:spatial-distribution}

\begin{figure}[!t]
\centering
\includegraphics[width=0.5\textwidth]{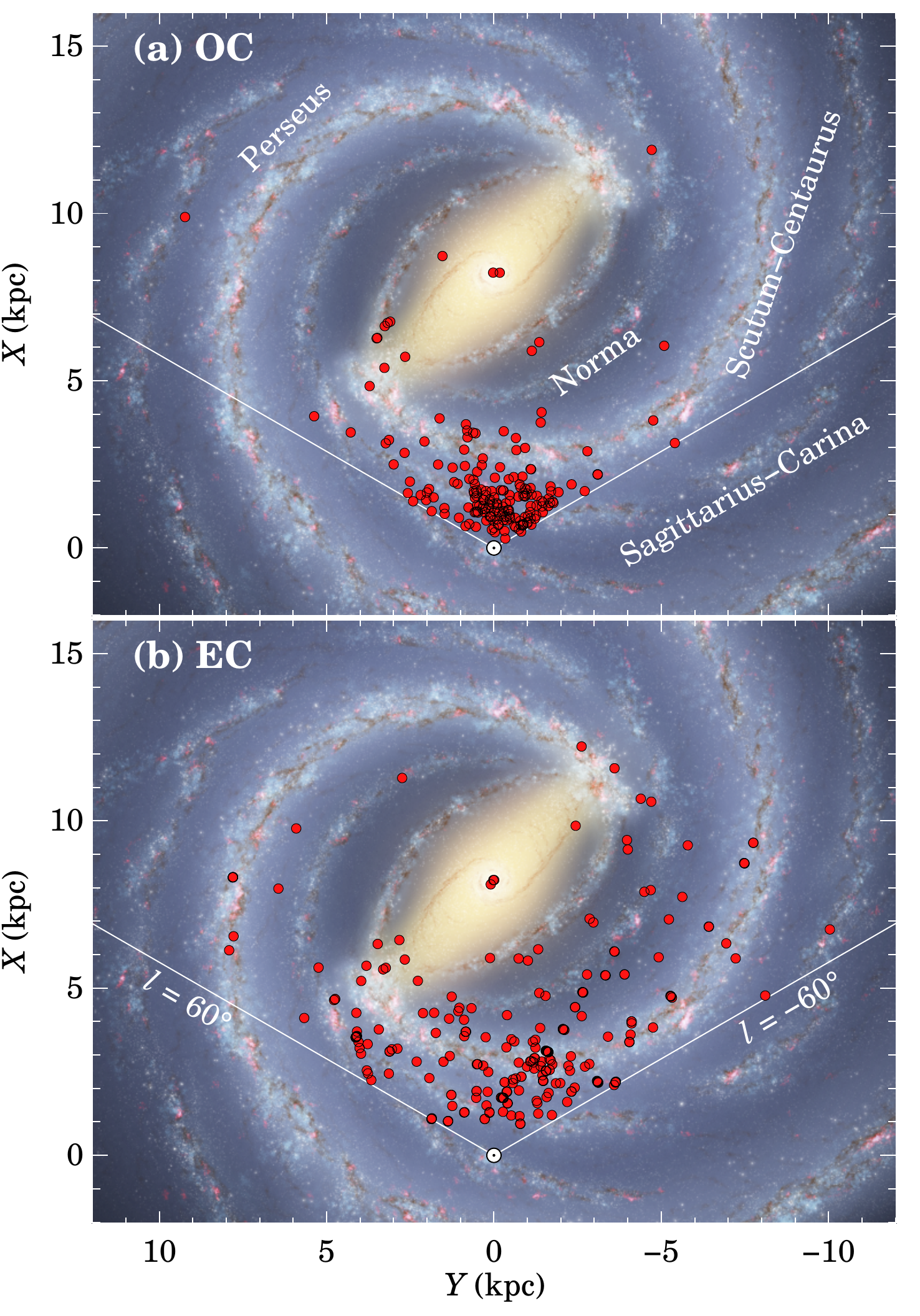}
\caption[Galactic distribution of the star cluster sample]{Galactic locations of (a) OCs and (b) ECs within the ATLASGAL range, superimposed over an artist's conception of the Milky Way (R. Hurt from the \emph{Spitzer} Science Center, in consultation with R. Benjamin), which was based on data obtained from the literature at radio, infrared, and visible wavelengths, and attempts to synthesize many of the key elements of the Galactic structure. The coordinate system is centered at the Sun position, indicated by the `$\sun$' symbol, and we have scaled the image such that $R_0 = 8.23$~kpc \citep{Genzel2010}. The two diagonal lines represent the ATLASGAL range in Galactic longitude ($|\ell| \le 60\degr$). In panel (a), we indicate the names of the spiral arms.}
\label{fig:galactic-distribution}
\end{figure}

In this Section, for the clusters in our sample with available distance estimates we study their spatial distribution in the Galaxy, and with respect to the Sun. Figure~\ref{fig:galactic-distribution} shows the Galactic distribution of the clusters separated in the (a) OC and (b) EC categories defined in the previous Section, on top of an artist's conception of the Milky Way viewed from the north Galactic pole (R. Hurt from the \emph{Spitzer} Science Center, in consultation with R. Benjamin). The image was constructed based on multiwavelength data obtained from the literature, and we have scaled it to $R_0 = 8.23$~kpc \citep[][see Section~\ref{sec:kin-distance}]{Genzel2010}. It is clear from the image that ECs probe deeper the inner Galaxy than the OC sample, which is concentrated within a few kpc from the Sun ($\lesssim 2$~kpc). This, of course is an observational effect mainly produced by the difficulty in detecting exposed clusters against the Galactic background, compared to ECs (see Section~\ref{sec:completeness}), and enhanced by the fact that some genuine OCs have no distance estimates and therefore cannot be included in the spatial distribution analysis (e.g., there are 123 clusters of type OC2 without available distance, half of which might be real). ECs are spread over larger distances from the Sun ($\lesssim 6$~kpc) and, although few of them can be detected beyond the Galactic center, a paucity of ECs is hinted within the Galactic bar, augmented by some apparent crowding close to both ends of the bar. The Galactic distribution of ECs is consistent with the spiral structure delineated on the background image; however, the large distance uncertainties ($\sim 0.5$~kpc on average, see Section~\ref{sec:distance-and-ages}), and the limited distance coverage, prevent the ECs from clearly defining the spiral arms by their own.

\begin{figure}[!t]
\centering
\includegraphics[width=0.48\textwidth]{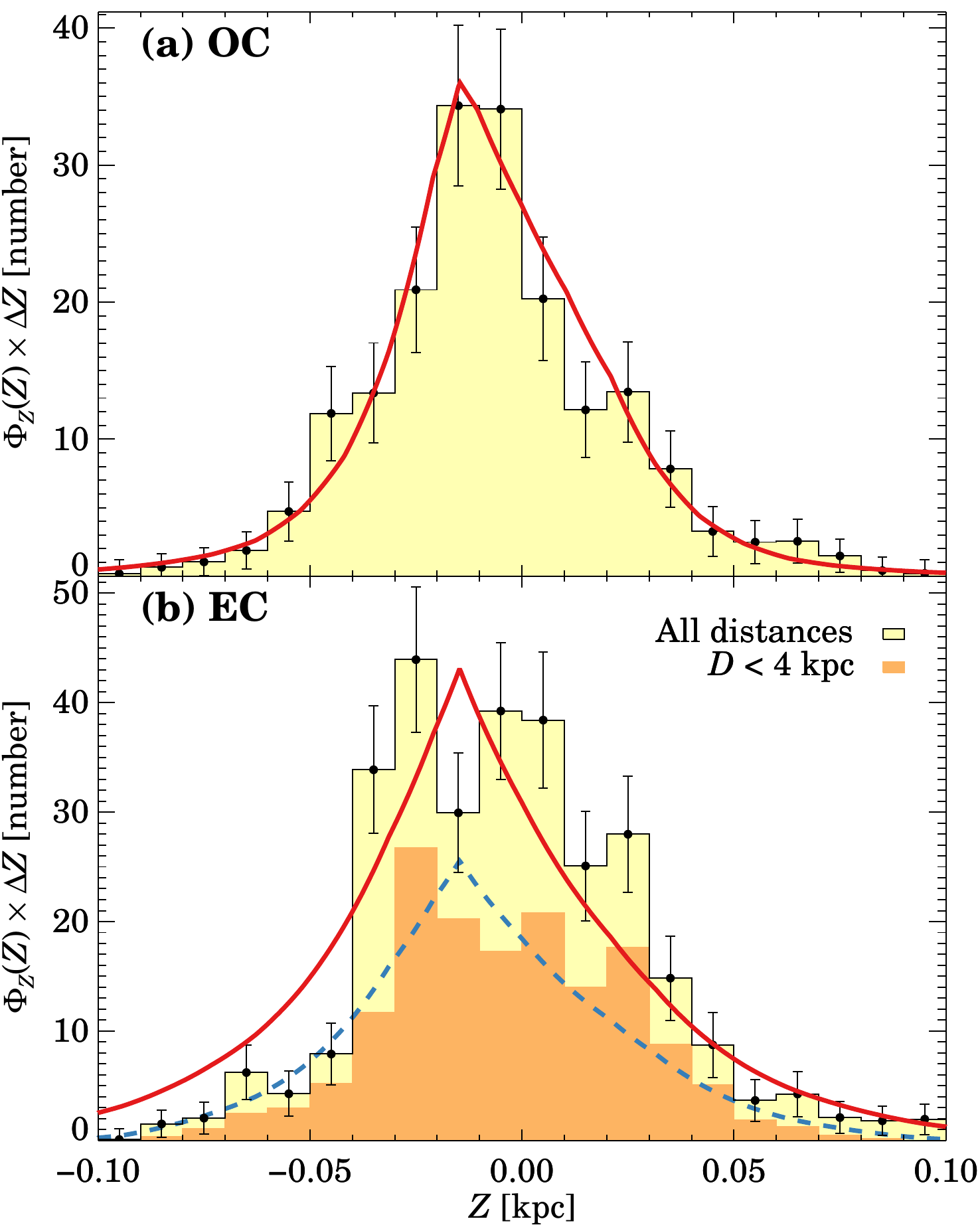}
\caption[Distribution of heights from the Galactic plane]{Histogram of heights from the Galactic plane, as measured from the Sun ($Z = z -z_0$), for (a) OCs and (b) ECs, using a bin width of $\Delta Z = 10$~pc and Poisson uncertainties. The overplotted solid curve in each panel represents: (a) the fitted $Z$-distribution $\Phi_Z(Z)$ from Equation~(\ref{eq:Phi(Z)}) with best-fit parameters $z_0 = 14.7 \pm 3.7$~pc and $z_{\rm h} = 42.5 \pm 9.9$~pc; (b) the predicted $Z$-distribution from Equation~(\ref{eq:Phi(Z)}), using the parameters fitted for the OC sample. In panel (b), the darker shaded region is the $Z$-histogram for ECs with distances $D < 4$~kpc, whereas the dashed curve indicates the corresponding distribution as predicted from Equation~(\ref{eq:Phi(Z)}) and the same parameters $z_0$ and $z_{\rm h}$.}
\label{fig:Z-distribution}
\end{figure}

\begin{figure}[!t]
\centering
\includegraphics[width=0.48\textwidth]{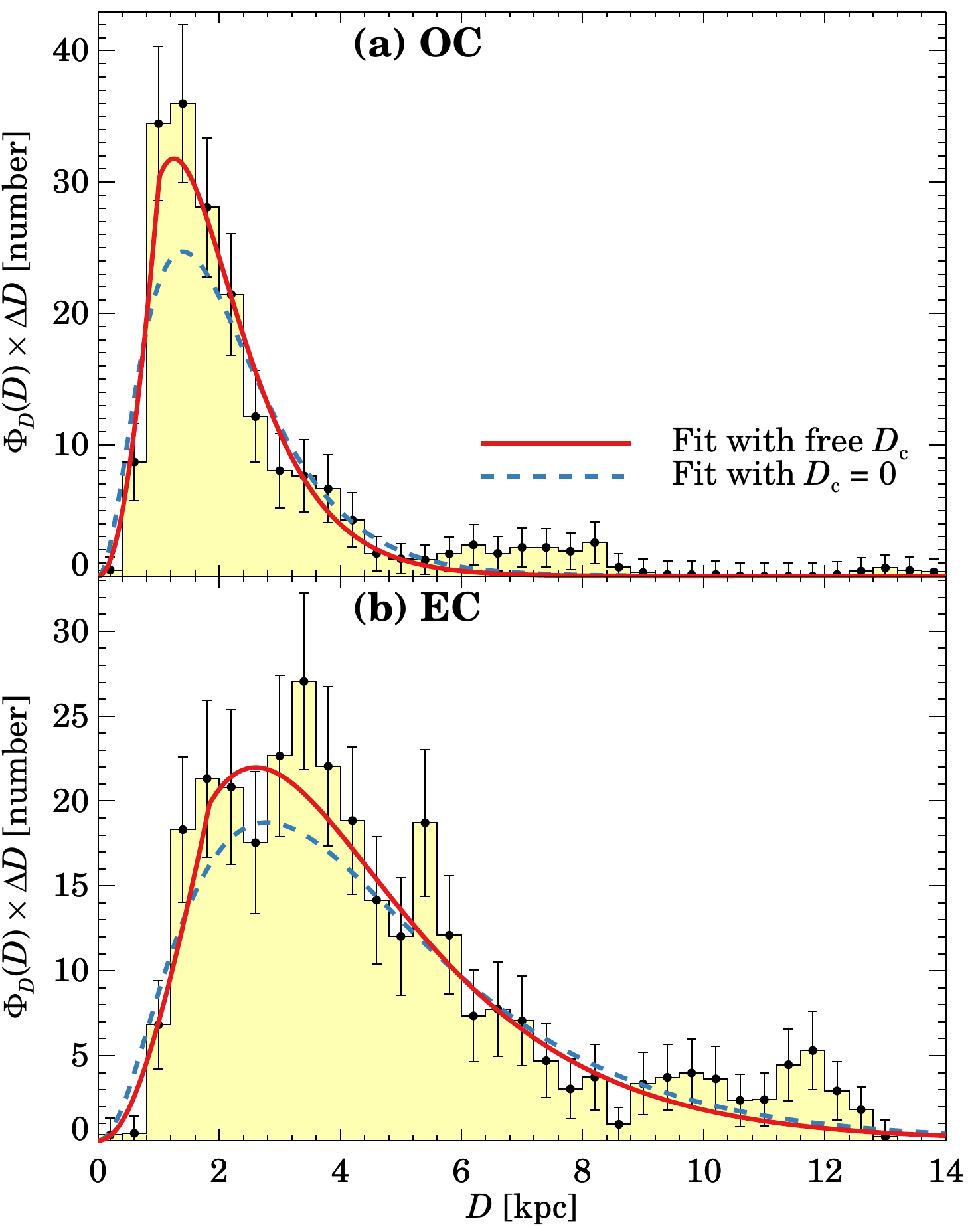}
\caption[Distribution of heliocentric distances]{Histogram of heliocentric distances, $D$, for (a) OCs and (b) ECs, using a bin width of $\Delta D = 0.4$~kpc and Poisson uncertainties. In each panel, the solid curve represents the fitted $D$-distribution $\Phi_D(D)$ from Equation~(\ref{eq:Phi(D)-observed}), with the completeness distance $D_{\rm c}$ as free parameter (see Equation~(\ref{eq:fc(D)})); the dashed curve shows the fit with fixed $D_{\rm c} = 0$ (see text for details). The best-fit parameters are given in Table~\ref{tab:parameters-ZD}.}
\label{fig:D-distribution}
\end{figure}

To really quantify how deep our OC and EC samples reach into the inner Galaxy, and to estimate the completeness fraction at a given distance, we need to study the observed heliocentric distance distribution of the clusters, and compare it to what is expected from making some basic assumptions. In the following, we denote by $D$ the distance of the cluster from the Sun, projected on the Galactic plane\footnote{In practice, we did not distinguish between the distance $d$ and the projected distance $D = d \cos b$. Since the maximum latitude within the ATLASGAL range is $|b| = 1.5\degr$, the difference is less than 0.03\%, far below the distance uncertainties.}, and by $z$ the height of the cluster above the Galactic plane. For simplicity, we also define $Z \equiv z - z_0$, where $z_0$ is the displacement of the Sun above the plane; this is actually what we obtain directly\footnote{In this paper, for simplicity we have assumed that the $b=0$ plane is parallel to the ``true'' Galactic plane, although in reality this is not the case (Goodman et al., in preparation). While this has a negligible effect on the distance distribution and the completeness, it may distort the derived height distribution when considering clusters at large distances from the Sun (see Section~\ref{sec:height-distribution}).} from the cluster distance $d$  and its Galactic latitude $b$, $Z = d \sin b$. The observed $Z$- and $D$-distributions are shown, respectively, in Figures~\ref{fig:Z-distribution} and \ref{fig:D-distribution}, for our cluster sample separated in OC and EC categories. In the construction of the histograms, we used fixed bins of $\Delta Z = 10$~pc and $\Delta D = 0.4$~kpc, but since the distance uncertainties are quite nonuniform, we have fractionally spread the ranges determined by the central values and their uncertainties over the covered bins. In other words, for a cluster with distance and uncertainty $D \pm \sigma_D$, we considered all the bins overlapping with the range $[D-\sigma_D,D+\sigma_D]$ and in each bin we added the fraction (with respect to the total width of the range, $2\sigma_D$) comprised by the corresponding overlap. The total OC and EC distance distributions were obtained by repeating this procedure for all the clusters. The $Z$-distributions were constructed using the same method, and the fitted curves plotted in Figures ~\ref{fig:Z-distribution} and \ref{fig:D-distribution} are explained in the following.

\subsubsection{Assumed model for the spatial distribution}

In general, we can assume that the spatial number-density of OCs or ECs in the Galaxy is described by a combination of two independent exponential-decay laws for the cylindrical coordinates $z$ and $R$, centered in the Galactic center: $\rho(R,z) = \rho_0 \,\varphi_R(R) \,\varphi_z(z)$, with $\varphi_R(R) = e^{-R/R_{\rm D}}$ and $\varphi_z(z) = e^{-|z|/z_{\rm h}}$. This is a common functional form used to characterize the Galactic distribution of stars \citep[see Section~1.1.2 of][]{BinneyTremaine2008}, and has already been applied in previous OC studies \citep{Bonatto2006,Piskunov2006}. One might want to consider the imprint of spiral arm structure in the azimuthal distribution of ECs, since they are still embedded in molecular clouds, but here we are interested in the distance and height longitude-averaged distributions, for which azimuthal substructure is less important. Furthermore, as noted above, our EC distances are not accurate enough to constrain the location of the spiral arms. If we transform the density $\rho(R,z)$ to a coordinate system centered at the Sun, and assume that we are observing the \emph{totality} of the clusters in the Galaxy within the ATLASGAL range ($|b| \le b_1$ and $|\ell| \le \ell_1$, with $b_1 \equiv 1.5\degr$ and $\ell_1 \equiv 60\degr$), the resulting density (not averaged in longitude $\ell$ yet) can be written as
\begin{equation}
\label{eq:rho(D,l,Z)} 
\rho_{\rm tot}(D,\ell,Z) = \left\{ \begin{array}{ll}
\rho_0 \, \varphi(D,\ell) \, \varphi_z(Z + Z_0) & \textrm{if}~~|Z| \le D \tan b_1 \\
0                                      & \textrm{else~,}
\end{array} \right.
\end{equation}
where
\begin{equation}
\label{eq:phi(D,l)} 
\varphi(D,\ell) \equiv \varphi_R\left(\sqrt{R_0^2 + D^2 - 2 R_0 D \cos \ell}\right)~.
\end{equation}
Now we can derive an analytical expression for the $D$-distribution of an ideally complete sample:
\begin{eqnarray}
\Phi_D^{\rm tot}(D) & \equiv & \int_{-\infty}^{\,\infty} \int_{-\ell_1}^{\,\ell_1} \rho_{\rm tot}(D,\ell,Z)\, D\,\rd \ell\,\rd Z\label{eq:Phi(D)-definition}\\
                    & =      & \Sigma_0\, f_{b_1}(D)\, D \int_{-\ell_1}^{\,\ell_1} \varphi(D,\ell) \,\rd \ell~, 
                    \label{eq:Phi(D)-general}
\end{eqnarray}
where $\Sigma_0 \equiv 2 z_{\rm h} \rho_0$ is the surface number-density on the Galactic disk for $R=0$, and we have defined the function $f_{b_1}(D)$ as
\begin{equation}
\label{eq:fb1(D)}
f_{b_1}(D) \equiv  \left\{ \begin{array}{ll}
e^{-z_0/z_{\rm h}}\, \sinh(D \tan b_1/ z_{\rm h}) & \textrm{if}~~D \le z_0 / \tan b_1 \\
1 - \cosh(z_0 / z_{\rm h})\, e^{-D \tan b_1 / z_{\rm h}} & \textrm{else~,}
\end{array} \right.
\end{equation}
which arises from the fact that the limited latitude coverage restricts the integration in $Z$ at each distance.

\subsubsection{Completeness fraction}

In practice, however, as already mentioned before and discussed in Section~\ref{sec:completeness}, we are unable to detect the totality of the clusters within the ATLASGAL range, due to the difficulty in star cluster identification towards the inner Galaxy. Indeed, the $D$-distributions that we really observe for OCs and ECs (see Figure~\ref{fig:D-distribution}) do not increase with distance up to the Galactic center ($D = R_0$), as we would expect from Equation~(\ref{eq:Phi(D)-general}); instead, they reach a maximum at a nearby distance and then decay considerably, especially for optical clusters. The observed $D$-distributions are dominated by the high incompleteness at increasingly larger distances from the Sun, and therefore, are insensitive to large scale structure on the Galactic disk such as the scale length $R_D$. Attempts to include $R_D$ in the parametric fit to the distance distributions described below resulted in heavily degenerated output parameters and practically no constraint on their values. We then eliminated the dependence of the model on $R_D$ by making the rough approximation that the underlying radial distribution of clusters is uniform, i.e., $\varphi_R(R) = 1$. This is supported by the fact that, due to the incompleteness, most clusters in our sample are within a few kpc from the Sun, where the variations in $\varphi_R(R)$ can be considered small relative to the completeness decay. The constants $\rho_0$ and $\Sigma_0$ must now be interpreted as Solar neighborhood values, and from Equation~(\ref{eq:Phi(D)-general}) the complete $D$-distribution becomes 
\begin{equation}
 \label{eq:Phi(D)-complete}
\Phi_D^{\rm tot}(D) = 2 \ell_1 \,\Sigma_0 \,f_{b_1}(D) \,D~.
\end{equation}
On the other hand, defining a fractional factor $f_{\rm c}(D)$ that quantifies the completeness of the cluster sample as a function of distance\footnote{Ideally, one should consider a completeness fraction dependent on Galactic longitude also, $f_{\rm c}(D,\ell)$, as we expect lower cluster detectability for low $|\ell|$, where the stellar background is higher. However, since we made the approximation $\varphi(D,\ell) = 1$, the integration in longitude would only affect the term $f_{\rm c}(D,\ell)$, and therefore the factor $f_{\rm c}(D)$ we used can be thought as a longitude-averaged completeness fraction.}, we can express the observed $D$-distribution $\Phi_D(D)$ as
\begin{equation}
 \label{eq:Phi(D)-observed}
\Phi_D(D) = 2 \ell_1 \,\Sigma_0 \,f_{\rm c}(D) \,f_{b_1}(D) \,D~.
\end{equation}

In order to assign a particular parametric shape to the completeness fraction, we chose an ansatz for $f_{\rm c}(D)$ based on previous statistical works of OCs in the whole sky. \citet{Bonatto2006} studied the WEBDA database\footnote{WEBDA is an on-line OC database originally developed by \citet{Mermilliod1996}, and available on \url{http://www.univie.ac.at/webda/}; the clusters of this database are included in the \citet{Dias2002} catalog.} at that time and found, by completeness simulations, that their analyzed OC sample is highly incomplete in the inner Galaxy, even within what they called the ``restricted zone'', defined as an annulus segment with Galactocentric distances $R$ in the range $[R_0 -1.3~{\rm kpc},R_0 +1.3~{\rm kpc}]$. The completeness fraction they determined decays almost immediately from $R = R_0$ to $R < R_0$ (see their Fig.~11; note that $R_0 = 8.0$~kpc in that work). However, \citet{Piskunov2006} claim that the \citet{Kharchenko2005-known,Kharchenko2005-new} OC catalogs constitute a complete sample up to about 0.85 kpc from the Sun. This is nicely illustrated in their Fig.~1, where a flat distribution of surface number-density of clusters is exhibited up to that distance, after which the distribution starts to decrease considerably. If the completeness fraction of their sample in the inner Galaxy were similar to that obtained by \citet{Bonatto2006}, the surface density distribution would be a decreasing function immediately from $D=0$~kpc rather than from $D=0.85$~kpc\footnote{We checked by numerical integration of $\Sigma(D) \propto \int_0^{2\pi} \varphi(D,\ell) \rd \ell$ that the raising of the surface density distribution in the inner Galaxy due to an exponential Galactic disk is practically imperceptible for $D < 1$~kpc, and therefore, a flat distribution cannot be the combined result of incompleteness and exponential disk structure.}. We think that this discrepancy is mainly caused by two effects: 1) the cluster sample studied by \citet{Bonatto2006} (654 objects with known distances) is less complete than, e.g., the current version of the \citet{Dias2002} catalog used in this work (1309 clusters with available distances), which is equivalent to the \citet{Kharchenko2005-known,Kharchenko2005-new} sample within 0.85 kpc; and 2) the ``restricted zone'' considered by \citet{Bonatto2006} covers a larger area than the circle defined by the completeness limit of \citet{Piskunov2006} (radius of 0.85 kpc centered at the Sun), and thus includes regions where the OC sample is indeed incomplete. In fact, we performed a quick test on the current \citet{Dias2002} catalog by constructing the Galactocentric radii distribution of clusters within 1~kpc from the Sun, and we obtained a shape that is not incompatible with a exponential law in the whole range, as opposed to the distribution derived by \citet[][their Fig.~9]{Bonatto2006}.

Based on the above discussion, the completeness fraction for our OC sample is likely $\sim 1$ up to a close distance from the Sun, $D_{\rm c}$, and then starts to decay significantly. We assume that the decay is exponential:
\begin{equation}
 \label{eq:fc(D)} 
f_{\rm c}(D) = \left\{ \begin{array}{ll}
1                      & \textrm{if}~~D \le D_{\rm c}\\
e^{-(D-D_{\rm c})/s_0} & \textrm{else~.}
\end{array} \right.
\end{equation}
This parametrization allows us to investigate the possibility that the sample is always incomplete, as for \citet{Bonatto2006}, by just imposing $D_{\rm c} = 0$. We employ the same functional form for the completeness fraction of ECs, but of course varying the parameters $D_{\rm c}$ and $s_0$.

\subsubsection{Fit for the height distribution}
\label{sec:height-distribution}

Before proceeding to fit Equation~(\ref{eq:Phi(D)-observed}) to the observed $D$-distributions, we first need some estimates for $z_{\rm h}$ and $z_0$ which are used to compute the factor $f_{b_1}(D)$. We obtain those estimates from the $Z$-distribution, which can be analytically written as
\begin{equation}
\label{eq:Phi(Z)}
\Phi_Z(Z) = e^{-|Z + z_0|/z_{\rm h}} \int_{|Z|/\tan b_1}^{\infty} \,\frac{\Phi_D(D)}{2 z_{\rm h}\,f_{b_1}(D)} \,\rd D~.
\end{equation}
The advantage in writing this equation explicitly in terms of $\Phi_D(D)$ is that we can directly use the observed $D$-distribution instead of its analytical expression (and compute the integral numerically), so that it is possible to fit the $Z$-distribution with only two free parameters, $z_0$ and $z_{\rm h}$, and independently of the fit for the distance distribution. All the fits were performed using the Levenberg-Marquardt least-squares minimization package \verb|mpfit| \citep{Markwardt2009}, implemented in IDL, and we have assumed Poisson uncertainties. The best fit of Equation~(\ref{eq:Phi(Z)}) to the observed $Z$-distribution of OCs is shown in Figure~\ref{fig:Z-distribution}(a) as a solid curve, and the corresponding fitted parameters are $z_0 = 14.7 \pm 3.7$~pc and $z_{\rm h} = 42.5 \pm 9.9$~pc. These values are in excellent agreement with the ones derived by \citet{Bonatto2006}, if we consider their scale height $z_{\rm h}$ within the Solar circle (which is the case for almost the totality of our OC sample).

The observed $Z$-distribution of ECs (Figure~\ref{fig:Z-distribution}(b)) is much more irregular than that of OCs, and therefore a proper fit is not possible. This is likely due to the fact that ECs are spread over a larger area than OCs, and therefore, present lower statistics in the Solar neighborhood and larger average errors in $Z$ ($Z \propto D$). In addition, ECs are usually grouped in complexes, as we will see in Section~\ref{sec:statistics} and can already be noted in Figure~\ref{fig:galactic-distribution}(b), where some particular locations appear crowded with many close objects, enhancing the non-uniformity of their spatial distribution. However, if we adopt the same parameters $z_0$ and $z_{\rm h}$ derived from the OC sample and compute the predicted distribution from Equation~(\ref{eq:Phi(Z)}) (naturally, using now the observed $\Phi_D(D)$ of ECs), the resulting curve is roughly consistent with the observed $Z$-distribution, as shown in Figure~\ref{fig:Z-distribution}(b) (solid line). The most systematic discrepancy can be identified for $Z < -40$~pc, where there is a significant deficit of observed clusters with respect to the predicted distribution, probably due to the difficulty in detecting ECs below the Galactic disk for large distances. Indeed, Figure~\ref{fig:Z-distribution}(b) also shows the observed $Z$-distribution for ECs with $D < 4$~kpc (darker inner histogram) and the corresponding prediction (dashed curve), and we can see that in this case the deficit of observed clusters below the Galactic plane is only marginal. Another explanation might be the fact that we have assumed that the $b=0$ plane is parallel to the Galactic disk, while in reality the combined effect of the offset of the Sun above the ``true'' Galactic plane, and of the Galactic center below the $b=0$ plane, slightly tilts the $b=0$ plane towards the south of the Galaxy (see Goodman et al., in preparation), so that clusters at large distances from the Sun and below the Galactic plane would appear at more negative values in the true $Z$-distribution. This could help to populate the bins in the range of the deficit of observed clusters, and would also explain why the deficit is less important for the distribution of clusters with $D < 4$~kpc.

\subsubsection{Fit for the distance distribution}

Using now values for $z_0$ and $z_{\rm h}$ obtained from the OC sample, which are also consistent with the EC height distribution, to compute the factor $f_{b_1}(D)$ defined in Equation~(\ref{eq:fb1(D)}), we fitted the analytical distribution $\Phi_D(D)$ from Equation~(\ref{eq:Phi(D)-observed}) to the observed $D$-distributions of OCs and ECs, with free parameters $\Sigma_0$, $D_{\rm c}$, $s_0$. The last two parameters are implicit in the completeness factor $f_{\rm c}(D)$ defined in Equation~(\ref{eq:fc(D)}). The best fits are overplotted as solid curves on the corresponding histograms of Figure~\ref{fig:D-distribution}, and the fitted parameters are given in Table~\ref{tab:parameters-ZD}. As can be already noted in the plots and confirmed by the reduced $\chi^2$ values (0.90 for OCs, and 1.48 for ECs), the assumed form of the completeness fraction (Equation~(\ref{eq:fc(D)})) is a good representation of the overall detectability of star clusters in the inner Galaxy. The few outliers in the observed distribution with respect to the fitted analytical function for OCs with distances $D \gtrsim 6$~kpc mainly correspond to exposed clusters recently discovered at infrared wavelengths. A similar tendency is hinted for ECs with $D \gtrsim 11$~kpc, although in this case these outliers are also consistent with the irregular nature of the distribution in general, which slightly deviates (at one-sigma level) from the fitted curve at other distance bins. However, some problems with the resolution of the KDA, resulting in ECs incorrectly assigned to the far distance, cannot be ruled out.

It is remarkable that, despite the lower statistics caused by restricting to the ATLASGAL range, the fitted completeness limit of our OC sample, $D_{\rm c} = 1.01 \pm 0.16$~kpc, is consistent with that derived by \citet{Piskunov2006} for their all-sky sample in the Solar neighborhood\footnote{Very recently, a significant effort in obtaining distances and other parameters of most of the known OCs and ECs has been published by \citet{Kharchenko2013}, who claim an overall completeness limit of 1.8~kpc. Since ECs are not dominant within a complete sample, the new limit represents an intrinsic improvement in the OC completeness.}. For ECs, both the completeness limit $D_{\rm c}$ and the completeness scale length $s_0$ are larger than the corresponding values of the OC distribution (see Table~\ref{tab:parameters-ZD}), quantitatively confirming that, from an observational point of view, the EC sample traces larger distances from the Sun than the ones traced by our OC sample.

\begin{table}[!t]
\renewcommand{\arraystretch}{1.1}
\caption[Best-fit parameters from the $Z$- and $D$-distributions.]{Best-fit parameters from the $Z$- and $D$-distributions of OCs and ECs.}
\label{tab:parameters-ZD}
\centering
\begin{tabular}{lll}
\hline\hline
Parameter               &  OC         & EC \\
\hline
$z_0$~(pc)              &  14.7 (3.7)  & \nodata\tablefootmark{a}\\
$z_{\rm h}$~(pc)        &  42.5 (9.9)  & \nodata\tablefootmark{a}\\
\hline
$\Sigma_0$~(kpc$^{-2}$) &  82.9 (12.9) &  19.5 (3.1)  \\ 
$s_0$~(kpc)             &  0.72 (0.05) &  1.81 (0.10) \\
$D_{\rm c}$~(kpc)       &  1.01 (0.16) &  1.84 (0.35) \\
\hline
$\Sigma_0'$~(kpc$^{-2}$)&  209.1 (33.3)& 40.3 (5.0) \\
$s_0'$~(kpc)            &  0.82 (0.04) & 1.99 (0.09)\\
\hline
\end{tabular}
\tablefoot{
$z_0$ is the displacement of the Sun above the plane, and $z_{\rm h}$ is the scale height; $\Sigma_0$ is the local surface number-density, $s_0$ is the length scale of the completeness, and $D_{\rm c}$ is the completeness distance (see Equation~(\ref{eq:fc(D)})); $\Sigma_0'$ and $s_0'$ represent values derived from an alternative fit with fixed $D_{\rm c} = 0$. Quantities between parentheses are the corresponding uncertainties.\\
\tablefoottext{a}{Fit not possible; assumed values from the OC sample.}
}
\end{table}

The fitted completeness limits for OCs and ECs are significantly above zero, practically discarding the possibility that the cluster samples are always incomplete in the inner Galaxy, as suggested by \citet{Bonatto2006} for OCs. To further test this option, we performed an alternative fit of Equation~(\ref{eq:Phi(D)-observed}) to the observed $D$-distributions, now fixing $D_{\rm c} = 0$. For each distribution in Figure~\ref{fig:D-distribution}, the resulting best fit is shown as a dashed line, and we immediately notice that this alternative fit is poorer than the one with $D_{\rm c}$ as free parameter, specially for OCs. Indeed, we applied a Kolmogorov-Smirnov test to all the fitted distribution functions in a distance range free of far-distance outliers ($D \le 6$~kpc for OCs, $D \le 9$~kpc for ECs), and we found that the $D_{\rm c} = 0$ fit can be rejected with a significance level of 5\% for OCs, and 6.5\% for ECs. We thus conclude that the OC and EC samples in the inner Galaxy are roughly complete up to a distance of $\sim 1$~kpc and $\sim 1.8$~kpc, respectively, as derived from the free-$D_{\rm c}$ fits.

\subsection{Discussion on the completeness}
\label{sec:completeness}

In general, the existence of a stellar cluster is observationally established by an excess surface density of stars over the background, so that its detectability depends on its richness, its angular size, the number of resolved individual members and their apparent brightness (which is directly related to the distance), the surface density of field stars, and the amount of extinction on the line of sight \citep{LadaLada2003}. Consequently, it is particularly difficult to identify a star cluster in the inner Galactic plane, where both the stellar background and the extinction are relatively high, or a very distant cluster, for which its members appear faint and could be confused as a few single stars due to limited angular resolution of the observations. In fact, we have shown in the previous Section that the current samples of OCs and ECs in the inner Galaxy are complete up to only a close distance from the Sun, and then the completeness heavily decreases as distance increases.

We have also seen that incompleteness affects the OC sample more severely than the ECs, i.e., the latter have a higher completeness limit and a less drastic decay in the completeness fraction. At first glance, this might seem contradictory since ECs are, by definition, embedded in molecular clouds and thus subject to a high degree of in situ dust extinction. However, at infrared wavelengths, ECs become easier to detect than exposed clusters because it is easier to distinguish them from the field population. Since ECs are usually associated with illuminated interstellar material, they can be identified by eye towards the locations of known nebulae or star-forming regions \citep[e.g.,][]{Dutra2003-2mass,Bica2003-2mass,Borissova2011}, even if the clusters are partially resolved or highly contaminated by extended emission. In other words, despite bright nebular emission can prevent young stars from being found by point source detection algorithms and therefore hide the host EC from automated searches, at the same time it can help to identify such a cluster when searched by eye against a high stellar background. For clusters with fainter or less irregular extended emission, automated searches can also take advantage of some distinctive characteristic of ECs (like the red-color criterion of our GLIMPSE search, see Section~\ref{sec:newglimpse}) to separate them from the background, which is in general not feasible for an evolved OC because its member stars present similar observational properties than the field population.

It is interesting to compare our distance distribution of ECs (Figure~\ref{fig:D-distribution}(b)) with that of individual \emph{Spitzer}-detected YSOs \citep{Robitaille2008}, as simulated by \citet{RobitailleWhitney2010} using a population synthesis model. They show that the synthetic YSOs that would have been detected by \emph{Spitzer} and included in the \citet{Robitaille2008} catalog correspond to massive objects with a mass distribution that peaks at $\sim 8 M_{\sun}$. The corresponding distance distribution of this model is presented in Fig.~1 of \citet{Beuther2012} for the $10\degr \le \ell \le 20\degr$ range. The plot reveals a high number of far YSOs up to distances of $\sim 14$~kpc, showing that, despite the high extinction, individual (massive) YSOs can be detected deep into the Galactic plane, as opposed to ECs. We therefore think that the low detectability of a far EC is mainly due to the faint apparent brightness of its low-mass population and confusion of its members, so that the whole cluster might be misidentified as an individual massive young star. At near-infrared wavelengths, however, extinction could still play an important role in hiding a far EC.

\subsection{Definition of a representative sample}
\label{sec:representative-sample}

We can quantify how many OCs and ECs we are missing within a certain distance from the Sun, using the analytical expressions for the observed distance distribution, $\Phi_D(D)$ (Equation~(\ref{eq:Phi(D)-observed})), and for the distance distribution that would be observed if we detected the totality of the clusters in the inner Galaxy, $\Phi_D^{\rm tot}(D)$ (Equation~(\ref{eq:Phi(D)-complete})), and using the fitted parameters given in Table~\ref{tab:parameters-ZD}.  We define the cumulative completeness fraction, $F_{\rm c}(D)$, as the ratio of the number of observed clusters with distances~$\le D$ to the number that would represent a complete sample within~$D$:
\begin{equation}
 \label{eq:cumulative-completeness}
F_{\rm c}(D) \equiv \frac{N_{\rm cl}(\le D)}{N_{\rm cl}^{\rm tot}(\le D)} = 
                    \frac{\displaystyle \int_0^D \Phi_D(D')\,\rd D'}
                         {\displaystyle \int_0^D \Phi_D^{\rm tot}(D')\,\rd D'}~. 
\end{equation}

Now we can define a \emph{representative} cluster sample as all objects with distances~$D \le D_{\rm rep}$ for which the fraction $F_{\rm c}(D_{\rm rep})$ is above a certain threshold in both the OC and EC samples (this naturally places the restriction on the OC sample alone, since it is more incomplete). We chose a threshold of 0.25, for which the distance has to be $D \le 3.15$~kpc. For simplicity, we just adopt $D_{\rm rep} = 3.0$~kpc, where $F_{\rm c}(D_{\rm rep}) = 0.28$ and $F_{\rm c}(D_{\rm rep}) = 0.79$ for the OC and EC samples, respectively. Note that although the selection of the threshold is somewhat arbitrary, if we keep in mind the above fractions, we only need a certain distance limit $D_{\rm rep}$ where the samples are not too incomplete and at the same time have a reasonable absolute number of objects to perform a statistical analysis.

In Column 4 of Table~\ref{tab:morph-types}, we list the number of clusters with $D \le 3.0$~kpc for each morphological type; the total number of ECs in the representative sample is 98. To count the number of OCs, according to our definition we need that the clusters are also confirmed (\verb|ref_Conf| not empty). The number of confirmed clusters with $D \le 3.0$~kpc is given in Column 5 for each morphological type, from which we obtain a total number of 146 OCs in the representative sample. With the fractions $F_{\rm c}(D_{\rm rep})$ computed before, it is also possible to estimate the number of clusters $N_{\rm cl}^{\rm tot}(\le D_{\rm rep})$ that we would observe within 3~kpc, if we had complete samples of OCs and ECs. The corresponding estimates are listed in Column 6, and were simply derived as $N_{\rm cl}(\le D_{\rm rep})/0.79$ for EC types, and $N_{\rm cl}^{\rm conf}(\le D_{\rm rep})/0.28$ for OC types. Note that the large number of OC2 clusters in this ideally complete sample is due to the fact that they cover a wide age range. The age distribution of our sample is analyzed in the next Section.

\subsection{Ages}
\label{sec:age-distribution}

We would expect that the ages of the stellar clusters increase along the morphological evolutionary sequence defined in Section~\ref{sec:evolutionary-sequence}. By dividing the cluster sample in such morphological types, we indeed obtained an increasing tendency in the corresponding ages distributions. However, we were unable to estimate an average age or age ranges for each individual type, given the low number of clusters with available ages that fall within each category, except for OC2. In the whole sample, for types EC1, EC2, OC0 and OC1 there are, respectively, only 9, 16, 15 and 9 objects with age estimates, whereas for OC2 clusters there are 160. Note that for types OC0 and OC1, the total number of objects is also low (see Table~\ref{tab:morph-types}), so that the main reason for the small number of age estimates is the low absolute statistics. On the other hand, for the much more numerous EC1 and EC2 morphological types (and possibly also part of the OC0 type), the lack of age estimates may simply be caused by the difficulties involved in obtaining these values.

It is still possible, however, to derive an upper limit for the ages of the ECs (EC1 and EC2 together), and also to study the age distribution of the whole OC population (OC0, OC1 and OC2 together), as described below.

\subsubsection{Upper limit age of ECs}
\label{sec:age-embedded}

The EC ages compiled from the literature were estimated using a variety of methods, including: comparison with theoretical isochrones on a Hertzsprung-Russell diagram constructed after spectroscopic classification in the near-infrared \citep[e.g.,][]{Furness2010}, use of the relation between the circumstellar disk fraction in the cluster and its age \citep[following][]{Haisch2001}, and comparison with synthetic clusters constructed by Monte Carlo simulations \citep{SteadHoare2011}, among others. We remark that from the 25 ECs with available age estimates, there are two objects that seem to be artificial outliers, with too old ages to be embedded, namely $7.5 \pm 2.6$~Myr and $25 \pm 7.5$~Myr \citep[respectively, clusters VVV~CL100 and VVV CL059 from][]{Borissova2011}\footnote{Note that the quoted uncertainties are from our catalog, which might be larger than the values given in the original paper because we adopted minimum errors for the age estimates (see Section~\ref{sec:distance-and-ages})\label{fn:age-errors}.}. These two objects are precisely the only ECs in our sample whose age was determined with the distance via isochrone fitting and the high uncertainty of this method for very young clusters is indeed acknowledged by the authors \citep{Borissova2011}. In a few other cases where isochrone fitting was used to derive the age of an EC, an independent measure of the distance was used as input in order to reduce the uncertainty \citep[e.g.,][]{Ojha2010}.

Excluding these two outliers from our sample, we found that 90\% (21 out of 23) of the ECs with available age estimates are younger than 3~Myr. Furthermore, given the high errors in this age range, even the remaining two clusters are consistent with being younger than 3~Myr, within the uncertainties: age of $3.3 \pm 2.1$~Myr for $[$BDS2003$]$~139 \citep{SteadHoare2011}, and $4.2 \pm 1.5$~Myr for $[$DBS2003$]$~118 \citep{Roman2007}\textsuperscript{\ref{fn:age-errors}}. We therefore adopt an upper limit of 3~Myr for the embedded phase, which represents a better constraint than the 5~Myr limit often quoted in the literature \citep[from][]{Leisawitz1989}. Since practically all available EC ages in our sample are $\lesssim 3$~Myr, the same result is obtained if we consider the representative sample ($D \leq D_{\rm rep} = 3$~kpc), despite the low statistics (10 out of 11 ECs are formally younger than 3 Myr, after removing one outlier).

\subsubsection{Age distribution of OCs}
\label{sec:age-distribution-fit}

\begin{figure}[!t]
\centering
\includegraphics[width=0.48\textwidth]{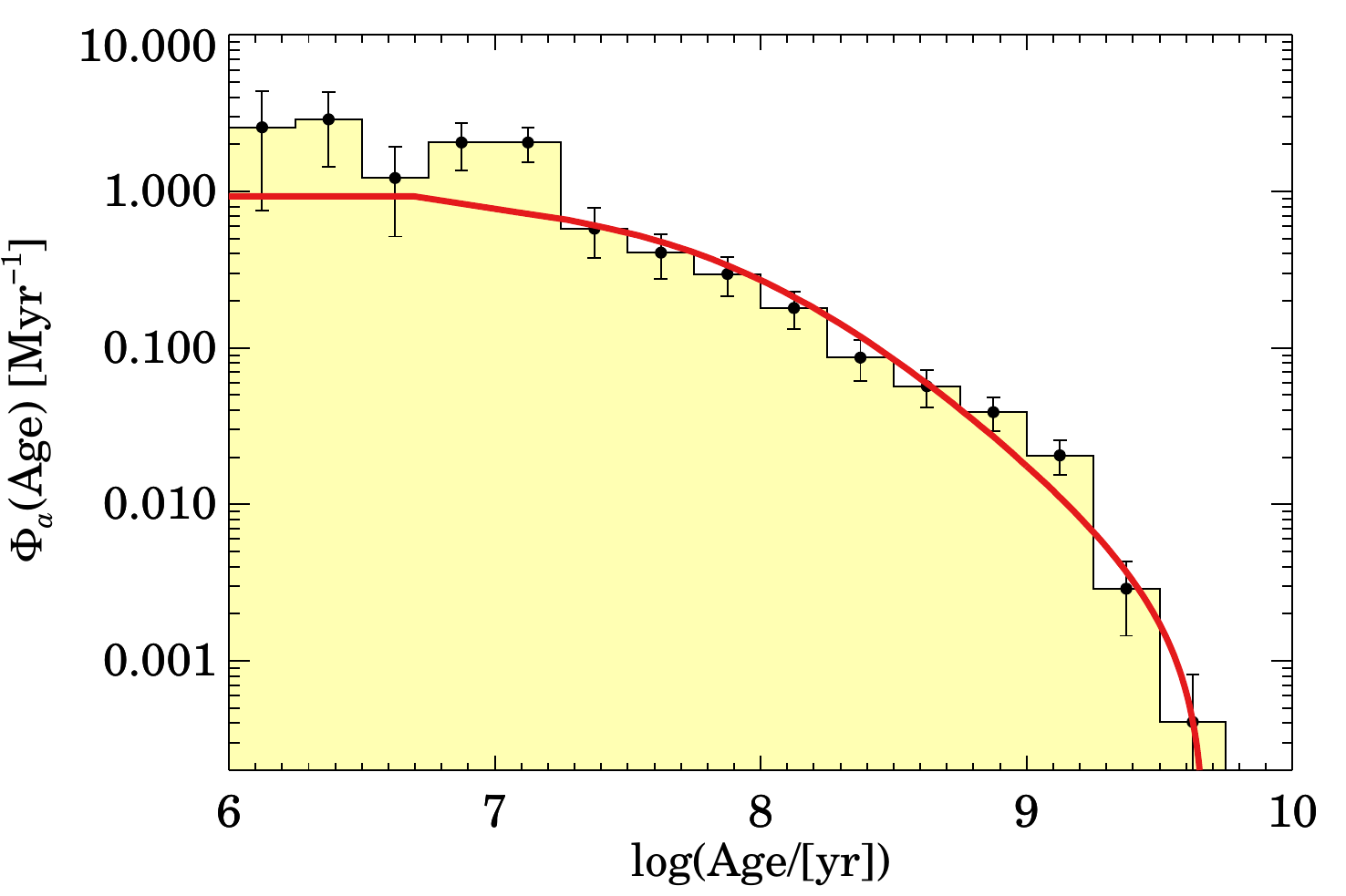}
\caption[Age distribution of OCs]{Age distribution of OCs within the representative sample ($D \leq 3$~kpc), using a logarithmic bin width of $\Delta \log({\rm Age}/{\rm yr}) = 0.25$ and Poisson uncertainties. The solid curve corresponds to the fitted age distribution from Equation~(\ref{eq:age-distribution}), following \citet{LamersGieles2006}, with best-fit parameters ${\rm CFR} = 0.93 \pm 0.09$~Myr$^{-1}$ and $M_{\rm max} = (4.46 \pm 0.85)\times 10^4~M_{\sun}$.}
\label{fig:age-distribution}
\end{figure}

The much higher number of OCs with available age estimates allowed us to study their age distribution, which is shown in Figure~\ref{fig:age-distribution} for the representative sample (a total of 143 OCs). Assuming a constant cluster formation rate (CFR), the decreasing number of OCs as time evolves is due to the effect of different disruption processes. \citet{LamersGieles2006} provide a theoretical parameterization of the survival time of initially bound OCs in the solar neighborhood, taking into account four main mechanisms: stellar evolution, tidal stripping by the Galactic gravitational field, shocking by spiral arms, and encounters with giant molecular clouds. They show that the observed age distribution $\Phi_a(a)$ for a constant CFR and a power-law cluster initial mass function with a slope of $-2$ can be written as
\begin{equation}
\label{eq:age-distribution}
\Phi_a(a) = C \left[\left(\frac{M_{\rm lim}(a)}{M_{\sun}}\right)^{-1} - 
                    \left(\frac{M_{\rm max}}{M_{\sun}}\right)^{-1} \right]~,
\end{equation}
where $a$ is the age, $C$ is a constant, $M_{\rm lim}(a)$ is the initial mass of a cluster that, at an age $a$, reaches a mass equal to the detection limit (assumed to be 100~$M_{\sun}$), and $M_{\rm max}$ is the maximum initial mass of clusters that are formed. It can be shown that the cluster formation rate within the initial mass range $[100\,M_{\sun},M_{\rm max}]$ is related with the factor $C$ by
\begin{equation}
\label{eq:cfr}
{\rm CFR} = C \left[\frac{1}{100} - 
                    \left(\frac{M_{\rm max}}{M_{\sun}}\right)^{-1} \right]~.
\end{equation}

We fitted $\Phi_a(a)$ from Equation~(\ref{eq:age-distribution}) to the observed age distribution of OCs in the representative sample, with free parameters $C$ and $M_{\rm max}$; the input function $M_{\rm lim}(a)$ was obtained by digitizing the dashed curve in Fig.~2 of \citet{LamersGieles2006}. We plot the resulting best fit as a solid curve in Figure~\ref{fig:age-distribution}, corresponding to the parameters ${\rm CFR} = 0.93 \pm 0.09$~Myr$^{-1}$ and $M_{\rm max} = (4.46 \pm 0.85)\times 10^4~M_{\sun}$. It is clear from the figure that there is an excess of observed young OCs with respect to the fitted theoretical distribution, whereas for older ages the fit is a pretty good representation of the data. The observed excess of young OCs could be the result of two effects. First, young OCs dominate at larger distances because they contain more luminous stars, so that within an incomplete sample the proportion of young OCs is relatively higher than that of older clusters \citep{Piskunov2006}. Second, since the parameterization of \citet{LamersGieles2006} considers the dissolution of initially bound OCs due to classical mechanisms, the observed over-population of young clusters might consists of associations, i.e., clusters which are already unbound due to disruption processes that are not accounted for by \citet{LamersGieles2006}. These associations will quickly dissolve into the field and, therefore, will not be able to populate the older age bins of the distribution in the future. 

While the age-dependent incompleteness is likely playing a role within our $D_{\rm rep} = 3$~kpc limit, it is interesting to investigate whether or not there is also a contribution from the presence of associations, for which we need to restrict the sample to smaller distances, where the incompleteness is not important. We found that the excess of observed young OCs still holds if we perform the fit for samples restricted to successively smaller distances, down to $D \leq 1.4$~kpc; nevertheless, the low statistics in the Solar neighborhood within the ATLASGAL range prevents us to perform this test on an even more restricted subsample of our catalog. We therefore fitted the model to all-sky samples of OCs, namely, the \citet{Dias2002} catalog and the \citet{Kharchenko2005-known,Kharchenko2005-new} sample, restricted to a certain limit in projected distance, $D$. For clusters with $D \leq 0.6$~kpc, in both samples, we recovered the results from \citet{LamersGieles2006}\footnote{This is totally expected for the Kharchenko et al. sample, since \citet{LamersGieles2006} used basically the same clusters. The only difference is that they did not include the objects newly detected by \citet{Kharchenko2005-new}. On the other hand, the fact that for the \citet{Dias2002} sample we obtain the same result implies that there are no systematic effects arising from differences between both samples, in particular regarding the age estimates.}, whose observed age distribution practically does not show the excess of young OCs with respect to the fitted curve (see their Fig.~3). If we restrict the samples to $D \leq 1.4$~kpc, however, the age distribution for the \citet{Dias2002} catalog presents a statistically significant over-population of young OCs, whereas for the \citet{Kharchenko2005-known,Kharchenko2005-new} sample the excess is only marginal.

Given that the \citet{Kharchenko2005-known,Kharchenko2005-new} sample is a subset of the \citet{Dias2002} catalog, this behavior means that the young excess in the sample with $D \leq 1.4$~kpc cannot \emph{purely} be due to the age-dependent incompleteness, since otherwise we would obtain a more noticeable effect in the less complete sample. Then, there must necessarily be a contribution from presence of associations. The excess is less significant for the Kharchenko et al. catalog and not noticeable for clusters in both samples with $D \leq 0.6$~kpc probably because there is an observational limitation in detecting associations at very close distances, due to their larger sizes. In summary, we think that the excess of young clusters in our representative OC sample ($D \leq 3.0$~kpc) with respect to the theoretical description of \citet{LamersGieles2006} is caused by a combination of age-dependent incompleteness and presence of associations.

The age distribution shown in Figure~\ref{fig:age-distribution} was constructed using a bin width large enough to ensure good statistics over the whole age range, but we can refine the grid to constrain better a certain feature, as long as the presentation remains statistically significant. By constructing the age distribution with smaller bin widths and doing the fitting again, we found that the transition after which the theoretical description fits well the data occurs at an age of $\log(a/{\rm yr}) \simeq 7.2$, i.e., $\sim 16$~Myr. Consistently, we have seen in Section~\ref{sec:classification-oc-ec} that the $\sim 16$~Myr limit is roughly the age before which an observed OC might be either an association or a physical OC, whereas observed OCs older than that are practically always bound and therefore are disrupted through ``classical'' mechanisms over a longer timescale.

\subsubsection{Young cluster dissolution}

Similarly to the estimation of the cumulative completeness fraction (see Section~\ref{sec:representative-sample}), we can use the analytical expressions for the distance distributions from Section~\ref{sec:spatial-distribution} to transform the absolute CFR in the representative sample to an incompleteness-corrected cluster formation rate per unit area, $\dot{\Sigma}$, representative of the inner Galaxy close to the Sun. It can be easily shown that the conversion is
\begin{equation}
\label{eq:cfr/area}
\dot{\Sigma} = \frac{{\rm CFR(D \leq D_{\rm rep})}}{\ell_1 D_{\rm eff}^2(D_{\rm rep})}~,
\end{equation}
where
\begin{equation}
\label{eq:Deff}
D_{\rm eff}^2(D) \equiv 2 \int_0^D f_{\rm c}(D') \,f_{b_1}(D') \,D'\,\rd D'~. 
\end{equation}
For the OC sample, $D_{\rm eff}(D_{\rm rep}) = 1.28$~kpc, which implies that the fitted cluster formation rate per unit area is $\dot{\Sigma}_{\rm fit} = 0.54 \pm 0.05$~Myr$^{-1}$~kpc$^{-2}$. This value can now be compared with the analogous parameter in the \citet{LamersGieles2006} fit for a complete all-sky sample within 0.6~kpc from the Sun, $\dot{\Sigma}_{\rm LG06} = 0.63$~Myr$^{-1}$~kpc$^{-2}$. Together with the maximum mass of $M_{\rm max} = 3 \times 10^4~M_{\sun}$ they obtain, we can see that both fits are consistent within the uncertainties, assuming that their errors are similar to ours (theirs are not provided). On the other hand, from the observed number of OCs in our representative sample with ages $\log(a/{\rm yr}) < 7.2$, we derive $\dot{\Sigma}_{\rm obs} = 1.18 \pm 0.22$~Myr$^{-1}$~kpc$^{-2}$ (using Poisson errors), which sets an upper limit of $\sim 0.5$ to the fraction of observed young OCs that are actually associations. The observed cluster formation rate corrected by age-dependent incompleteness is some value between $\dot{\Sigma}_{\rm fit}$ and $\dot{\Sigma}_{\rm obs}$ that can be parametrized as $\dot{\Sigma}_{\rm obs}^{\rm corr} = \dot{\Sigma}_{\rm obs} - f_{\rm adi}(\dot{\Sigma}_{\rm obs} - \dot{\Sigma}_{\rm fit})$, where $f_{\rm adi}$ is a factor in the range $[0,1]$ ($f_{\rm adi} = 0$ for no age-dependent incompleteness, and $f_{\rm adi} = 1$ for no intrinsic young excess).

To obtain a realistic estimate of the fraction of young clusters that will dissolve or merge with other(s) agglomerate(s), and therefore will not become physical OCs by their own, we also need an equivalent estimate for the formation rate of ECs. For that, we can simply take the local surface density $\Sigma_0$ obtained from fitting the distance distribution of ECs (Table~\ref{tab:parameters-ZD}), and divide it by their upper limit age of 3~Myr, resulting in $\dot{\Sigma}_{\rm EC} = 6.50 \pm 1.03$~Myr$^{-1}$~kpc$^{-2}$. This EC formation rate, however, is not directly comparable to that of OCs, since within 3 kpc from the Sun we are likely detecting ECs with masses below the detection limit of 100~$M_{\sun}$ adopted by \citet{LamersGieles2006} for OCs, as shown, e.g., by \citet{LadaLada2003}, whose EC catalog includes objects with masses down to 20~$M_{\sun}$, with a large number of clusters with masses in the range $[50,100]~M_{\sun}$. Fortunately, we found that the uncertainty in the fraction of ECs with masses above 100~$M_{\sun}$, $f_{>100\,M_{\sun}}$, is not dominant and does not prevent us to compute a good estimate of the young dissolution fraction.

If we assume that $f_{>100\,M_{\sun}}$ is in the range $[0.1,1]$, we obtain that the fraction of ECs and young exposed clusters, $f_{\rm diss}$, that will not become physical OCs is
\begin{equation}
\label{eq:fdiss}
 f_{\rm diss} = 1 - \frac{\dot{\Sigma}_{\rm fit}}{\dot{\Sigma}_{\rm obs} - f_{\rm adi}(\dot{\Sigma}_{\rm obs} - \dot{\Sigma}_{\rm fit}) + f_{>100\,M_{\sun}}\dot{\Sigma}_{\rm EC}} = 88 \pm 8 \%~,
\end{equation}
where the uncertainty has been numerically computed assuming Gaussian random variables, except for $f_{>100\,M_{\sun}}$ and $f_{\rm adi}$ which were drawn from uniform probability distributions in the corresponding domains ($[0,1]$ range for $f_{\rm adi}$, see above). The value is in excellent agreement with that obtained by \citet{LadaLada2003}. However, the explanation proposed by these authors, that this high fraction is produced by the dissolution of ECs after fast gas expulsion, has been modified (or extended) considerably in recent years. As we have reviewed in the Introduction, depending on the physical conditions of each individual system and its environment, several other phenomena can contribute to the high observed number of ECs relative to physical OCs, namely: dissolving associations from birth, merging of young subclusters, and young cluster dispersion due to tidal shocks from environment or due to fast relaxation for small-$N$ systems.

\subsection{Correlations}
\label{sec:statistics}

\begin{table*}[!t]
\renewcommand{\arraystretch}{1.1}
\caption[Statistics for each morphological type.]{Statistics for each morphological type (in percentages).}
\label{tab:statistics}
\centering
\begin{tabular}{cccccccc}
\hline\hline
Type & PAH or Bub. & Trigg. & Edge & IRDC\tablefootmark{a} & \ion{H}{ii} & UC\ion{H}{ii} & Complex  \\
(1) & (2)          & (3)    & (4)  & (5)  & (6)         & (7)           & (8)      \\
\hline
EC1 &    59 (8) &   0 (0.8) & 3 (1.5) &  52 (13) &    62 (9) &    18 (4) &  52 (8) \\
EC2 &    87 (9) & 8.2 (2.1) & 0 (0.5) &   11 (5) &    69 (8) & 5.6 (1.7) &  63 (7) \\
OC0 &   50 (12) &    12 (5) & 0 (1.8) &  0 (5.9) &   55 (12) &   0 (1.8) & 52 (12) \\
OC1 &   50 (18) & 9.1 (6.7) & 0 (4.5) & 0 (16.7) &   59 (21) &   0 (4.5) & 45 (17) \\
OC2 & 1.4 (0.7) &   0 (0.3) & 0 (0.3) &  0 (0.7) & 0.7 (0.5) &   0 (0.3) & 1 (0.6) \\
\hline
\end{tabular}
\tablefoot{
Within each morphological type, the given number is the percentage in the whole sample of clusters associated with PAH emission or IR bubbles (Column 2), clusters with signposts of triggered star formation on the surroundings (Column 3), clusters located at the edge of an IR bubble (Column 4), clusters associated with IRDCs (Column 5), clusters associated with \ion{H}{ii} regions including ultra compact ones (Column 6), clusters associated with ultra compact \ion{H}{ii} regions alone (Column 7), and finally clusters that are part of a complex of several clusters (Column 8). Numbers between parentheses are the corresponding Poisson uncertainties, with a minimum error of $\pm 1$ clusters for null values.\\
\tablefoottext{a}{Percentages are from the representative sample (clusters with $D \leq 3$~kpc).}}
\end{table*}

In this Section, we look for correlations between the morphological types defined in Section~\ref{sec:evolutionary-sequence} and other information compiled in our cluster catalog, such as the MIR morphology and association with known objects. The percentages of clusters that satisfy the studied criteria within each morphological type are presented in Table~\ref{tab:statistics}. Column 2 gives the percentage of clusters that appear to be exciting PAH emission through UV radiation from their stars, as traced by bright diffuse 8~\micron\ emission (12~\micron\ for WISE) or the presence of IR bubbles (MIR morphology \verb|bub-cen|, \verb|bub-cen-trig|, or \verb|pah|, see Section~\ref{sec:atlasgal-and-mir}). Column 3 lists the fraction of clusters that seem to be triggering further star formation at the edge of the associated IR bubble (MIR morphology \verb|bub-cen-trig| alone), whereas Column 4 indicates the fraction of clusters that are located at the edge of an IR bubble (MIR morphology \verb|bub-cen-edge|). Columns 5, 6 and 7 give, respectively, the percentage of objects that are associated with IRDCs, \ion{H}{ii} regions of any type, and UC\ion{H}{ii} regions alone. Finally, Column 8 lists the fraction of clusters that are part of a complex of several clusters (see Section~\ref{sec:complexes}). In this table we present the statistics calculated for the whole cluster sample, because we obtained the same results for the representative sample, within the uncertainties (assumed to be Poisson errors). The only exception is the association with infrared dark clouds, for which we give the fractions within the representative sample. This is expected since an IRDC can only be identified at a relatively near distance because, to be detectable, it has to manifest itself as a dark extinction feature in front of the diffuse Galactic background. We also computed the statistics restricted to clusters with GLIMPSE data available, in order to minimize possible systematic errors arising from the lower resolution and sensitivity of the WISE images (see Section~\ref{sec:MIR-morphology}), but since only 7\% of the clusters have no GLIMPSE data, we obtained identical results than those presented in Table~\ref{tab:statistics}.

We note from the table that the presence of stellar feedback as traced by PAH emission and \ion{H}{ii} regions is very important in the first four stages of the evolutionary sequence. When excluding UC\ion{H}{ii} regions, we found that both indicators of feedback are roughly equivalent, i.e., the same clusters present both tracers. That a few clusters have PAH emission but no \ion{H}{ii} region is probably due to the incompleteness of the current sample of \ion{H}{ii} regions. Alternately, in some cases we might be dealing with lower mass clusters whose UV radiation is strong enough to excite the PAH molecules, but not to produce a detectable region of ionized gas \citep{Allen2007}. On the other hand, the few \ion{H}{ii} regions without PAH emission are probably more evolved, or UC\ion{H}{ii} regions not identified as such. However, it is remarkable that although the identification of an ultra compact region was only based on the literature, such objects are much more frequently associated with the first morphological type, which presumably covers the youngest clusters. The almost null correlation of OC2 clusters with indicators of stellar feedback is consistent with the fact that these clusters are mostly classical OCs and already gas-free.

Concerning triggered star formation, we see that only EC2, OC0, and OC1 clusters are able to produce it, in roughly 10\% of the cases. EC1 clusters are not able because they are too embedded and have not yet started to sweep up the surrounding material; in turn, their formation might be triggered itself by another cluster or massive star, but in only a very small fraction (see Column 4). We warn, however, that our diagnoses of triggered star formation are purely based on morphology, so that its real existence in these cases is definitely not conclusive.

Infrared dark clouds are mostly associated with the first morphological type, confirming that they trace the earliest phases of star cluster formation. Interestingly, we found that the presence of IRDCs and PAH emission are almost mutually exclusive: within the representative sample, both tracers combined practically account for the totality of EC1 clusters, with almost null intersection. In other words, IRDCs and PAH emission trace, respectively, an earlier and later stage within the deeply embedded phase (type EC1). A simple interpretation for this behavior is that at some point IRDCs are ``illuminated'' by the radiation of the recently formed ECs, before their actual disruption, so that they become undetectable as extinction features in the mid-infrared but still prominent in the submm dust continuum emission traced by ATLASGAL.

Although we have not identified the totality of complexes of physically related clusters in our sample, Table~\ref{tab:statistics} shows a clear tendency for ECs to be grouped in complexes. In contrast, OCs are much more isolated (the type OC2 dominates the OC population). Only those OCs that are still associated with some molecular gas (types OC0, OC1) present a similar degree of grouping with other clusters as ECs. This is consistent with the fact that star formation occurs in giant molecular cloud complexes with a hierarchical structure, in which star-forming regions with a relatively higher stellar density would be observationally identified as ECs. Many of them will dissolve, while others, if close enough, will undergo a merging process as a result of dynamical evolution, all in a timescale shorter than $\sim 15$~Myr (see Section~\ref{sec:age-distribution}). The final outcome, after the parent molecular cloud is destroyed, might therefore be very few or even an unique physical OC, which will appear relatively in isolation.

\section{Conclusions}
\label{sec:conclusions}

We have statistically studied all ECs and OCs known so far in the inner Galactic plane and their correlation with dense molecular gas, taking particular advantage of the improved cluster sample over the past decade and the ATLASGAL submm continuum survey, which traces cold dust and dense molecular gas. The main results and conclusions presented in this paper are summarized as follows.

\begin{enumerate}

\item We compiled a merged full-sky list of 3904 ECs and OCs in the Galaxy, collected from several optical and infrared cluster catalogs in the literature, dealing properly with cross-identifications.

\item As part of the above compilation, we performed our own search for ECs on the mid-infrared GLIMPSE survey, complementing the catalog of 92 exposed and less-embedded clusters detected by \citet{Mercer2005} on the same data. Our method basically consisted on visual inspection of three-color images around positions previously selected as potential YSO overdensities, which correspond to enhancements on a stellar density map of the GLIMPSE point source catalog filtered by a red color criterion. With this technique, we found 75 new clusters.

\item The sample of 695 ECs and OCs within the ATLASGAL Galactic range ($|\ell| \le 60\degr$ and $|b| \le 1.5\degr$) was studied in more detail, particularly regarding the correlation with submm emission. We constructed an extensive catalog (available in electronic form at the CDS) with all the relevant information on these objects, including: the characteristics of the submm and mid-infrared emission; correlation with IRDCs, IR bubbles, and \ion{H}{ii} regions; distances (kinematic and/or stellar) and ages; and membership in big molecular complexes.

\item Based on the morphology of the submm emission and, for exposed clusters, on the agreement of the clump kinematic distances and cluster stellar distances, we defined an evolutionary sequence with decreasing correlation with ATLASGAL emission: deeply embedded clusters (EC1), partially embedded clusters (EC2), emerging exposed clusters (OC0), totally exposed clusters still physically associated with molecular gas in their surrounding neighborhood (OC1), and all the remaining exposed clusters, with no correlation with ATLASGAL emission (OC2).

\item The morphological evolutionary sequence correlates well with other observational indicators of evolution. In particular, we found that IR bubbles/PAH emission and \ion{H}{ii} regions are both equivalently important in the first four stages of the evolutionary sequence, suggesting that ionization is one of the main feedback mechanisms in our cluster sample. IRDCs are significant mostly in the first type (EC1), tracing a very early phase prior to the stage in which the EC starts to ``illuminate'' the host molecular clump while still embedded (EC1 clusters with PAH emission). The presence of big complexes containing several clusters is, again, relevant in the first four morphological types, which is consistent with the fact that star formation occurs in giant molecular clouds, and that older OCs (OC2) are just the bound survivors of a very complex process of merging and dissolution of young agglomerates.

\item We observationally defined an EC as any cluster with morphological types EC1 or EC2; OCs were defined as all the remaining types, OC0, OC1, and OC2, but were required to be confirmed by follow-up studies, in order to minimize the contamination by spurious candidates.

\item We found that our observational definition of OC agrees with the physical one (a bound exposed cluster, referred to in this work as a \emph{physical OC}) for ages greater than $\sim 16$~Myr. In our sample, some OCs younger than this limit can actually be associations.

\item By fitting the observed heliocentric distance distribution for OCs and ECs within the ATLASGAL range, we found that our OC and EC samples are roughly complete up to a distance of $\sim 1$~kpc and $\sim 1.8$~kpc, respectively. Beyond these limits, the completeness of the OC and EC samples decay exponentially with scale lengths of $\sim 0.7$~kpc and $\sim 1.8$~kpc, respectively.

\item We argued that ECs probe deeper the inner Galactic plane than OCs because, at infrared wavelengths, ECs can be more easily distinguished from the field population than OCs. On the other hand, a very distant EC is hardly detected due to the combined effect of extinction, the faint apparent brightness of its low-mass population and confusion of its members.

\item From a subsample of 23 ECs with available age estimates, we derived an upper limit of 3~Myr for the duration of the embedded phase.

\item We studied the OC age distribution within 3~kpc from the Sun, which was used to fit the theoretical parametrization of \citet{LamersGieles2006} of different disruption mechanisms for bound OCs. We found an excess of observed young OCs with respect to the fit, thought to be a combined effect of age dependent incompleteness and presence of associations for ages $\lesssim 16$~Myr.

\item We derived formation rates of 0.54, 1.18, and 6.50~Myr$^{-1}$~kpc$^{-2}$ for bound OCs, all observed young OCs, and ECs, respectively, which translates into a EC dissolution fraction of $88 \pm 8\%$. This high fraction is thought to be produced by a combination of the following effects: dissolving associations from birth; merging of young subclusters; and young cluster dispersion due to fast gas expulsion, tidal shocks from environment, or fast relaxation for small-$N$ systems.

\end{enumerate}

The new generation of all-sky near-infrared surveys, such as the UKIDSS Galactic Plane Survey \citep{Lucas2008} and VISTA Variables in the V\'{i}a L\'actea \citep[VVV,][]{Minniti2010}, will constitute valuable tools to discover new OCs and ECs in the Galactic plane and to start filling in the highly incomplete parts of the plane beyond 1 or 2 kpc from the Sun (for OCs and ECs, respectively). In the future, we plan to update our cluster database for the inner Galaxy to include the new discoveries. Furthermore, the improved sensitivity and resolution of these surveys relative to 2MASS will allow studies of the stellar population of ECs which appear too crowded and/or faint in the 2MASS data. Very importantly, this will increase the number of young clusters with available estimates of their physical properties, such as ages and masses. In particular, stellar masses can be combined with estimates of gas masses (e.g., from ATLASGAL) to derive star formation efficiencies and investigate possible trends with the age and the presence of feedback, placing important constraints on star formation theories.

\begin{acknowledgements}

We thank the referee for making useful suggestions that improved the clarity of the paper, and Thomas Robitaille for reading the manuscript and providing helpful comments. We acknowledge the useful discussions with Pavel Kroupa, Maria Messineo (about the GLIMPSE search for ECs), and Marion Wienen (about kinematic distances). We also benefited from the email discussions with D.~Froebrich (about its catalog of clusters), A. Mois\'es (about NIR spectrophotometric distances), and M. Gieles (about Equation~(\ref{eq:tcross-units-GP11})).

This research is based on: data from the ATLASGAL project, which is a collaboration between the Max-Planck-Gesellschaft (MPIfR and MPIA), the European Southern Observatory and the Universidad de Chile; observations made with the \emph{Spitzer Space Telescope}, which is operated by the Jet Propulsion Laboratory, California Institute of Technology under a contract with NASA; data products from the 2MASS, which is a joint project of the University of Massachusetts and the Infrared Processing and Analysis Center/California Institute of Technology, funded by the National Aeronautics and Space Administration and the National Science Foundation; and data products from the WISE, which is a joint project of the University of California, Los Angeles, and the Jet Propulsion Laboratory/California Institute of Technology, funded by the National Aeronautics and Space Administration.

This work has made use of the SIMBAD database, operated at CDS, Strasbourg, France, the NASA's Astrophysics Data System, and the VizieR database of astronomical catalogs \citep{Ochsenbein2000}. This paper has made use of information from the Red MSX Source survey database at \url{www.ast.leeds.ac.uk/RMS} which was constructed with support from the Science and Technology Facilities Council of the UK.

E.F.E.M was supported for part of this research through a stipend from the International Max Planck Research School (IMPRS) for Astronomy and Astrophysics at the Universities of Bonn and Cologne. This was work partially carried out in the Max Planck Research Group \emph{Star formation throughout the Milky Way Galaxy} at the Max Planck Institute for Astronomy (MPIA).

\end{acknowledgements}

\bibliography{references}

\newpage

\Online

\begin{appendix}
\section{Cluster lists in the literature}
\label{sec:catalogs-long}

In this appendix, we describe the diverse catalogs and references used for our cluster compilation, separated in three categories according to the wavelength at which the clusters are detected: optical, NIR and MIR clusters. Furthermore, we present a brief discussion of the contamination by false cluster candidates. Again, as for Table~\ref{tab:catalogs}, the number of clusters quoted within the text represent values after removing these spurious objects and some globular clusters (listed in Table~\ref{tab:spurious}), unless explicitly mentioned.

\subsection{Optical clusters}
\label{sec:optical}

\citet{Dias2002} provide the most complete catalog of optically visible OCs and candidates, containing revised data compiled from old catalogs and from isolated papers recently published. The list is regularly updated on a dedicated webpage\footnote{\url{http://www.astro.iag.usp.br/~wilton/}}, with additional clusters seen in the optical and revised fundamental parameters from new references. We used the version 3.1 (from November, 2010), which contains 2117 objects, of which 99.7\% have estimated angular diameters, and 59.4\% have simultaneous reddening, distance and age determinations. Kinematic information is also given for a fraction of clusters, 22.9\% of the list have both radial velocity and proper motion data. It should be noted that this catalog aims at collecting not only the OCs first \emph{detected} in the optical, but also most of (ideally, all) the clusters which were detected in the infrared and are \emph{visible} in the optical. For example, 293 objects from the 998 2MASS-detected clusters of \citet{Froebrich2007} were included in the last version of the catalog, based on by-eye inspection of the Digitized Sky Survey (DSS) images.

We also included in our compilation the list of new galactic OC candidates by \citet{Kronberger2006}, who did a visual inspection of DSS and 2MASS images towards selected regions, and a subsequent analysis of the 2MASS color-magnitude diagrams of the candidates. The clusters were divided in different lists, some of them with fundamental parameters determined, and are all included in the \citet[ver. 3.1]{Dias2002} catalog, except most of the stellar fields classified as \emph{suspected} OC candidates (their Table 2e), which adds 130 objects to the optical cluster sample.

\subsection{NIR clusters}

Stellar clusters detected by NIR imaging, mainly from surveys of individual star-forming regions, are compiled from the literature by \citet{Porras2003}, \citet{LadaLada2003}, and \citet{Bica2003-lit}. The first two catalogs are exclusively limited to nearby regions (distances less than 1 kpc and $\simeq 2$~kpc, respectively); \citet{Bica2003-lit} did not use that restriction, but their list is only representative for nearby distances too ($\lesssim 2$~kpc). It is not surprising that the three compilations overlap considerably, as is shown in Table~\ref{tab:catalogs}. All together, these catalogs contribute 297 additional objects with respect to the optical cluster sample.

However, most of the NIR clusters correspond to recent discoveries using the 2MASS survey. More than 300 new clusters were found by visual inspection of a huge number of 2MASS $J$, $H$, and specially $K_s$ images \citep{DutraBica2000,DutraBica2001,Bica2003-2mass,Dutra2003-2mass}. In the pioneer work of \citet{DutraBica2000}, 58 star clusters and candidates were originally detected by doing a systematic visual search on a field of $5\degr \times 5\degr$ centered close to the Galactic Center, and towards the directions of \ion{H}{ii} regions and dark clouds for $|\ell| \le 4\degr$; though most of them were observed later at higher angular resolution, and 36 turned out to be spurious detections mainly due to the high contamination from field stars in this area (see Section~\ref{sec:spurious}). Additional 42 objects were discovered by \citet{DutraBica2001}, who searched for ECs around the central positions of optical and radio nebulae in the Cygnus X region and other specific regions of the sky \citep[they are included in the literature compilation by][]{Bica2003-lit}. They extended the method for the whole Milky Way \citep[][southern and equatorial/northern Galaxy, respectively]{Dutra2003-2mass,Bica2003-2mass}, inspecting a sample of 4450 nebulae collected from the literature, and they found a total of 337 new clusters.

In addition to the visual inspection technique, a large number of 2MASS star clusters have been discovered by automated searches, which are based on the selection of enhancements on stellar surface density maps constructed with the point source catalog. The early works of \citet{Ivanov2002} and \citet{Borissova2003} led to 14 detections (the ones not present in any of the catalogs mentioned above are counted in the ``Not cataloged (NIR)'' row of Table~\ref{tab:catalogs}); similarly, \citet{Kumar2006} found 54 ECs of which 20 are new detections, focusing the search around the positions of massive protostellar candidates. More recently, \citet{Froebrich2007} searched for 2MASS clusters along the entire Galactic Plane with $|b| \le 20\degr$, automatically looking for star density enhancements, and manually selecting all remaining objects possessing the same visual appearance in the star density maps as known star clusters. They identified a total of 1788 star cluster candidates, 1021 of which resulted to be new discoveries and were presented as a catalog; an estimate of the contamination suggested that about half of these new candidates are real star clusters. A considerable number of objects from the \citet{Froebrich2007} list have been analyzed in more detail by a variety of authors, and they were compiled by \citet{Froebrich2008}. For these objects and the ones recently studied by \citet{Froebrich2010} (comprising a total of 68 clusters), we use the refined coordinates and diameters instead of the original ones. The follow-up studies compiled by \citet{Froebrich2008} also unveil 22 spurious clusters and one globular cluster (see Table~\ref{tab:spurious}). A similar automatic 2MASS search done by \citet{Glushkova2010} in the $|b| < 24\degr$ range, which includes the verification of the obtained star density enhancements by the analysis of color-magnitude diagrams and radial density distributions, produced a list of $\sim 100$ new clusters \citep[most of them included in the last version of the catalog by][]{Dias2002}, providing physical parameters for a total of 168 new and previously discovered objects.

Expectations for the near future are that the new generation of all-sky NIR surveys, such as the United Kingdom Infrared Deep Sky Survey (UKIDSS) and VISTA Variables in the V\'{i}a L\'actea (VVV), will give rise to the discovery of many more stellar clusters, thanks to their improved limiting magnitude and angular resolution compared to 2MASS. A cluster search using these data has already been performed by \citet{Borissova2011}, who found 96 previously unknown stellar clusters by visually inspecting multiwavelength NIR images of the VVV survey in the covered disk area ($295\degr \le \ell \le 350\degr$ and $|b| \le 2\degr$), towards directions of star formation signposts (masers, radio, and infrared sources). The objects listed in their catalog were required to present distinguishable sequences on the color-color and color-magnitude diagrams, after applying a field-star decontamination algorithm, in order to minimize the presence of false detections. Automated cluster searches in the UKIDSS and VVV surveys are being done by the corresponding teams.\footnote{According to unpublished data, there seem to be more than 300 new clusters detected so far by the UKIDSS team. An independent automated search on UKIDSS, leading to the discovery of 167 additional clusters and multiple star forming regions, has already been published by \citet{Solin2012}, after the last update of our cluster compilation was done.}

In our star cluster compilation, we also included recent NIR studies towards specific star-forming regions, or individual star clusters, which are not listed in the previous catalogs. In their NIR survey of 26 high-mass star-forming regions, \citet{Faustini2009} identified the presence of 23 clusters, 16 of which are new discoveries. Additional individual new objects are counted as ``Not cataloged clusters (NIR)'' in Table~\ref{tab:catalogs}.

\subsection{MIR clusters}

As a result of the high sensitivity of the GLIMPSE mid-infrared survey, \citet{Mercer2005} managed to find 92 new star clusters (2 of which are globular clusters) using an automated algorithm applied to the GLIMPSE point source catalog and archive, and a visual inspection of the image mosaics to search for ECs (the GLIMPSE Galactic range at that time was $10\degr \le |\ell| \le 65\degr$ and $|b| \le 1\degr$, excluding the inner part of the GLIMPSE~II survey). The automatic detection method consisted of the construction of a renormalized star density map, which accounts for the varying background, the estimation of the clusters' spatial parameters by fitting 2D Gaussians to the point sources with an expectation-maximization algorithm, and finally the removal of false detections by using a Bayesian criterion. This technique yielded 91 cluster candidates, 59 of which were new discoveries. Most of the clusters were detected applying a bright magnitude cut at 3.6~\micron\ before the construction of the stellar density map. Additional 33 new ECs were identified by the visual inspection, which were missed by the automated method.

However, simple by-eye examination of some GLIMPSE color images led us to conclude that there are still some ECs missing in the \citet{Mercer2005} list. Because of this (and also to cover the GLIMPSE~II area) we performed a new semi-automatic search in the whole GLIMPSE data, focused in the ECs, which resulted in increasing the number of MIR clusters to a total of 164 objects\footnote{Including 3 additional GLIMPSE clusters from the literature counted as `Not cataloged clusters (MIR)'' in Table~\ref{tab:catalogs}}. The search is described in Section~\ref{sec:newglimpse}.

\subsection{Spurious cluster candidates}
\label{sec:spurious}

The majority of the new IR star cluster catalogs compiled here are based on algorithmic or by-eye detections of stellar density enhancements on images of IR Galactic surveys, and do not provide information whether the identified objects are really composed of physically related stars or are instead produced by chance alignments on the line of sight. Due to the patchy interstellar extinction, an apparent stellar overdensity can simply correspond to a low extinction region with high extinction surroundings. In addition, bright spatially extended emission might be incorrectly classified as unresolved star clusters embedded in nebulae. Confirmation of a real cluster can be achieved through deeper, high-resolution IR photometry or through spectroscopic observations of the candidate stellar members \citep[e.g.,][]{Dutra2003-ntt,Borissova2005,Borissova2006,Messineo2009,Hanson2010,Davies2011-Mc81}, which in some cases enables the estimation of physical parameters. Though an important number of such studies have been carried out during the past decade, they still cover a small fraction of the total sample of cluster candidates to be confirmed, mainly because these objects represent relatively new discoveries and the observations needed for a more detailed analysis are very time-consuming.

Nevertheless, we can roughly estimate the contamination by spurious detections in our sample of cluster candidates in a statistical way. For example, by comparison of the basic characteristics (Galactic distribution, detection method and morphology) of the cluster candidates with those of known clusters rediscovered by their method, \citet{Froebrich2007} found that about 50\% of their catalog entries correspond to false clusters. Detailed follow-up studies of unbiased subsets of objects from this catalog, only restricted to certain areas, have determined similar contamination fractions \citep[][and references therein]{Froebrich2008}. Another example is the \citet{DutraBica2000} catalog, where 52 (out of 58) candidates have been observed using higher resolution NIR imaging \citep{Dutra2003-ntt,Borissova2005}, resulting in 36 previously unresolved alignments of a few bright stars (probably in most cases unrelated) which resemble compact clusters at the 2MASS resolution. This would imply a $\sim 70\%$ contamination by spurious detections, but we note that, since this catalog is based on a systematic search for sources projected close to the Galactic center, it is particularly affected by a higher number of background/foreground stars and more intervening dust, which all help to mimic (or hide) star clusters.

The subsequent 2MASS by-eye searches performed by this team \citep{DutraBica2001,Dutra2003-2mass,Bica2003-2mass} cover the whole Galactic plane and, furthermore, they are focused on radio/optical nebulae which generally correspond to \ion{H}{ii} regions, increasing the chance to find real stellar clusters. Typical spurious clusters associated with radio/optical nebulae represent one or a couple of bright stars plus extended emission \citep[e.g.,][]{Borissova2005}. We caution that, however, as the number of stars in these embedded multiple systems is larger, under the assumption that the stars are physically related, the consideration of a particular candidate as spurious or possible cluster is more dependent on how we define an EC. Under the definition used throughout this work (see Section~\ref{sec:cluster-definition}), since we do not impose any constraint on the number of members, we expect a minimal contamination by false detections for clusters associated with molecular gas\footnote{For consistency with earlier studies, however, we anyway excluded from our sample a few EC candidates that have been considered spurious in the literature.}. For exposed clusters, on the contrary, the probability that a cluster candidate consists of only unrelated stars on the same line of sight is much higher. Based on the above discussion, we estimate an overall spurious contamination rate of $\sim 50\%$ for exposed clusters that have not been confirmed by follow-up studies.

In Table~\ref{tab:spurious} we list the spurious candidates within the compiled cluster catalogs that were not included in our final sample. This table comprises the false detections found by \citet{Dutra2003-ntt} and \citet{Borissova2005}, and the candidates from the \citet{Froebrich2007} catalog listed as ``not a cluster'' by the literature compilation of follow-up studies by \citet{Froebrich2008}. The other objects are a few globular clusters and false clusters or duplications found in this work, primarily from the literature revision of the cluster sample in the ATLASGAL range (see Appendix~\ref{sec:huge-table-details}).

\begin{onecolumn}
\begin{longtab}
\renewcommand{\arraystretch}{1.1}
\begin{longtable}[p]{lclll}
\caption{List of spurious clusters, duplicated entries, and globular clusters within the catalogs used in this work.}
\label{tab:spurious}\\
\hline\hline
Name & Flag\tablefootmark{a} & Catalog\tablefootmark{b}  & Ref. &  Comments\\
\hline
\endfirsthead
\caption[]{continued.}\\
\hline\hline
Name & Flag\tablefootmark{a} & Catalog\tablefootmark{b}  & Ref. &  Comments\\
\hline
\endhead
\hline
\endfoot
$[$DB2000$]$ 2                      &  S &    03 &    1 &                   \\
$[$DB2000$]$ 3                      &  S &    03 &    1 &                   \\
$[$DB2000$]$ 4                      &  S &    03 &    1 &                   \\
$[$DB2000$]$ 7                      &  S & 01,03 &    2 &                   \\
$[$DB2000$]$ 8                      &  S &    03 &  1,2 &                   \\
$[$DB2000$]$ 9                      &  S &    03 &    1 &                   \\
$[$DB2000$]$ 13                     &  S &    03 &    1 &                   \\
$[$DB2000$]$ 14                     &  S &    03 &    1 &                   \\
$[$DB2000$]$ 15                     &  S &    03 &    1 &                   \\
$[$DB2000$]$ 16                     &  S &    03 &    1 &                   \\
$[$DB2000$]$ 19                     &  S &    03 &    1 &                   \\
$[$DB2000$]$ 20                     &  S &    03 &    1 &                   \\
$[$DB2000$]$ 21                     &  S &    03 &    1 &                   \\
$[$DB2000$]$ 22                     &  S &    03 &    1 &                   \\
$[$DB2000$]$ 23                     &  S &    03 &    1 &                   \\
$[$DB2000$]$ 24                     &  S &    03 &    1 &                   \\
$[$DB2000$]$ 29                     &  S &    03 &    1 &                   \\
$[$DB2000$]$ 30                     &  S &    03 &    1 &                   \\
$[$DB2000$]$ 33                     &  S &    03 &    1 &                   \\
$[$DB2000$]$ 34                     &  S &    03 &    1 &                   \\
$[$DB2000$]$ 36                     &  S &    03 &    1 &                   \\
$[$DB2000$]$ 37                     &  S &    03 &    1 &                   \\
$[$DB2000$]$ 38                     &  S &    03 &    1 &                   \\
$[$DB2000$]$ 39                     &  S &    03 &    1 &                   \\
$[$DB2000$]$ 40                     &  S & 01,03 &    2 &                   \\
$[$DB2000$]$ 41                     &  S &    03 &    2 &                   \\
$[$DB2000$]$ 43                     &  S &    03 &    1 &                   \\
$[$DB2000$]$ 44                     &  S &    03 &    1 &                   \\
$[$DB2000$]$ 46                     &  S &    03 &    1 &                   \\
$[$DB2000$]$ 47                     &  S &    03 &    1 &                   \\
$[$DB2000$]$ 48                     &  S &    03 &    1 &                   \\
$[$DB2000$]$ 53                     &  S &    03 &    1 &                   \\
$[$DB2000$]$ 54                     &  S &    03 &    1 &                   \\
$[$DB2000$]$ 56                     &  S &    03 &    2 &                   \\
$[$DB2000$]$ 57                     &  S &    03 &    1 &                   \\
$[$DB2000$]$ 58                     &  S & 01,03 &    2 &                   \\
NGC 6334 VI                         &  S &    04 &    3 &                   \\
$[$DBS2003$]$ 83                    &  S &    05 &    2 &                   \\
$[$DBS2003$]$ 84                    &  S &    05 &    2 &                   \\
$[$DBS2003$]$ 95                    &  D &    05 &    4 & \tablefootmark{d} \\
$[$DBS2003$]$ 170                   &  S &    05 &    2 &                   \\
$[$DBS2003$]$ 172                   &  S &    05 &    5 &                   \\
$[$BDS2003$]$ 101                   &  S &    06 &    2 &                   \\
$[$BDS2003$]$ 103                   & GC &    06 &    2 &                   \\
$[$BDS2003$]$ 105                   &  S &    06 &    2 &                   \\
$[$BDS2003$]$ 150                   &  D &    06 &    4 & \tablefootmark{e} \\
$[$MCM2005b$]$ 3                    & GC &    09 &  6,7 &                   \\
$[$MCM2005b$]$ 5                    & GC &    09 &    8 &                   \\
$[$FSR2007$]$ 2                     &  S &    11 &    9 &                   \\
$[$FSR2007$]$ 23                    &  S & 01,11 &    9 &                   \\
$[$FSR2007$]$ 41                    &  S &    11 &   10 &                   \\
$[$FSR2007$]$ 91                    &  S &    11 &   10 &                   \\
$[$FSR2007$]$ 94                    &  S & 01,11 &    9 &                   \\
$[$FSR2007$]$ 114                   &  S &    11 &   10 &                   \\
$[$FSR2007$]$ 128                   &  S &    11 &   10 &                   \\
$[$FSR2007$]$ 744                   &  S & 01,11 &   11 &                   \\
$[$FSR2007$]$ 776                   &  S &    11 &   11 &                   \\
$[$FSR2007$]$ 801                   &  S &    11 &   11 &                   \\
$[$FSR2007$]$ 841                   &  S &    11 &   11 &                   \\
$[$FSR2007$]$ 894                   &  S & 01,11 &   11 &                   \\
$[$FSR2007$]$ 927                   &  S & 01,11 &   11 &                   \\
$[$FSR2007$]$ 956                   &  S & 01,11 &   11 &                   \\
$[$FSR2007$]$ 1527                  &  S &    11 &    9 &                   \\
$[$FSR2007$]$ 1635                  &  S &    11 &   10 &                   \\
$[$FSR2007$]$ 1647                  &  S &    11 &   10 &                   \\
$[$FSR2007$]$ 1659                  &  S &    11 &    9 &                   \\
$[$FSR2007$]$ 1685                  &  S &    11 &   10 &                   \\
$[$FSR2007$]$ 1695                  &  S &    11 &   10 &                   \\
$[$FSR2007$]$ 1754                  &  S &    11 & 10,9 &                   \\
$[$FSR2007$]$ 1767                  &  S & 01,11 &    9 &                   \\
$[$FSR2007$]$ 1735                  & GC &    11 & 12,9 &                   \\
Ruprecht 166                        &  S &    01 &   13 &                   \\
Lynga 3                             &  S &    01 &   14 &                   \\
NGC 6334                            &  S &    01 &    4 & \tablefootmark{f} \\
NGC 6357                            &  D &    01 &    4 & \tablefootmark{g} \\
SAI 24\tablefootmark{c}             &  D &    01 &    4 & \tablefootmark{h} \\
$[$FSR2007$]$ 101\tablefootmark{c}  &  D &    01 &    4 & \tablefootmark{i} \\
$[$FSR2007$]$ 124\tablefootmark{c}  &  S &    01 &    4 & \tablefootmark{j} \\
$[$FSR2007$]$ 178\tablefootmark{c}  &  D &    01 &    4 & \tablefootmark{i} \\
$[$FSR2007$]$ 198\tablefootmark{c}  &  D &    01 &    4 & \tablefootmark{i} \\
$[$FSR2007$]$ 869\tablefootmark{c}  &  D &    01 &    4 & \tablefootmark{k} \\
$[$FSR2007$]$ 923\tablefootmark{c}  &  D &    01 &    4 & \tablefootmark{i} \\
$[$FSR2007$]$ 974\tablefootmark{c}  &  D &    01 &    4 & \tablefootmark{i} \\
$[$FSR2007$]$ 1471\tablefootmark{c} &  D &    01 &    4 & \tablefootmark{i} \\
\end{longtable}
\tablefoot{We exclude in this list: [FSR2007]~119 and [FSR2007]~584 from the \citet{Froebrich2008} list, reconsidered by \citet{Froebrich2010} as possible very old cluster and embedded young cluster, respectively; [DBS2003] 174 from \citet{Borissova2005}, since we discovered an associated compact cluster of YSOs in the GLIMPSE images.\\
\tablefoottext{a}{Flag indicates if the cluster is spurious (S), a duplicated entry in the corresponding catalog (D), or a globular cluster (GC).}
\tablefoottext{b}{Catalog ID as given in Table~\ref{tab:catalogs}.}
\tablefoottext{c}{Only affects the corresponding entry in the \citet[ver.~3.1]{Dias2002} catalog.}\\
Comments: \tablefoottext{d}{Significantly overlaps $[$DBS2003$]$ 96.}
\tablefoottext{e}{Significantly overlaps $[$BDS2003$]$ 151, and does not show an independent overdensity.}
\tablefoottext{f}{NGC 6334 is not a single cluster but a molecular complex containing many young star clusters (already included in our sample).}
\tablefoottext{g}{= Pismis 24.}
\tablefoottext{h}{= Collinder 34.}
\tablefoottext{i}{Duplicated name.}
\tablefoottext{j}{Wrong coordinates with respect to the original catalog.}
\tablefoottext{k}{= Koposov 63.}
}
\tablebib{(1) \citet{Dutra2003-ntt};
 (2) \citet{Borissova2005};
 (3) \citet{Straw1989};
 (4) This work;
 (5) \citet{Borissova2006};
 (6) \citet{StraderKobulnicky2008};
 (7) \citet{Kurtev2008};
 (8) \citet{Longmore2011-GC};
 (9) \citet{Froebrich2008};
 (10) \citet{Bica2008};
 (11) \citet{BonattoBica2008};
 (12) \citet{Froebrich2007-GC};
 (13) \citet{PiattiClaria2001};
 (14) \citet{Carraro2006}.
}
\end{longtab}
\end{onecolumn}

\twocolumn
\section{Construction of the cluster catalog}
\label{sec:huge-table-details}

Here, we report in detail the construction of our cluster catalog within the ATLASGAL Galactic range ($|\ell| \le 60\degr$ and $|b| \le 1.5\degr$), including explanations for all the assumptions and procedures made when compiling the used information, as well as descriptions for all the columns provided in the catalog. The catalog and a list of cited references are electronically available at the CDS, and an excerpt is given in Appendix~\ref{sec:catalog-excerpt}.

\subsection{Designations, position and angular size}
\label{sec:basic-information}

The basic information for each cluster is directly obtained from the original cluster catalogs compiled (see Section~\ref{sec:catalogs-summary}). The column \verb|ID| is a record number from 1 to 695 with the clusters sorted by Galactic longitude. The cluster designation, based on the original catalog, is listed in the column \verb|Name|, which was chosen, in general, to be consistent with the SIMBAD database identifier. Other common names, or designations from other catalog(s) (for clusters originally present in more than one catalog), are given in the column \verb|OName|. In the column \verb|Cat|, we provide the original cluster catalog(s) from which each object was extracted, using the reference ID defined in Table~\ref{tab:catalogs}.

The position of each object is based on the equatorial coordinates listed in the original catalog(s). For multiple catalogs, we averaged the listed positions and angular sizes to obtain the final values given here, ignoring in some cases certain references that were considered less accurate or redundant (which are listed between parentheses in the column \verb|Cat|). The galactic coordinates are given in \verb|GLON| and \verb|GLAT|, whereas the equatorial coordinates (J2000.0) are listed in \verb|RAJ2000| and \verb|DECJ2000|. The column \verb|Diam| is the angular diameter in arcseconds.
 

\subsection{ATLASGAL emission}
\label{sec:clumpfind}
\VerbatimFootnotes
From the ATLASGAL survey images, we extracted submaps centered at the cluster locations and with a field of view of $\max\{30\arcmin, 2*\verb|Diam|\}$ to search for submm dust continuum emission tracing molecular gas likely associated with the clusters, and to then characterize its morphology. The first computation needed to determine the presence of real emission in those fields is a proper estimation of the local rms noise level, $\sigma$, for which we used an iterative sigma-clipping procedure\footnote{We use the routine \verb|meanclip| from the IDL Astronomy User's Library.} with a threshold of $2\sigma$ and a convergence criterion of 1\% (iteration stops when the non-sky pixels are a fraction lower than 1\% of the total of sky pixels of the previous iteration). With these chosen parameters, the computed values of $\sigma$ agree well with quick estimates of the noise over emission-free regions identified by eye in some test fields. The average noise level is $\sigma = 45$~mJy/beam, and 95\% of the total of fields have $\sigma$ in the range $[30,60]$~mJy/beam.

Using the computed rms noise level of each field, we identified clumps of emission by applying the decomposition algorithm \emph{Clumpfind} \citep{Williams1994} in its IDL implementation for 2D data, \verb|clfind2d|.  This routine requires only two input parameters: 1) the intensity threshold, which determines the minimum emission to be included in the decomposition; and 2) the stepsize which sets the contrast needed between two contiguous features to be identified as different clumps. We chose threshold = stepsize = $3\sigma$, after visualizing the decomposition on some test fields and requiring that the obtained clumps were roughly similar to those that would be identified by the human eye. We slightly modified the IDL code of \verb|clfind2d| to improve the clump decomposition and to avoid false detections. Originally, the code developed by \citet{Williams1994} deals with blended emission by splitting it into its corresponding clumps using a simple friends-of-friends method, but instead the current implementation breaks up the emission by assigning the blended pixels to the clump with the nearest peak, which produces some disconnected clumps, i.e., pixels of the same clump not connected by a continuous path. We thus changed the peak distance criterion by the \emph{minimum distance to a clump} to assign blended emission to the existing clumps, which noticeably minimizes the effect of disconnected clumps and resembles the friends-of-friends method. A second modification to the code was to require that the clumps have angular sizes larger than the beam in both image directions, in order to reject ``snake''-shaped clumps marginally above the threshold which correspond to minor image artifacts rather than real astronomical emission.

The employed algorithm assigns into clumps all the emission above the given threshold and with an extent larger than the beam. We computed the angular distance from the cluster center of the nearest detected ATLASGAL emission pixel to have a quick first impression of the presence of molecular gas. Such values are listed in the column \verb|Clump_sep|, normalized to the cluster angular radius (when no emission is detected in the whole ATLASGAL submap, a lower limit is given).

We also performed a careful visual inspection of every ATLASGAL submap, using an IDL script to overplot the positions of all star clusters of our sample within the field, and the submm clumps detected before, as well as any interesting object, such as the positions of measured molecular line velocities (see Section~\ref{sec:line-velocities}). In another window, the script displays a smaller field of view ($\sim 10\arcmin$) with the cluster itself seen by whole set of IR images (2MASS and GLIMPSE, including three-color images) overlaid with ATLASGAL contours, in order to morphologically compare the IR and the submm emissions. The column \verb|Clump_flag| is a two-digit flag which indicates whether or not the cluster appears physically related to the nearest submm clump detected by \emph{Clumpfind}, as seen by the inspection of these images. The first digit of \verb|Clump_flag| can take the values: 0, when the nearest ATLASGAL clump does not seem to be associated with the cluster; 1, when it does seem to be clearly associated, specially for the cases of star clusters deeply embedded within centrally condensed ATLASGAL clumps; and 2, when the physical connection is less clear but still likely, in most cases when the clump appears to belong to the same star-forming region than the stellar cluster, connected by some diffuse MIR emission. The second digit of \verb|Clump_flag| provides information about the line velocity available for each object and will be described in Section~\ref{sec:line-velocities}.

The column \verb|Morph| is a text flag composed of two parts separated by a period. The first part gives further information about the morphology of the detected ATLASGAL emission throughout the immediate star cluster area, including the cases: \verb|emb|, \verb|p-emb|, \verb|surr|, \verb|few|, \verb|few*|, \verb|exp|, and \verb|exp*|, which are explained in Section~\ref{sec:atlasgal-and-mir}. The second part indicates the MIR morphology and will be described in the next Section.


\subsection{MIR morphology and association with known objects}
\label{sec:MIR-morphology}

The mid-infrared morphology of a stellar cluster can also provide some clues about its evolutionary stage and presence of feedback, in particular the intensity and distribution of the 8.0~\micron\ emission. We indicate in the second part of the column \verb|Morph| (after the period) details about the 8.0~\micron\ morphology of each cluster, after visually inspecting GLIMPSE three-color images made with the 3.6 (blue), 4.5 (green) and 8.0 \micron\ (red) bands, as part of the process described in the previous Section. This flag includes the cases: \verb|bub-cen|, \verb|bub-cen-trig|, \verb|bub-edge|, and \verb|pah|, which are explained in Section~\ref{sec:atlasgal-and-mir}.

All IR bubbles associated with star clusters and recognized in this work are identified in the table column \verb|Bub|. We give the bubble names from the catalogs by \citet{Churchwell2006,Churchwell2007} when the objects are listed there, otherwise an identifier based on the cluster \verb|ID| is provided. We also list in this column IR bubbles that are located in the neighborhood of the clusters but that do not appear clearly associated with them or do not represent any of the scenarios defined above (e.g., bubble in the same star-forming region but not directly interacting with the cluster). Similarly, on the GLIMPSE three-color images and on the 8.0~\micron\ images  we identified the presence of an infrared dark cloud in which the cluster appears to be embedded (see Fig.~\ref{fig:EC-examples}, \emph{top}). These objects are listed in the column \verb|IRDC| using a name based on the cluster \verb|ID| when the IRDC has not been cataloged so far, or the designations from the catalogs by \citet{Simon2006} and \citet{PerettoFuller2009} if it was identified there before. Unlike the IR bubbles, since we do not provide information of the IRDCs within the \verb|Morph| flag, we only list in the column \verb|IRDC| those objects that exhibit possible physical connection with the cluster. Many of the IRDCs reported by \citet{PerettoFuller2009} are only small dark fluctuations over a bright background and do not constitute cluster-forming clumps.

We note that, since the ATLASGAL Galactic range is wider than the GLIMPSE coverage, 7\% of the cluster sample have no GLIMPSE data available, and this is indicated in the column \verb|no_GL| ($\verb|no_GL| = 1$ when there is no GLIMPSE data, otherwise $\verb|no_GL| = 0$). In those cases, we used WISE three-color images made with the 3.4 (blue), 4.6 (green) and 12 \micron\ (red) filters, to identify all the features described above. Prominent PAH bands are covered by the 12~\micron\ filter; indeed, by comparing both sets of 3-color images for clusters with GLIMPSE data available, we found that bright PAH 8.0~\micron\ emission illuminated by the clusters is unambiguously detected at 12~\micron. Similarly, most of the extended IRDCs identified at 8.0~\micron\ can also be seen at 12~\micron. However, because of saturation and the relatively low resolution, more detailed structures such as the presence of IR bubbles, smaller IRDCs, or possible triggered star formation are much harder to identify than in the GLIMPSE images. 

In addition, we searched in the literature for the presence of \ion{H}{ii} regions associated with the clusters, and they are listed in the column \verb|HII_reg| with designations compatible with SIMBAD or common names used in the literature for large molecular complexes (see the references for complexes, \verb|ref_Complex|, explained in Section~\ref{sec:complexes}). Particular designations used here which do not exist in SIMBAD and do not belong to complexes are those starting with: ``HRDS'', indicating the \ion{H}{ii} regions discovered recently by \citet{Anderson2011} using radio recombination line (RRL) observations; and ``RMS'', which represent possible \ion{H}{ii} regions corresponding to radio continuum sources found by the RMS survey (see Section~\ref{sec:line-velocities} for a description of the on-line search we performed in such database; the objects listed here were taken from the ``Radio Catalogue Search Results'' section of the webpage of each individual RMS source investigated). It is worth noting that, for the \ion{H}{ii} regions primarily found using SIMBAD, we carefully checked their nature in the literature by requiring the presence of radio continuum emission or RRLs, since some sources are misclassified as \ion{H}{ii} regions in SIMBAD. Two important consulted references of RRL observations were \citet{CaswellHaynes1987} (sources with prefix [CH87]) and \citet{Lockman1989} (sources with prefix [L89b]). We also specified two flags at the end of some names to indicate two particular situations: the flag ``(UC)'', when the source is classified as an ultra compact \ion{H}{ii} region in the literature; and the flag ``(bub)'', when the \ion{H}{ii} region appears associated with the listed IR bubble, but not directly with the star cluster. However, we note that classification as an UC \ion{H}{ii} region may not be accurate, considering that detailed interferometric and large-scale observations are needed to really unveil the spatial distribution and evolutionary status of a particular \ion{H}{ii} region.


\subsection{Kinematic distance}
\label{sec:kin-distance}

As stated in Section~\ref{sec:distance-and-ages}, many of the ATLASGAL clumps at the locations or in the vicinity of the stellar clusters have measurements of molecular line LSR velocities. By assuming a Galactic rotation model, we can transform these velocities in kinematic distance estimates for the clumps and, therefore, for the corresponding clusters when they were assumed to be physically associated.

\subsubsection{Line velocities}
\label{sec:line-velocities}

We used four main references of line velocities, which were systematically searched on the ATLASGAL submaps (positions overlaid there), in the following priority order: 1) follow-up NH$_3\,(1,1)$ observations towards bright ATLASGAL sources \citep[][for northern sources; and Wienen et al., in preparation, for southern ones]{Wienen2012}; 2) similar targets observed in the N$_2$H$^+\,(1-0)$ line (Wyrowski et al., in preparation); 3) the CS$\,(2-1)$ Galactic survey by \citet{Bronfman1996} towards \emph{IRAS} sources with colors typical of compact \ion{H}{ii} regions; and 4) velocities of massive YSO candidates from the Red MSX Source (RMS) survey \citep{Urquhart2008} available on-line\footnote{\url{http://www.ast.leeds.ac.uk/cgi-bin/RMS/RMS_SUMMARY_PAGE.cgi}}, corresponding mainly to targeted observations in the $(1-0)$ and $(2-1)$ transitions of $^{13}$CO, or literature velocities compiled there. The priority sequence was primarily based on the number of ATLASGAL clumps available in each of the lists, in order to make the velocity assignments more uniform; the RMS survey was put at the end because the $^{13}$CO traces less dense gas than the other three molecules, which are unambiguously linked to the ATLASGAL emission. We note that, however, when the same clump is found in more than one list, the velocity differences are negligible compared to the error assumed for the computation of the kinematic distance (7~\kms, see below). The adopted LSR velocity is listed in the column \verb|Vlsr| (in \kms) of the catalog. We give the corresponding reference in the column \verb|ref_Vlsr|, and the source name in \verb|name_Vlsr| (SIMBAD compatible or the one used in the original paper). If no velocity was available from any of the four main lists mentioned before, additional velocity references were found by doing a coordinate query in SIMBAD.

In some cases, we did not find any velocity for the closest detected ATLASGAL clump, but we did for another possibly associated clump or for the \ion{H}{ii} region. This information is indicated in the second digit of the flag \verb|Clump_flag|, which can take the values: 0, when no velocity is available; 1, when the listed velocity is from the nearest ATLASGAL clump or from a clump directly adjacent to it; 2, when the clump with the velocity is not the nearest but is within the cluster area (used in cases of optical clusters with large angular size); 3, when the velocity is from an ATLASGAL clump which is apparently associated with the cluster as seen in the images, but is independent of the nearest one; and 4, when we list the RRL velocity of the related \ion{H}{ii} region. Considering the value of \verb|Clump_flag| as an unique integer number, i.e., combining the first digit which gives information about the closest ATLASGAL clump (see Section~\ref{sec:clumpfind}) with the second digit explained here, the kinematic distance computed from \verb|Vlsr| can be assigned to the star cluster if $\verb|Clump_flag| \geq 03$.

\subsubsection{Rotation curve}
\label{sec:rotation-curve}

Once all the available LSR velocities had been collected, kinematic distances were calculated using a Galactic rotation curve. The widely employed rotation curve fitted by \citet{BrandBlitz1993} was based on a sample of \ion{H}{ii} regions and reflection nebulae with known stellar distances, and their associated molecular clouds, which have the velocity information. Most of these sources are located in the outer Galaxy, out to a Galactocentric radius $R$ of about 17~kpc. They added to the sample the \ion{H}{i} tangent point velocities available at that time to cover the inner Galaxy, (i.e., for $R < R_0$, where $R_0 \sim 8$~kpc is the distance from the Sun of the Galactic center). However, since they used a global functional form to simultaneously fit the inner and the outer Galaxy, this curve does not properly match the data for $R < R_0$, as is shown, e.g., in Figures 6 and 7 of \citet{Levine2008}. These authors constructed an updated rotation curve for the inner Galaxy using recent high-resolution \ion{H}{i} tangent point data. The linear function fitted by them to $R \leq 8$~kpc resulted to be steeper than the \citet{BrandBlitz1993} curve in that range, and better reproduces the increase of the rotation velocity with increasing $R$. Given that most of our studied sources are within the solar circle ($R < R_0$), we decided to adopt the \citet{Levine2008}\footnote{\citet{Levine2008} provide a rotation curve as a function of both Galactocentric radius, $R$, and height off the Galactic plane, $z$. Here we $z$-averaged their rotation curve, so that it only depends on $R$.} rotation curve for $R/R_0 \leq 0.78$, which is the point where it intersects the \citet{BrandBlitz1993} curve. For $R/R_0 > 0.78$, we adopted the \citet{BrandBlitz1993} curve to cover large Galactocentric radii. We used this intersection point instead of the whole range available in \citet{Levine2008} to ensure continuity of the overall rotation curve assumed.

It is worth mentioning that the fourth quadrant part of the same \ion{H}{i} data used by \citet{Levine2008} were previously analyzed by \citet{McClureDickey2007} who fitted their own rotation curve. As already suspected by \citet{Levine2008}, the systematic shift of $\sim 7$~\kms\ between the two curves (see their Figure 7) is due to the differences in determining the terminal velocities from the data. We note that the erfc fitting method \citep[used by ][]{McClureDickey2007} is conceptually equivalent to consider the half-power point of the tangent velocity profile. Fitting instead the theoretical function derived by \citet{Celnik1979}, which is a better approximation of the tangent velocity profile, it is found that the half-power point is shifted by $\sim 0.7\sigma_v$ from the real terminal velocity (where $\sigma_v$ is the typical velocity dispersion; see the proof in that paper). We thus favor the rotation curve by \citet{Levine2008}, since they fitted \citet{Celnik1979} profiles to derive the tangent point velocities.

We did not use the more recent rotation curve by \citet{Reid2009} mainly because it is based on maser parallax distances of only 18 star-forming regions, which cover just the first and second quadrant, so that the obtained rotation curve is not fully representative of our Galactic range and, as the authors acknowledge, cannot conclusively be distinguished from a flat curve (which is their assumed form at the end). In addition, their recommended fit assumes that the massive star-forming gas orbits slower the Galaxy than expected for circular rotation, which has been questioned by some subsequent studies \citep{Baba2009,McMillanBinney2010}.

\subsubsection{Derivation of the kinematic distances}
\label{sec:derivation-kdistance}

Both rotation curves used here \citep{BrandBlitz1993,Levine2008} were originally constructed assuming the standard IAU values for the Galactocentric radius and the orbital velocity of the Sun, $R_0 = 8.5$~kpc and $\Theta_0 = 220$~\kms, respectively. Nevertheless, it can be easily shown that the solution for $x = R/R_0$ derived by applying these curves and a particular LSR velocity is practically independent of the choice of $(R_0, \Theta_0)$ \citep[fully independent for the case of a linear rotation curve constructed from tangent point velocities, as for][]{Levine2008}, and that any scaling of the curve parameters to match updated values of $(R_0, \Theta_0)$ is equivalent to adopt the original parameters in all the parts of the equations. The only thing we need afterwards is an accurate value for $R_0$, to transform from the dimensionless solution $x$ to the physical Galactocentric radius $R$. Moreover, it can be also shown that the solution does not depend on the exact definition of the LSR, provided that the rotation curves and the input data use the same solar motion (generally standard in radiotelescopes), and that any possible correction is only important in the direction of the Galactic rotation, $V_{\sun}$ (which is also true; see Table 5 of \citealt{Reid2009}, and \citealt{Schonrich2010}), so that if applied it would be canceled out in the equations.

We then applied the original rotation curves and the velocities \verb|Vlsr| with no correction, to solve for $x = R/R_0$. To finally obtain $R$, we adopted $R_0 = 8.23$~($\pm 0.20$)~kpc from \citet{Genzel2010}, who computed the weighted mean of all recent \emph{direct} estimations of the Galactic center distance from the Sun. We exclude from the kinematic distance estimation those sources with $R < 2.4$~kpc (only 2\% of the cases), which is the point were the approaching and receding parts of the rotation curve constructed by \citet[using coarser resolution \ion{H}{i} data, but covering smaller $R$]{MarascoFraternali2011} start to show significant differences likely due to non-circular motions in the region of the Galactic bar. The \citet{Levine2008} curve covers radii $R \geq 3$~kpc, which means that we implicitly extrapolated it to $R = 2.4$~kpc when we solved the equation for $x$.

There is a simple geometrical relation between the obtained Galactocentric radius $R$ and the kinematic distance, but within the solar circle (in our sample, 99\% of all kinematic distance estimations) an unique value of $R$ results in two possible distances equally spaced on either side of the tangent point, which are referred to as the near and far distances. This is known as the kinematic distance ambiguity (KDA) problem. Fortunately, as discussed in Section~\ref{sec:KDA-resolution}, there exist a number of methods that have been applied in the literature for an important fraction of the sample to solve the KDA, which allowed us to assign an unique kinematic distance in the 92\% of the cases. We list the 424 derived kinematic distances in the table column \verb|KDist| (in kpc); when the KDA is not solved, both near and far distances are given separated by `/'. Uncertainties in these distances, provided in the column \verb|e_KDist|, have been determined by shifting the LSR velocities by $\pm 7$~\kms\ to account for random motions, following \citet{Reid2009}, who suggest this value as the typical virial velocity dispersion of a massive star-forming region. We acknowledge, however, that the error in the kinematic distance can be larger due to randomly oriented peculiar motions of up to 20 or 30~\kms\ with respect to Galactic rotation, as shown, e.g., by the hydrodynamical simulations by \citet{Baba2009}. Similarly, such large systematic velocities have been found from maser parallax observations, leading to up to a factor 2 wrong kinematic distances \citep[e.g.,][]{Xu2006,Kurayama2011}. However, in some such cases it has been found also that the star-forming region does follow circular rotation \citep[e.g.,][]{Sato2010-W51}. With the assumed velocity dispersion of $\sigma_v = 7$~\kms, there are some critical cases where we can only assign an upper limit for the near distance ($|\verb|Vlsr|| < \sigma_v$), or a lower limit for the far distance (\verb|Vlsr| within $\sigma_v$ from the forbidden velocity), and that are properly indicated in the table column \verb|KDist|.

\subsubsection{Resolution of the kinematic distance ambiguity}
\label{sec:KDA-resolution}

The solutions for the distance ambiguity found in the literature are given in the table column \verb|KDA|, which informs whether the source with available velocity (listed in \verb|name_Vlsr|) is located on the near (\verb|KDA| = \verb|N|) or far side (\verb|KDA| = \verb|F|), or just at the tangent point (\verb|KDA| = \verb|T|). A companion question mark indicates a doubtful assignation, e.g., from low-quality flags in the original reference, but this happens for only 2\% of the solutions. The most common methods for resolution of the distance ambiguity are (examples of references are given below): 1) radio recombination lines in conjunction with \ion{H}{i} absorption toward  \ion{H}{ii} regions, called the \ion{H}{i} Emission/Absorption method (\ion{H}{i} E/A); and 2) \ion{H}{i} self-absorption (\ion{H}{i} SA) and molecular line emission towards molecular clouds and massive YSOs. We considered any source with \verb|Vlsr| within $\sigma_v = 7$~\kms\ of the terminal velocity as consistent with being at the tangent point, and in general we assigned a \verb|KDA| = \verb|T|. However, for some of these sources, there still exist reliable\footnote{Considering that the source is near the tangent point and some method/solution combinations are not longer valid. Examples of reliable solutions are: an associated stellar distance, a far solution from the \ion{H}{i} E/A method, or a near solution from the \ion{H}{i} SA method.} KDA solutions that can further constrain the kinematic distance to a either the near (for which \verb|KDA| = \verb|NT|) or the far distance (\verb|KDA| = \verb|FT|).

The following references for resolved KDAs were checked systematically (positions overplotted on the ATLASGAL submaps) : \citet[presence/absence of optical counterparts + \ion{H}{i} E/A for a few sources]{CaswellHaynes1987}, \citet[application of a spiral arms model of the IV quadrant]{Faundez2004}, \citet[\ion{H}{i} E/A + \ion{H}{i} SA]{AndersonBania2009}, \citet[\ion{H}{i} SA]{Roman-Duval2009}, and the RMS survey \citep{Urquhart2008}. For the RMS survey, which is an ongoing project, we took the KDA solutions from an on-line search we performed for every possibly associated source on ``The RMS Database Server''\footnote{\url{http://www.ast.leeds.ac.uk/cgi-bin/RMS/RMS_DATABASE.cgi}; we did the search on August, 2011.}; these solutions arise from dedicated application of \ion{H}{i} absorption methods \citep{Urquhart2011,Urquhart2012}, from the literature, or from grouping of sources close in the phase space where there is at least one with resolved KDA. Additional KDA solutions were found through the SIMBAD coordinate query of each source, or from the reference from which the final cluster distance was adopted (e.g., a more accurate method such as maser parallax, see Section~\ref{sec:complexes}). All used references are listed as integer numbers in the column table \verb|ref_KDA|. An `\verb|*|' following the number means that the source in the corresponding reference with resolved KDA is not located at the same position of the source from which we took the velocity, but is nearby in the phase space (close position and similar velocity) indicating that is likely connected. A reference between parentheses means that it contradicts the KDA solution adopted in this work (see below). Non-numeric flags in the column \verb|ref_KDA| indicate complementary criteria used here to solve the distance ambiguity:
\begin{itemize}[label=\textbullet]
 \item \verb|C|: we adopt the KDA solution for the whole associated complex (see Section~\ref{sec:complexes}), or from a particular source in the complex. 
 \item \verb|D|: source associated with an IRDC, favoring the near distance \citep[see the arguments given by][]{Jackson2008}
 \item \verb|O|: out of the solar circle, i.e., no ambiguity in the kinematic distance.
 \item \verb|S|: adopted KDA solution consistent with the stellar distance (see Section~\ref{sec:physical-parameters}) 
 \item \verb|z|: near distance adopted, since if located at the far distance the source would be too high above the Galactic plane. We adopted a height value of $|z| = 200$~pc to exclude the far distance, following \citet{Blitz1991}.
\end{itemize}

If the assumption of two or more references or criteria delivered contradictory solutions for the KDA, in general we adopted the more recent, or the one using a more accurate method. Although this decision is somehow arbitrary, there are some reasonable guidelines that can be applied, e.g., we favor the consistency with stellar distance or with the complex (flags S and C), and we adopted the solution from the \ion{H}{i} E/A method when conflicting with the \ion{H}{i} SA method, since the first has been found to be more robust \citep{AndersonBania2009}. In any case, the KDA solutions from different references usually agree; discrepant ones are only the 12\% of the total of resolutions and should not affect the statistical results of this work.


\subsection{Stellar distance and age}
\label{sec:physical-parameters}

A direct estimation of the distance to a cluster, i.e., from the member stars, is particularly useful when the accuracy is better than that of the kinematic distance from the gas (e.g., when a large sample of stars is used), or when the cluster is fully exposed and there is no nebula that can be associated to it. Using data from the original cluster catalogs and new references found in SIMBAD for each object, we compiled values for the stellar distance (in kpc; table column \verb|SDist|) and its uncertainty (column \verb|e_SDist|), as well as the age and its error (in Myr; columns \verb|Age| and \verb|e_Age|, respectively) computed by studies of the cluster stellar population. The corresponding references of the adopted parameters are listed in the columns \verb|ref_SDist| and \verb|ref_Age|. For the optical clusters in the \citet[see Section~\ref{sec:optical}]{Dias2002} catalog, we generally used the original parameters given there, unless new estimates based on a better method (or data) provided a real improvement in accuracy. A more rigorous approach for multiple references of the same cluster would be similar to that taken in \citet{PaunzenNetopil2006}, and is beyond the scope of this work. However, these authors concluded that their literature-averaged parameters have the same statistical significance as the data from the \citet{Dias2002} catalog, so that for the purposes of our work a correct estimation of the uncertainties (see below) is much more important than careful averaging. Out of the 216 clusters from the \citet{Dias2002} catalog present in our sample, 131 objects come with determinations of both age and distance (+4 clusters with only the distance). We adopted these parameters for most of clusters (110 with original values, and 21 with new ones), and added parameters for 25 more. To keep track of all these changes, the original references used in the \citet{Dias2002} catalog are listed in the column \verb|ref_Dias|. 

The uncertainties in the cluster fundamental parameters are often ignored or underestimated in the literature; in particular, they are not provided in the \citet{Dias2002} catalog. We therefore collected all available errors from the corresponding references and, to prevent underestimation, we imposed uniform \emph{minimum} uncertainties in the derived parameters. We also assumed these values as errors when they were not given in the literature. For the stellar distance, the minimum uncertainty was carefully chosen depending on the method used to calculate it, in order to correctly compare it with the kinematic distance (e.g., to decide which of both distances is finally adopted, see Section~\ref{sec:complexes}). All most common methods for cluster distance determination use stellar photometry, so that the corresponding uncertainty is dominated by the errors from the absolute magnitude calibration and from the extinction estimation \citep[e.g.,][]{Pinheiro2010}. For the extinction, in addition to the statistical error intrinsic to the method, there is a systematic error produced by possible variations in the extinction law \citep[e.g.,][]{Fritz2011, Moises2011}, which is often not considered in the literature and might be particularly relevant in the NIR regime. In the optical, we can consider that the typical extinction law assumed ($R_V \simeq 3.1$, appropriate for diffuse local gas) is not subject to important variations, since the observed stars are relatively close to the Sun and not heavily embedded in the associated molecular clouds (if any), otherwise they would not be visible at these wavelengths. In the NIR, the extinction law can be described by a power law, $A_\lambda \propto \lambda^{-\beta}$, and the variations can be accounted for with different values for the exponent $\beta$. Using the typical spread in $\beta$ obtained by \citet{Fritz2011} in their compilation, we found that the corresponding uncertainty in the $K$-band extinction is $\sigma(A_K) \simeq 0.2 \,A_K$.

In the following, we list the main methods for stellar distance determinations of the used references, and the corresponding minimum uncertainties adopted in this work:

\begin{itemize}[label=\textbullet]
 \item Optical main-sequence (MS) or isochrone fitting \citep[e.g.,][]{Kharchenko2005-known, Loktin2001}: In this case, we follow \citet{PhelpsJanes1994} who estimated an uncertainty in distance modulus of $\sigma(m-M) \sim 0.32$, from a detailed analysis of the typical error in fitting a template main sequence to the optical color-magnitude diagram. This is equivalent to an error of $\sim 15\%$ in distance. Due to the fact that, from the point of view of the distance uncertainty, fitting a MS is analogous to fitting an isochrone, we also adopted a minimum error of $\sim 15\%$ for the isochrone method. Furthermore, this is consistent with the spread in distance modulus found by \citet[see their Table 2]{GrocholskiSarajedini2003} in their comparison of different isochrone models.

\item NIR isochrone fitting \citep[e.g.,][]{Tadross2008,Glushkova2010}: We adopted the same minimum distance error as for optical isochrone fitting, 15\%. Extinction law variations might be present, but since the type of clusters for which isochrone fitting is possible are not severely extinguished (they are generally not young), the corresponding uncertainty in $A_K$ due to these variations is also low (recall $\sigma(A_K) \simeq 0.2 \,A_K$).

\item Optical spectrophotometric distance \citep[e.g.,][]{Herbst1975}: Here, we assumed an absolute magnitude calibration uncertainty of $\sigma(M_V) \simeq 0.5$, consistent with the typical spread of massive OB star calibration scales \citep[e.g.,][]{Martins2005}, and an error in spectral type determination of 1 subtype, equivalent to $\pm 0.3$ magnitudes in $M_V$ for the \citet{Martins2005} calibration. Adding both contributions in quadrature gives an overall 
uncertainty of $\sim 0.58$ magnitudes in distance modulus, or $\sim 27\%$ in distance.

\item NIR spectrophotometric distance \citep[e.g.,][]{Moises2011}: For calibration and spectral type errors, we adopted the same overall uncertainty of $\sim 0.58$ magnitudes in distance modulus as for the optical method (absolute magnitudes are usually converted from the optical to the NIR using tabulated intrinsic colors with little error). We added in quadrature an uncertainty to account for possible extinction law variations: assuming a typical extinction of $A_K \simeq 1.5$, $\sigma(A_K) \simeq 0.2 \,A_K \simeq 0.3$. The final error in distance modulus is $\sim 0.66$ magnitudes, equivalent to $\sim 30\%$ in distance.

\item Average of spectrophotometric distances from many stars \citep[e.g.,][]{Moises2011, Pinheiro2010}: Redefining the errors here would mean a complete re-computation of the average distance, since the minimum errors should be imposed in every individual star. Fortunately, in general the uncertainty of the average is dominated by the variance of the sample rather than by the individual errors. We thus kept the original quoted uncertainty in this case.

\item Kinematic distance from average stellar radial velocity \citep[e.g.,][]{Davies2008}: For consistency with gas kinematic distances, here we recomputed the stellar kinematic distance using the cluster LSR velocity, a velocity dispersion of 7~\kms\ (in all cases higher than the quoted error in the cluster velocity) and the rotation curve as described in Section~\ref{sec:kin-distance}. This special case is indicated with the flag `(K)' after the reference number in the column \verb|ref_SDist|.

\item 10$^{\rm th}$ brightest star method \citep{Dutra2003-ntt, Borissova2005}: We do not use the stellar distances derived by applying this technique, because they are very uncertain. The errors can easily reach a factor 10 or more in distance \citep{Borissova2005}, which thus places no constraints on the cluster location at Galactic scales.

\end{itemize}

For the cluster ages, we simply adopted uniform minimum errors based on the corresponding age range, following \citet{BonattoBica2011}: 35\% for $\verb|Age| < 20$~Myr, 30\% for 20~Myr~$\leq \verb|Age| < 100$~Myr, 20\% for 100~Myr~$\leq \verb|Age| < 2$~Gyr, and 50\% for $\verb|Age| \geq 2$~Gyr. The most common method for age determination is isochrone fitting \citep[e.g.,][]{Loktin2001}. For a few clusters with stars studied spectroscopically, the age can be estimated using the evolutionary types of the identified stars and knowledge about their typical ages and lifetimes \citep[e.g.,][]{Messineo2009}. For a total of 209 clusters age estimates can be found in the literature (30\% of our sample).

For some clusters of our sample for which no fundamental parameters are available, there are still some studies in the literature that present what can be considered as \emph{confirmations} of the star cluster nature of the objects, i.e., the possibility of an erroneous identification as a cluster can be practically discarded. These references are given in the column \verb|ref_Conf| of the catalog, and usually report higher resolution or/and sensitivity imaging NIR observations in which the star cluster is unequivocally revealed \citep[e.g.,][]{Dutra2003-ntt,Borissova2005,Kumar2004}. They also comprise detailed studies towards star-forming regions which are too young to really constrain the cluster physical parameters by isochrone fitting, but where it is still possible to recognize YSO candidates within the cluster as color excess sources in color-color and color-magnitude diagrams \citep[e.g.,][]{RomanAbraham2006}. The objects with both determined age and stellar distance can also be considered as confirmed stellar clusters, because the derivation of parameters usually requires the identification of the cluster sequence or stellar spectroscopy. We thus listed again the references for age and distance in the column \verb|ref_Conf|, including in some cases additional references presenting further cluster analysis.
 

\subsection{Complexes, subclusters, and adopted distance}
\label{sec:complexes}

Young star clusters are normally not found in isolation but within bigger complexes of gas, stars and other clusters, as a result of the fact that star formation occurs in giant molecular clouds with a hierarchical structure. If a group of stellar clusters in our sample was found to form a physically associated complex according to their positions and radial velocities, we identified it in the column \verb|Complex| of the catalog. When the complex was identified in the literature, we here list its name \citep[e.g., the giant molecular cloud W51;][]{Kang2010}. References for complex identification and analysis are provided in the column \verb|ref_Complex|. Small complexes of clusters not previously established in the literature but whose morphology in the IR images (field of view of $\sim 10\arcmin$) suggests that they belong to the same star-forming region are indicated by \verb|Complex| = MC-$i$, where $i$ is a record number. Bigger complexes of stellar clusters not found in the literature and visually identified within the ATLASGAL fields (of $\sim 30\arcmin$) through the proximity of their members in the phase-space are marked by \verb|Complex| = KC-$j$, where $j$ is another record number. We warn that, however, since the complexes were recognized as part of the visual inspection of the maps, or were found in the literature, not all possible physical groupings of star clusters are provided here. For that, a subsequent statistical analysis is needed, which will be presented in a forthcoming paper. We also identified in the IR images a few cases where there is a pair of star clusters even closer, usually sharing part of their population, which can be considered as subclusters of an unique merging (or merged) entity. Those subclusters are indicated in the table column \verb|SubCl| with an identical record number.

For all the clusters of our sample, the final adopted distances with their corresponding errors are listed in the table columns \verb|Dist| and \verb|e_Dist| (in kpc), respectively, and were chosen to be the available distance estimate with the lowest uncertainty, corresponding in some cases to a determination from the literature which was more accurate than \verb|SDist| and \verb|KDist|. Clusters within a particular complex were assumed to be all located at the same distance. The origin of the adopted distance is properly indicated in the column \verb|ref_Dist|, and can be one of the following:

\begin{itemize}[label=\textbullet]
 \item \verb|K|: kinematic distance adopted, $\verb|Dist|=\verb|KDist|$.
 \item \verb|S|: stellar distance adopted, $\verb|Dist|=\verb|SDist|$.
 \item \verb|Ref:|$n$: adopted distance from literature reference with identification number $n$.
 \item \verb|KC|: complex distance computed kinematically from an average position and velocity, using the values compiled here for all the clusters within the complex with available (and not repeated) \verb|Vlsr|, and the rotation curve used in Section~\ref{sec:kin-distance}.         
 \item \verb|SC|: complex distance computed by averaging the stellar distances (\verb|SDist|) of the member clusters.       
 \item \verb|C(Ref:|$n$\verb|)|: distance for the whole complex adopted from literature reference with identification number $n$.
 \item \verb|CV(Ref:|$n$\verb|)|: complex distance computed kinematically from an average position and velocity given by the reference with identification number $n$, and the rotation curve used in this work.
 \item \verb|C(ID:|$m$\verb|)|: adopted for the whole complex the distance given for the cluster with $\verb|ID|=m$ (used when a particular cluster within a complex has a very accurate distance estimation). 
\end{itemize}


\subsection{Additional comments}
\label{sec:comments}

Specific comments about the stellar cluster itself, or its compiled fundamental parameters (stellar distance and age) are provided in column \verb|Comments1|. We give additional remarks about the ATLASGAL emission, the associated complex or other objects, or about the finally adopted distance in column \verb|Comments2|. For comments, the quoted literature is indicated by the code \verb|Ref:|$n$, where $n$ is the identification number of the used reference.


\section{Excerpt of the cluster catalog}
\label{sec:catalog-excerpt}

This appendix gives an excerpt of the cluster catalog whose construction is explained in Appendix~\ref{sec:huge-table-details}. The totality of the catalog, together with a list of cited references, is electronically available at the CDS. Here, we present all the catalog columns (except columns \verb|Comments1| and \verb|Comments2| which are sometimes too wide for the paper version) for 50 (out of 695) stellar clusters. Only for presentation, here the columns are distributed in five tables (Tables~\ref{tab:big-catalog1} to \ref{tab:big-catalog5}), but the on-line version of the catalog is a single table.  The names of the columns are the same as defined in Appendix~\ref{sec:huge-table-details}, and they are briefly described in the following (the corresponding Sections of the paper in which they are explained in more detail are given in parentheses):

\begin{itemize}[label=\textbullet]
 \item \verb|ID| : identification number (Section~\ref{sec:basic-information})
 \item \verb|Name| : main name (Section~\ref{sec:basic-information})
 \item \verb|OName| : other designation (Section~\ref{sec:basic-information})
 \item \verb|Cat| : catalogs from which each cluster was extracted (Section~\ref{sec:basic-information})
 \item \verb|GLON| : Galactic longitude (Section~\ref{sec:basic-information})
 \item \verb|GLAT| : Galactic latitude (Section~\ref{sec:basic-information})
 \item \verb|RAJ2000| : right ascension (Section~\ref{sec:basic-information})
 \item \verb|DEC2000| : declination (Section~\ref{sec:basic-information})
 \item \verb|Diam| : angular size (Section~\ref{sec:basic-information})
 \item \verb|Dist| : adopted distance (Section~\ref{sec:complexes})
 \item \verb|e_Dist| : distance error (Section~\ref{sec:complexes})
 \item \verb|ref_Dist| : distance reference (Section~\ref{sec:complexes})
 \item \verb|Age| : age (Section~\ref{sec:physical-parameters})
 \item \verb|e_Age| : age error (Section~\ref{sec:physical-parameters})
 \item \verb|ref_Age| : age reference (Section~\ref{sec:physical-parameters})
 \item \verb|Morph_type| : morphological type (Section~\ref{sec:evolutionary-sequence})
 \item \verb|Morph| : morphological flag (Section~\ref{sec:atlasgal-and-mir})
 \item \verb|Clump_sep| : projected distance to the nearest ATLASGAL emission pixel (Section~\ref{sec:clumpfind})
 \item \verb|Clump_flag| : gives information about the correlation with ATLASGAL and line velocity available (Sections \ref{sec:clumpfind} and \ref{sec:line-velocities})
 \item \verb|name_Vlsr| : source name for line velocity (Section~\ref{sec:line-velocities})
 \item \verb|Vlsr| : gas line velocity (Section~\ref{sec:line-velocities})
 \item \verb|ref_Vlsr| : reference for line velocity (Section~\ref{sec:line-velocities})
 \item \verb|KDist| : kinematic distance (Section~\ref{sec:derivation-kdistance})
 \item \verb|e_KDist| : error in the kinematic distance (Section~\ref{sec:derivation-kdistance})
 \item \verb|KDA| : solution of the kinematic distance ambiguity (Section~\ref{sec:KDA-resolution})
 \item \verb|ref_KDA| : reference for the KDA solution (Section~\ref{sec:KDA-resolution})
 \item \verb|SDist| : stellar distance (Section~\ref{sec:physical-parameters})
 \item \verb|e_SDist| : error in the stellar distance (Section~\ref{sec:physical-parameters})
 \item \verb|ref_Sdist| : reference for the stellar distance (Section~\ref{sec:physical-parameters})
 \item \verb|ref_Dias| : reference for stellar parameters adopted in the \citet{Dias2002} catalog (Section~\ref{sec:physical-parameters})
 \item \verb|ref_Conf| : reference for cluster confirmation (as real cluster) or further studies (Section~\ref{sec:physical-parameters})
 \item \verb|HII_reg| : associated \ion{H}{ii} region (Section~\ref{sec:MIR-morphology})
 \item \verb|Bub| : associated infrared bubble (Section~\ref{sec:MIR-morphology})
 \item \verb|IRDC| : associated infrared dark cloud (Section~\ref{sec:MIR-morphology})
 \item \verb|no_GL| : indicates when there is no GLIMPSE data available (Section~\ref{sec:MIR-morphology})
 \item \verb|SubCl| : groups subclusters (Section~\ref{sec:complexes})
 \item \verb|Complex| : groups spatially associated clusters (Section~\ref{sec:complexes})
 \item \verb|ref_Complex| : reference for complex identification (Section~\ref{sec:complexes})
\end{itemize}

\begin{table*}
\renewcommand{\arraystretch}{1.1}
\caption{Excerpt of the cluster catalog (Columns 1--8).}
\label{tab:big-catalog1}
\centering
\begin{tabular}{rlllrrcc}
\hline\hline
\verb|ID| & \verb|Name| & \verb|OName| & \verb|Cat| & \verb|GLON| & \verb|GLAT| 
& \verb|RAJ2000| & \verb|DEC2000|  \\
 &  &  &  & ($\degr$) & ($\degr$) & ( $^{\rm h}$: $^{\rm m}$: $^{\rm s}$) & ( $\degr$: $'$: $''$) \\
\hline
477 & Ruprecht 139                  & \nodata                       & 01         & 6.410  & $-$0.236 & 18:01:03.0 & $-$23:32:00 \\
478 & G3CC 52                       & \nodata                       & 17         & 6.797  & $-$0.256 & 18:01:57.6 & $-$23:12:26 \\
479 & NGC 6514                      & \nodata                       & 01,(07)    & 7.086  & $-$0.287 & 18:02:42.0 & $-$22:58:18 \\
480 & Teutsch 72                    & \nodata                       & 02         & 7.236  & $-$0.238 & 18:02:50.2 & $-$22:49:00 \\
481 & NGC 6546                      & \nodata                       & 01         & 7.328  & $-$1.382 & 18:07:22.0 & $-$23:17:48 \\
482 & NGC 6531                      & \nodata                       & 01         & 7.677  & $-$0.355 & 18:04:13.0 & $-$22:29:24 \\
483 & Teutsch 14b                   & \nodata                       & 02         & 7.906  & $-$0.044 & 18:03:32.1 & $-$22:08:17 \\
484 & Teutsch 14a                   & \nodata                       & (01),02    & 7.916  & $-$0.036 & 18:03:31.3 & $-$22:07:32 \\
485 & ESO 589$-$26                  & \nodata                       & 01         & 7.952  & 0.328    & 18:02:14.0 & $-$21:54:54 \\
486 & $[$BDS2003$]$ 110             & \nodata                       & 06         & 8.032  & 0.413    & 18:02:05.0 & $-$21:48:12 \\
487 & $[$BDS2003$]$ 111             & \nodata                       & 06         & 8.060  & 0.425    & 18:02:06.0 & $-$21:46:21 \\
488 & ASCC 93                       & \nodata                       & 01         & 8.329  & $-$1.050 & 18:08:13.0 & $-$22:15:36 \\
489 & G3CC 53                       & \nodata                       & 17         & 8.492  & $-$0.633 & 18:06:59.3 & $-$21:54:55 \\
490 & $[$BDS2003$]$ 3               & \nodata                       & 06         & 8.663  & $-$0.342 & 18:06:15.0 & $-$21:37:27 \\
491 & $[$FSR2007$]$ 31              & SAI 125                       & (01),11,13 & 8.899  & $-$0.270 & 18:06:28.4 & $-$21:22:60 \\
492 & vdBergh 113                   & \nodata                       & 01         & 9.110  & $-$0.718 & 18:08:36.0 & $-$21:25:00 \\
493 & G3CC 54                       & \nodata                       & 17         & 9.221  & 0.166    & 18:05:31.3 & $-$20:53:21 \\
494 & $[$FSR2007$]$ 35              & \nodata                       & (01),11    & 9.686  & 0.764    & 18:04:16.0 & $-$20:11:27 \\
495 & SGR 1806$-$20 Cluster         & $[$BDB2003$]$ G010.00$-$00.24 & 04         & 9.995  & $-$0.241 & 18:08:39.0 & $-$20:24:39 \\
496 & $[$BDB2003$]$ G010.16$-$00.36 & \nodata                       & 04         & 10.161 & $-$0.363 & 18:09:27.0 & $-$20:19:30 \\
497 & $[$FSR2007$]$ 39              & \nodata                       & (01),11    & 10.249 & 0.320    & 18:07:05.3 & $-$19:55:01 \\
498 & $[$BDS2003$]$ 112             & \nodata                       & 06         & 10.307 & $-$0.144 & 18:08:56.0 & $-$20:05:30 \\
499 & $[$BDS2003$]$ 113             & \nodata                       & 06         & 10.322 & $-$0.153 & 18:08:60.0 & $-$20:04:57 \\
500 & $[$BDB2003$]$ G010.62$-$00.38 & \nodata                       & 04,17      & 10.623 & $-$0.383 & 18:10:28.6 & $-$19:55:50 \\
501 & NGC 6554                      & \nodata                       & 01         & 11.762 & 0.648    & 18:08:59.0 & $-$18:26:06 \\
502 & Markarian 38                  & \nodata                       & 01         & 11.983 & $-$0.931 & 18:15:17.0 & $-$19:00:00 \\
503 & Turner 2                      & \nodata                       & 01         & 12.352 & $-$1.243 & 18:17:11.0 & $-$18:49:27 \\
504 & Turner 3                      & \nodata                       & 01         & 12.360 & $-$1.342 & 18:17:34.0 & $-$18:51:50 \\
505 & $[$BDS2003$]$ 6               & \nodata                       & 06         & 12.425 & $-$1.114 & 18:16:51.0 & $-$18:41:52 \\
506 & Turner 4                      & \nodata                       & 01         & 12.455 & $-$1.174 & 18:17:08.0 & $-$18:41:60 \\
507 & $[$BDS2003$]$ 7               & \nodata                       & 06         & 12.636 & 0.606    & 18:10:55.0 & $-$17:41:25 \\
508 & $[$MDF2011$]$ cl2             & \nodata                       & 15         & 12.728 & $-$0.216 & 18:14:08.0 & $-$18:00:15 \\
509 & Collinder 469                 & \nodata                       & 01         & 12.733 & $-$0.871 & 18:16:34.0 & $-$18:18:42 \\
510 & $[$MCM2005b$]$ 1              & \nodata                       & 09         & 12.752 & $-$0.144 & 18:13:55.0 & $-$17:56:55 \\
511 & $[$MFD2008$]$ Cluster         & SAI 126                       & (13)       & 12.743 & $-$0.009 & 18:13:24.0 & $-$17:53:31 \\
512 & $[$BDB2003$]$ G012.80$-$00.19 & \nodata                       & 04         & 12.801 & $-$0.199 & 18:14:13.0 & $-$17:55:55 \\
513 & $[$MDF2011$]$ cl1             & \nodata                       & 15         & 12.814 & $-$0.232 & 18:14:22.0 & $-$17:56:10 \\
514 & NGC 6603                      & \nodata                       & 01         & 12.860 & $-$1.306 & 18:18:26.0 & $-$18:24:24 \\
515 & $[$BDS2003$]$ 114             & \nodata                       & 06         & 12.908 & $-$0.263 & 18:14:40.0 & $-$17:52:07 \\
516 & $[$BDS2003$]$ 115             & \nodata                       & 06         & 13.185 & 0.046    & 18:14:05.0 & $-$17:28:40 \\
517 & $[$FSR2007$]$ 46              & \nodata                       & (01),11    & 13.350 & 0.068    & 18:14:20.0 & $-$17:19:19 \\
518 & NGC 6561                      & \nodata                       & 01         & 13.434 & 1.159    & 18:10:30.0 & $-$16:43:30 \\
519 & Mol 45 Cluster                & \nodata                       & 12         & 13.656 & $-$0.599 & 18:17:24.1 & $-$17:22:12 \\
520 & $[$BDS2003$]$ 116             & \nodata                       & 06         & 13.879 & 0.283    & 18:14:36.0 & $-$16:45:17 \\
521 & $[$MCM2005b$]$ 2              & \nodata                       & 09         & 13.995 & $-$0.126 & 18:16:20.0 & $-$16:50:51 \\
522 & G3CC 55                       & \nodata                       & 17         & 14.113 & $-$0.571 & 18:18:12.4 & $-$16:57:18 \\
523 & NGC 6613                      & \nodata                       & 01         & 14.183 & $-$1.011 & 18:19:58.0 & $-$17:06:06 \\
524 & NGC 6596                      & \nodata                       & 01         & 14.307 & $-$0.288 & 18:17:33.0 & $-$16:38:60 \\
525 & G3CC 56                       & \nodata                       & 17         & 14.341 & $-$0.642 & 18:18:55.2 & $-$16:47:15 \\
526 & Mol 50 Cluster                & \nodata                       & 12         & 14.892 & $-$0.403 & 18:19:07.6 & $-$16:11:21 \\

\hline
\end{tabular}
\end{table*}

\begin{table*}
\renewcommand{\arraystretch}{1.1}
\caption{Excerpt of the cluster catalog (Columns 9--17).}
\label{tab:big-catalog2}
\centering
\begin{tabular}{rrcclccccl}
\hline\hline
\verb|ID| & \verb|Diam| & \verb|Dist| & \verb|e_Dist| & \verb|ref_Dist| & \verb|Age| & \verb|e_Age| & \verb|ref_Age| & \verb|Morph_type| & \verb|Morph| \\
 & ($''$) & (kpc) & (kpc) &  & (Myr) & (Myr) &  &  &  \\
\hline
477 & 720  & 0.55    & 0.08    & S         & 1122    & 517     & 115     & OC2 & few                \\
478 & 50   & 2.70    & 0.50    & C(Ref:48) & \nodata & \nodata & \nodata & EC1 & emb                \\
479 & 1680 & 2.70    & 0.50    & C(Ref:48) & \nodata & \nodata & \nodata & OC0 & few*               \\
480 & 180  & 2.70    & 0.50    & C(Ref:48) & \nodata & \nodata & \nodata & OC1 & exp.bub-cen        \\
481 & 840  & 0.94    & 0.14    & S         & 70.6    & 21.2    & 128     & OC2 & exp                \\
482 & 840  & 1.21    & 0.18    & S         & 11.7    & 4.1     & 128     & OC2 & exp                \\
483 & 30   & 1.72    & 0.41    & C(ID:484) & \nodata & \nodata & \nodata & OC2 & exp                \\
484 & 132  & 1.72    & 0.41    & S         & 100     & 20      & 37      & OC2 & exp                \\
485 & 150  & \nodata & \nodata & \nodata   & \nodata & \nodata & \nodata & OC2 & exp                \\
486 & 62   & \nodata & \nodata & \nodata   & \nodata & \nodata & \nodata & OC2 & exp                \\
487 & 88   & \nodata & \nodata & \nodata   & \nodata & \nodata & \nodata & OC2 & exp                \\
488 & 1944 & 2.50    & 0.38    & S         & 16.6    & 7.6     & 116     & OC0 & few*               \\
489 & 126  & 1.51    & 0.75    & Ref:136   & \nodata & \nodata & \nodata & EC2 & p-emb              \\
490 & 63   & 4.45    & 0.48    & K         & \nodata & \nodata & \nodata & EC2 & p-emb.bub-cen      \\
491 & 244  & 1.60    & 0.24    & S         & 1100    & 220     & 36      & OC2 & exp*               \\
492 & 840  & 3.47    & 0.52    & S         & 31.6    & 14.6    & 115     & OC2 & few                \\
493 & 42   & \nodata & \nodata & \nodata   & \nodata & \nodata & \nodata & EC1 & emb                \\
494 & 115  & \nodata & \nodata & \nodata   & \nodata & \nodata & \nodata & OC2 & exp                \\
495 & 42   & 8.86    & 1.61    & S         & 4.00    & 1.40    & 80      & OC2 & exp                \\
496 & 69   & 2.77    & 1.07    & SC        & 0.60    & 0.21    & 89      & EC2 & p-emb.bub-cen      \\
497 & 59   & 3.50    & 0.53    & S         & 1000    & 200     & 86      & OC2 & exp                \\
498 & 74   & 2.77    & 1.07    & SC        & \nodata & \nodata & \nodata & EC2 & p-emb.bub-cen-trig \\
499 & 75   & 2.77    & 1.07    & SC        & \nodata & \nodata & \nodata & EC2 & p-emb.bub-cen-trig \\
500 & 53   & 2.77    & 1.07    & SC        & \nodata & \nodata & \nodata & EC1 & emb.pah            \\
501 & 1200 & \nodata & \nodata & \nodata   & \nodata & \nodata & \nodata & OC2 & few                \\
502 & 120  & 1.47    & 0.22    & S         & 7.62    & 2.67    & 128     & OC2 & exp                \\
503 & 456  & 1.19    & 0.18    & S         & 100     & 20      & 228     & OC2 & exp                \\
504 & 120  & 1.79    & 0.27    & S         & 29.0    & 8.7     & 237     & OC2 & exp                \\
505 & 48   & 4.15    & 0.43    & K         & \nodata & \nodata & \nodata & EC2 & p-emb.pah          \\
506 & 210  & 2.33    & 0.35    & S         & 10.0    & 3.5     & 237     & OC2 & exp                \\
507 & 108  & \nodata & \nodata & \nodata   & \nodata & \nodata & \nodata & OC2 & exp                \\
508 & 60   & 3.79    & 0.48    & KC        & \nodata & \nodata & \nodata & EC2 & p-emb.pah          \\
509 & 180  & 1.48    & 0.22    & S         & 63.0    & 18.9    & 128     & OC2 & exp                \\
510 & 96   & 3.79    & 0.48    & KC        & \nodata & \nodata & \nodata & OC1 & exp.bub-cen        \\
511 & 210  & 3.79    & 0.48    & KC        & 4.25    & 1.49    & 144     & OC2 & few                \\
512 & 48   & 3.79    & 0.48    & KC        & \nodata & \nodata & \nodata & EC1 & emb.pah            \\
513 & 108  & 3.79    & 0.48    & KC        & \nodata & \nodata & \nodata & EC2 & p-emb.bub-cen      \\
514 & 360  & 3.60    & 0.54    & S         & 200     & 100     & 16      & OC2 & exp                \\
515 & 59   & 3.79    & 0.48    & KC        & \nodata & \nodata & \nodata & EC1 & emb                \\
516 & 108  & 4.53    & 0.36    & K         & \nodata & \nodata & \nodata & EC2 & p-emb.bub-cen-trig \\
517 & 380  & \nodata & \nodata & \nodata   & \nodata & \nodata & \nodata & OC2 & few                \\
518 & 900  & 3.40    & 0.51    & S         & 8.32    & 3.83    & 115     & OC2 & exp                \\
519 & 48   & 11.61   & 0.37    & K         & \nodata & \nodata & \nodata & EC1 & emb                \\
520 & 59   & 4.44    & 0.36    & K         & \nodata & \nodata & \nodata & EC2 & p-emb.pah          \\
521 & 84   & 3.67    & 0.47    & K         & \nodata & \nodata & \nodata & EC2 & p-emb.bub-cen-trig \\
522 & 57   & 1.12    & 0.13    & C(ID:525) & \nodata & \nodata & \nodata & EC1 & emb                \\
523 & 300  & 1.30    & 0.19    & S         & 16.7    & 5.8     & 128     & OC2 & exp                \\
524 & 600  & 1.10    & 0.17    & S         & 398     & 183     & 115     & OC2 & exp*               \\
525 & 124  & 1.12    & 0.13    & Ref:212   & \nodata & \nodata & \nodata & EC2 & p-emb.pah          \\
526 & 48   & 4.91    & 0.30    & K         & \nodata & \nodata & \nodata & EC1 & emb.pah            \\

\hline
\end{tabular}
\end{table*}

\begin{table*}
\renewcommand{\arraystretch}{1.1}
\caption{Excerpt of the cluster catalog (Columns 18--25).}
\label{tab:big-catalog3}
\centering
\begin{tabular}{rrclccccc}
\hline\hline
\verb|ID| & \verb|Clump_sep| & \verb|Clump_flag| & \verb|name_Vlsr| & \verb|Vlsr| & \verb|ref_Vlsr| & \verb|KDist| & \verb|e_KDist| & \verb|KDA| \\
 & (\verb|Diam|/2) & & & (\kms) & & (kpc) & (kpc) & \\
\hline
477 & 0.28    & 00 & \nodata                    & \nodata & \nodata & \nodata    & \nodata   & \nodata \\
478 & 0.00    & 11 & G6.80$-$0.25               & 21.33   & 249     & 3.86       & 0.76      & N       \\
479 & 0.13    & 02 & G6.92$-$0.25               & 21.58   & 249     & 3.86       & 0.75      & N       \\
480 & 1.70    & 20 & \nodata                    & \nodata & \nodata & \nodata    & \nodata   & \nodata \\
481 & $>$1.87 & 00 & \nodata                    & \nodata & \nodata & \nodata    & \nodata   & \nodata \\
482 & 1.23    & 00 & \nodata                    & \nodata & \nodata & \nodata    & \nodata   & \nodata \\
483 & 6.75    & 00 & \nodata                    & \nodata & \nodata & \nodata    & \nodata   & \nodata \\
484 & 1.40    & 00 & \nodata                    & \nodata & \nodata & \nodata    & \nodata   & \nodata \\
485 & 6.74    & 00 & \nodata                    & \nodata & \nodata & \nodata    & \nodata   & \nodata \\
486 & 13.01   & 00 & \nodata                    & \nodata & \nodata & \nodata    & \nodata   & \nodata \\
487 & 9.14    & 00 & \nodata                    & \nodata & \nodata & \nodata    & \nodata   & \nodata \\
488 & 0.53    & 02 & G8.48$-$0.98               & 16.00   & 249     & 2.73       & 0.96      & N       \\
489 & 0.00    & 10 & \nodata                    & \nodata & \nodata & \nodata    & \nodata   & \nodata \\
490 & 0.00    & 11 & IRAS 18032$-$2137          & 33.50   & 43      & 4.45       & 0.48      & N       \\
491 & 1.07    & 01 & IRAS 18035$-$2126          & 40.20   & 43      & 4.83/11.43 & 0.40/0.40 & \nodata \\
492 & 0.34    & 00 & \nodata                    & \nodata & \nodata & \nodata    & \nodata   & \nodata \\
493 & 0.00    & 10 & \nodata                    & \nodata & \nodata & \nodata    & \nodata   & \nodata \\
494 & 5.99    & 00 & \nodata                    & \nodata & \nodata & \nodata    & \nodata   & \nodata \\
495 & 7.70    & 00 & \nodata                    & \nodata & \nodata & \nodata    & \nodata   & \nodata \\
496 & 0.00    & 11 & G010.1615$-$00.3623        & 15.80   & 187     & 2.40       & 0.90      & N       \\
497 & 24.34   & 00 & \nodata                    & \nodata & \nodata & \nodata    & \nodata   & \nodata \\
498 & 0.00    & 11 & G10.30$-$0.15              & 12.83   & 249     & 1.99       & 0.97      & N       \\
499 & 0.00    & 11 & G10.32$-$0.16              & 12.02   & 249     & 1.87       & 1.00      & N       \\
500 & 0.00    & 11 & G010.6311$-$00.3864        & $-$2.20 & 187     & $<$0.81    & \nodata   & N       \\
501 & 0.62    & 02 & G011.9019+00.7265          & 24.80   & 187     & 3.10/13.01 & 0.64/0.64 & \nodata \\
502 & 11.26   & 00 & \nodata                    & \nodata & \nodata & \nodata    & \nodata   & \nodata \\
503 & 1.96    & 01 & G12.43$-$1.11              & 39.75   & 249     & 4.15       & 0.43      & N       \\
504 & 13.01   & 01 & G12.43$-$1.11              & 39.75   & 249     & 4.15       & 0.43      & N       \\
505 & 0.00    & 11 & G12.43$-$1.11              & 39.75   & 249     & 4.15       & 0.43      & N       \\
506 & 1.93    & 01 & G12.43$-$1.11              & 39.75   & 249     & 4.15       & 0.43      & N       \\
507 & 5.04    & 00 & \nodata                    & \nodata & \nodata & \nodata    & \nodata   & \nodata \\
508 & 0.00    & 11 & G12.72$-$0.22              & 34.09   & 249     & 3.73       & 0.49      & N       \\
509 & 6.39    & 00 & \nodata                    & \nodata & \nodata & \nodata    & \nodata   & \nodata \\
510 & 1.95    & 24 & W33 C                      & 35.20   & 23      & 3.80       & 0.48      & N       \\
511 & 0.64    & 00 & \nodata                    & \nodata & \nodata & \nodata    & \nodata   & \nodata \\
512 & 0.00    & 11 & G012.8062$-$00.1987        & 34.40   & 187     & 3.73       & 0.49      & N       \\
513 & 0.22    & 11 & G012.8062$-$00.1987        & 34.40   & 187     & 3.73       & 0.49      & N       \\
514 & $>$4.99 & 00 & \nodata                    & \nodata & \nodata & \nodata    & \nodata   & \nodata \\
515 & 0.00    & 11 & G12.91$-$0.26              & 36.73   & 249     & 3.88       & 0.46      & N       \\
516 & 0.00    & 11 & G13.18+0.06                & 48.55   & 249     & 4.53       & 0.36      & N       \\
517 & 0.41    & 00 & \nodata                    & \nodata & \nodata & \nodata    & \nodata   & \nodata \\
518 & $>$1.99 & 00 & \nodata                    & \nodata & \nodata & \nodata    & \nodata   & \nodata \\
519 & 0.00    & 11 & G13.66$-$0.60              & 47.01   & 249     & 11.61      & 0.37      & F       \\
520 & 0.00    & 11 & G13.87+0.28                & 48.52   & 249     & 4.44       & 0.36      & N       \\
521 & 0.00    & 14 & $[$L89b$]$ 13.998$-$00.128 & 36.00   & 127     & 3.67       & 0.47      & N       \\
522 & 0.00    & 11 & G14.12$-$0.57              & 20.11   & 249     & 2.33       & 0.70      & N       \\
523 & 7.35    & 00 & \nodata                    & \nodata & \nodata & \nodata    & \nodata   & \nodata \\
524 & 1.09    & 01 & G14.31$-$0.19              & 38.82   & 249     & 3.82       & 0.44      & N       \\
525 & 0.00    & 11 & G14.33$-$0.64              & 21.97   & 249     & 2.48       & 0.67      & N       \\
526 & 0.00    & 11 & G14.89$-$0.40              & 61.23   & 249     & 4.91       & 0.30      & N       \\

\hline
\end{tabular}
\end{table*}

\begin{table*}
\renewcommand{\arraystretch}{1.1}
\caption{Excerpt of the cluster catalog (Columns 26--32).}
\label{tab:big-catalog4}
\centering
\begin{tabular}{rlcccccl}
\hline\hline
\verb|ID| & \verb|ref_KDA| & \verb|SDist| & \verb|e_SDist| & \verb|ref_Sdist| & \verb|ref_Dias| & \verb|ref_Conf| & \verb|HII_reg| \\
 & & (kpc) & (kpc) & & & & \\
\hline
477 & \nodata               & 0.55    & 0.08    & 115     & 115     & 115     & \nodata                      \\
478 & C,D                   & \nodata & \nodata & \nodata & \nodata & \nodata & \nodata                      \\
479 & C                     & 0.82    & 0.12    & 128     & 128     & 193     & M20                          \\
480 & \nodata               & \nodata & \nodata & \nodata & \nodata & \nodata & \nodata                      \\
481 & \nodata               & 0.94    & 0.14    & 128     & 128     & 128     & \nodata                      \\
482 & \nodata               & 1.21    & 0.18    & 128     & 128     & 128     & \nodata                      \\
483 & \nodata               & \nodata & \nodata & \nodata & \nodata & \nodata & \nodata                      \\
484 & \nodata               & 1.72    & 0.41    & 37      & 37      & 37      & \nodata                      \\
485 & \nodata               & \nodata & \nodata & \nodata & \nodata & \nodata & \nodata                      \\
486 & \nodata               & \nodata & \nodata & \nodata & \nodata & \nodata & \nodata                      \\
487 & \nodata               & \nodata & \nodata & \nodata & \nodata & \nodata & \nodata                      \\
488 & D,z                   & 2.50    & 0.38    & 116     & 116     & 116     & \nodata                      \\
489 & \nodata               & \nodata & \nodata & \nodata & \nodata & \nodata & \nodata                      \\
490 & 186                   & \nodata & \nodata & \nodata & \nodata & \nodata & $[$L89b$]$ 8.666$-$00.351    \\
491 & \nodata               & 1.60    & 0.24    & 36      & \nodata & 36      & \nodata                      \\
492 & \nodata               & 3.47    & 0.52    & 115     & 115     & 115     & \nodata                      \\
493 & \nodata               & \nodata & \nodata & \nodata & \nodata & \nodata & \nodata                      \\
494 & \nodata               & \nodata & \nodata & \nodata & \nodata & \nodata & \nodata                      \\
495 & \nodata               & 8.86    & 1.61    & 15      & \nodata & 15,80   & \nodata                      \\
496 & S,187,(216)           & 3.55    & 1.87    & 148     & \nodata & 30,89   & G10.2$-$0.3                  \\
497 & \nodata               & 3.50    & 0.53    & 86      & \nodata & 86      & \nodata                      \\
498 & 187,(216)             & \nodata & \nodata & \nodata & \nodata & 25      & G10.3$-$0.1                  \\
499 & S,162,(216)           & 2.39    & 1.30    & 148     & \nodata & 25      & G10.3$-$0.1                  \\
500 & 162,(187)             & \nodata & \nodata & \nodata & \nodata & \nodata & $[$L89b$]$ 10.617$-$00.384   \\
501 & \nodata               & \nodata & \nodata & \nodata & \nodata & \nodata & \nodata                      \\
502 & \nodata               & 1.47    & 0.22    & 128     & 128     & 128     & \nodata                      \\
503 & 187,z                 & 1.19    & 0.18    & 228     & 228     & 228     & \nodata                      \\
504 & 187,z                 & 1.79    & 0.27    & 237     & 237     & 237     & \nodata                      \\
505 & 187,z                 & \nodata & \nodata & \nodata & \nodata & \nodata & RMS G012.4317$-$01.1112 (UC) \\
506 & 187,z                 & 2.33    & 0.35    & 237     & 237     & 237     & \nodata                      \\
507 & \nodata               & \nodata & \nodata & \nodata & \nodata & \nodata & \nodata                      \\
508 & C                     & \nodata & \nodata & \nodata & \nodata & \nodata & W33 f                        \\
509 & \nodata               & 1.48    & 0.22    & 128     & 128     & 128     & \nodata                      \\
510 & C                     & \nodata & \nodata & \nodata & \nodata & \nodata & W33 c                        \\
511 & \nodata               & 3.60    & 0.70    & 144     & \nodata & 141,144 & \nodata                      \\
512 & 224,187,162           & \nodata & \nodata & \nodata & \nodata & \nodata & W33 g                        \\
513 & C                     & \nodata & \nodata & \nodata & \nodata & \nodata & W33 h                        \\
514 & \nodata               & 3.60    & 0.54    & 16      & 16      & 16      & \nodata                      \\
515 & C,224,187,(162),(216) & \nodata & \nodata & \nodata & \nodata & \nodata & W33A (UC)                    \\
516 & 162,72                & \nodata & \nodata & \nodata & \nodata & \nodata & $[$L89b$]$ 13.186+00.045     \\
517 & \nodata               & \nodata & \nodata & \nodata & \nodata & \nodata & \nodata                      \\
518 & \nodata               & 3.40    & 0.51    & 115     & 115     & 115     & \nodata                      \\
519 & 98,(D),(149)          & \nodata & \nodata & \nodata & \nodata & 76      & \nodata                      \\
520 & 54,216                & \nodata & \nodata & \nodata & \nodata & \nodata & $[$L89b$]$ 13.875+00.282     \\
521 & 72                    & \nodata & \nodata & \nodata & \nodata & \nodata & $[$L89b$]$ 13.998$-$00.128   \\
522 & C,136,D               & \nodata & \nodata & \nodata & \nodata & \nodata & \nodata                      \\
523 & \nodata               & 1.30    & 0.19    & 128     & 128     & 128     & \nodata                      \\
524 & 187*                  & 1.10    & 0.17    & 115     & 115     & 115     & \nodata                      \\
525 & 212,187,216,D         & \nodata & \nodata & \nodata & \nodata & \nodata & G14.33$-$00.64               \\
526 & 149                   & \nodata & \nodata & \nodata & \nodata & 76      & Mol 50 (UC)                  \\

\hline
\end{tabular}
\end{table*}

\begin{table*}
\renewcommand{\arraystretch}{1.1}
\caption{Excerpt of the cluster catalog (Columns 33--38).}
\label{tab:big-catalog5}
\centering
\begin{tabular}{rllcclc}
\hline\hline
\verb|ID| & \verb|Bub| & \verb|IRDC| & \verb|no_GL| & \verb|SubCl| & \verb|Complex| & \verb|ref_Complex| \\
\\
\hline
477 & \nodata     & \nodata               & 0 & \nodata & \nodata       & \nodata \\
478 & \nodata     & MSXDC G006.81$-$00.25 & 0 & \nodata & Trifid Nebula & 125     \\
479 & CN88        & \nodata               & 0 & 6       & Trifid Nebula & 125     \\
480 & CN95        & \nodata               & 0 & 6       & Trifid Nebula & 125     \\
481 & \nodata     & \nodata               & 1 & \nodata & \nodata       & \nodata \\
482 & \nodata     & \nodata               & 0 & \nodata & \nodata       & \nodata \\
483 & \nodata     & \nodata               & 0 & 7       & MC-09         & \nodata \\
484 & \nodata     & \nodata               & 0 & 7       & MC-09         & \nodata \\
485 & \nodata     & \nodata               & 0 & \nodata & \nodata       & \nodata \\
486 & \nodata     & \nodata               & 0 & \nodata & \nodata       & \nodata \\
487 & \nodata     & \nodata               & 0 & \nodata & \nodata       & \nodata \\
488 & \nodata     & \nodata               & 0 & \nodata & \nodata       & \nodata \\
489 & \nodata     & MSXDC G008.47$-$00.61 & 0 & \nodata & \nodata       & \nodata \\
490 & CN120       & \nodata               & 0 & \nodata & \nodata       & \nodata \\
491 & \nodata     & \nodata               & 0 & \nodata & \nodata       & \nodata \\
492 & \nodata     & \nodata               & 0 & \nodata & \nodata       & \nodata \\
493 & CN127       & SDC G009.220+0.169    & 0 & \nodata & \nodata       & \nodata \\
494 & \nodata     & \nodata               & 0 & \nodata & \nodata       & \nodata \\
495 & \nodata     & \nodata               & 0 & \nodata & \nodata       & \nodata \\
496 & CN143       & \nodata               & 0 & \nodata & W31           & 118,14  \\
497 & \nodata     & \nodata               & 0 & \nodata & \nodata       & \nodata \\
498 & CN148       & \nodata               & 0 & 8       & W31           & 118,14  \\
499 & CN148       & \nodata               & 0 & 8       & W31           & 118,14  \\
500 & N2          & \nodata               & 0 & \nodata & W31           & 118,14  \\
501 & \nodata     & \nodata               & 0 & \nodata & \nodata       & \nodata \\
502 & \nodata     & \nodata               & 0 & \nodata & \nodata       & \nodata \\
503 & \nodata     & \nodata               & 1 & \nodata & \nodata       & \nodata \\
504 & \nodata     & \nodata               & 1 & \nodata & \nodata       & \nodata \\
505 & N5          & \nodata               & 0 & \nodata & \nodata       & \nodata \\
506 & \nodata     & \nodata               & 1 & \nodata & \nodata       & \nodata \\
507 & \nodata     & \nodata               & 0 & \nodata & \nodata       & \nodata \\
508 & \nodata     & \nodata               & 0 & \nodata & W33           & 23,225  \\
509 & \nodata     & \nodata               & 0 & \nodata & \nodata       & \nodata \\
510 & Bub(ID:510) & \nodata               & 0 & \nodata & W33           & 23,225  \\
511 & \nodata     & \nodata               & 0 & \nodata & W33           & 23,225  \\
512 & \nodata     & \nodata               & 0 & \nodata & W33           & 23,225  \\
513 & Bub(ID:513) & \nodata               & 0 & \nodata & W33           & 23,225  \\
514 & \nodata     & \nodata               & 1 & \nodata & \nodata       & \nodata \\
515 & \nodata     & \nodata               & 0 & \nodata & W33           & 23,225  \\
516 & N10         & \nodata               & 0 & \nodata & \nodata       & \nodata \\
517 & \nodata     & \nodata               & 0 & \nodata & \nodata       & \nodata \\
518 & \nodata     & \nodata               & 0 & \nodata & \nodata       & \nodata \\
519 & \nodata     & MSXDC G013.68$-$00.60 & 0 & \nodata & \nodata       & \nodata \\
520 & \nodata     & \nodata               & 0 & \nodata & \nodata       & \nodata \\
521 & N14         & \nodata               & 0 & \nodata & \nodata       & \nodata \\
522 & \nodata     & MSXDC G014.15$-$00.55 & 0 & \nodata & KC-24         & \nodata \\
523 & \nodata     & \nodata               & 0 & \nodata & \nodata       & \nodata \\
524 & \nodata     & \nodata               & 0 & \nodata & \nodata       & \nodata \\
525 & \nodata     & SDC G014.333$-$0.646  & 0 & \nodata & KC-24         & \nodata \\
526 & \nodata     & \nodata               & 0 & \nodata & \nodata       & \nodata \\

\hline
\end{tabular}
\end{table*}

\end{appendix}

\end{document}